\documentclass[preprint,3p,times]{elsarticle} 
\biboptions{sort&compress}

\usepackage[latin1]{inputenc}
\pdfoutput=1
\usepackage{bm}
\usepackage{multirow, amsmath, amssymb, amsfonts, array}
\usepackage{color}
\usepackage{bm,graphicx,cancel}
\usepackage{calc}
\usepackage{dcolumn}
\usepackage{mathrsfs}
\usepackage[normalem]{ulem} %temporary
\usepackage{xspace}
\usepackage{spverbatim}
\usepackage{fancyvrb}
\usepackage{xcolor}
\usepackage{hyperref}

\newcommand{\rmd}{\mathrm{d}}
\newcommand{\GeV}{{\rm\ GeV}}

\newcommand{\TeV}{{\rm\ TeV}}
\newcommand{\sigmav}{\langle \sigma v \rangle} 
\newcommand{\maddm}{\texttt{MadDM}\xspace} 
\newcommand{\madgraph}{\texttt{MG5\_aMC}\xspace} 
\newcommand{\pythia}{\texttt{Pythia 8}\xspace}
\newcommand{\PPPC}{\texttt{PPPC4DMID}\xspace}
\newcommand{\MN}{\texttt{MultiNest}\xspace}
\newcommand{\PyMN}{\texttt{PyMultiNest}\xspace}
\newcommand{\dragon}{\texttt{DRAGON}\xspace}
\newcommand{\eg}{{\it e.g.}\xspace}
\newcommand{\ie}{{\it i.e.}\xspace}

\def\bvec#1{\textrm{\boldmath $#1 $}}

\def\slashb#1{\setbox0=\hbox{$#1$}#1\hskip-\wd0\dimen0=5pt\advance
        \dimen0 by-\ht0\advance\dimen0 by\dp0\lower0.5\dimen0\hbox
          to\wd0{\hss\sl/\/\hss}}
\begin{document}

\journal{}

\begin{frontmatter}

\title{MadDM v.3.0: a Comprehensive Tool for Dark Matter Studies}
\author[CP3]{Federico Ambrogi}
\author[CP3]{Chiara Arina}
\author[CP3,B12]{Mihailo Backovi\'c}
\author[RWTH]{Jan Heisig}
\author[CP3]{Fabio Maltoni}
\author[CP3]{Luca Mantani}
\author[CP3]{\\Olivier Mattelaer}
\author[KS,BNL]{Gopolang Mohlabeng}

\address[CP3]{Centre for Cosmology, Particle Physics and Phenomenology (CP3), Universit\'e catholique de Louvain, B-1348 Louvain-la-Neuve, Belgium}
\address[B12]{B12, Chemin du Cyclotron 6, B-1348 Louvain-la-Neuve, Belgium}
\address[RWTH]{Institute for Theoretical Particle Physics and Cosmology, RWTH Aachen University, D-52056 Aachen, Germany}
\address[KS]{Department of Physics and Astronomy, University of Kansas, Lawrence KS 66045, USA}
\address[BNL]{Physics Department, Brookhaven National Laboratory, Upton, New York 11973, USA}

\begin{abstract}
We present \maddm v.3.0, a numerical tool to compute particle dark matter observables in generic new physics models. The new version features a comprehensive and automated framework for dark matter searches at the interface of collider physics, astrophysics and cosmology and is deployed as a plugin of the \texttt{MadGraph5\_aMC@NLO} platform, inheriting most of its features. With respect to the previous version, \maddm v.3.0 can now provide predictions for indirect dark matter signatures in astrophysical environments, such as the annihilation cross section at present time and the energy spectra of prompt photons, cosmic rays and neutrinos resulting from dark matter annihilation. \maddm indirect detection features support both $2\to2$ and $2 \to n$ dark matter annihilation processes. In addition, the ability to compare theoretical predictions with experimental constraints is extended by including the Fermi-LAT likelihood for gamma-ray constraints from dwarf spheroidal galaxies  and by providing an interface with the nested sampling algorithm \PyMN to perform high dimensional parameter scans efficiently. We validate the code for a wide set of dark matter models by comparing the results from \maddm v.3.0 to existing tools and results in the literature.
\end{abstract}


\end{frontmatter}

\section{Introduction}

There is compelling evidence for the existence of dark matter from observations of the cosmic microwave background (CMB)~\cite{Ade:2015xua} and from astrophysical measurements~\cite{Bertone:2016nfn}. This evidence, however, is still indirect and stems only from the gravitational interaction of dark matter, leaving its true nature unknown. Within the generic  hypothesis that dark matter is a particle,  a multitude of experimental approaches have been undertaken to detect it; from dark matter searches using the underground detectors~\cite{Akerib:2016vxi,Aprile:2017iyp,Amole:2017dex} (direct detection), to observations of gamma-ray, cosmic-ray and neutrino fluxes in astrophysical environments (indirect detection),~\eg, Refs.~\cite{Fermi-LAT:2016uux,PhysRevLett.117.091103,dampe,Aartsen:2013dxa}, and searches for missing energy signals at colliders~\cite{Albert:2017onk} (production). Yet, despite the enormous experimental effort,  a signal of particle dark matter remains elusive and our understanding of dark matter properties  limited.

In the absence of a clear dark matter signal, a plethora of theoretical dark matter models are currently viable and span many orders of magnitude in dark matter masses and interaction strengths. This makes it difficult to efficiently study all possible scenarios. In order to at least carve out the dark matter models which are inconsistent with experimental observations, it is necessary to combine the maximum amount of available experimental information (\ie the comprehensive dark matter studies). Including  constraints from many dark matter detection strategies, from particle physics, to astrophysics and cosmology, is  technically challenging and computationally intensive. This has created a demand for numerical tools that would facilitate and streamline computations, making them efficient and model independent. In the recent years,  several tools and frameworks have been proposed and developed that compute dark matter model predictions for relic density, direct and indirect detection, among which is \maddm. Initially, most of the attention was focused on supersymmetric models with the neutralino as dark matter candidate~\cite{Belanger:2001fz,Baer:2002fv,Gondolo:2004sc,Arbey:2009gu}. \texttt{MicrOMEGAs}~\cite{Belanger:2006is} has been the first tool for dark matter studies to allow for the computation of dark matter predictions for generic dark matter models. Nowadays, \texttt{MicrOMEGAs}~\cite{Belanger:2018ccd} supplies astrophysical and cosmological dark matter predictions and allows the comparison of dark matter signals with collider bounds for a generic model given a point in the available parameter space. More recently,  the \texttt{GAMBIT} collaboration released the first version of a module to compute dark matter observables~\cite{Workgroup:2017lvb} (interfacing with \texttt{MicrOMEGAs} and \texttt{DarkSUSY}) and to compare with experimental likelihoods, as well as a module for sampling the model parameter space~\cite{Workgroup:2017htr}. Finally,  \texttt{DarkSUSY} has very recently released a new code version that goes beyond supersymmetry and allows the implementation of generic dark matter models~\cite{Bringmann:2018lay}.

\begin{figure*}[t!]
\centering
\includegraphics[width=0.8\textwidth,trim=50mm 38mm 50mm 26mm, clip]{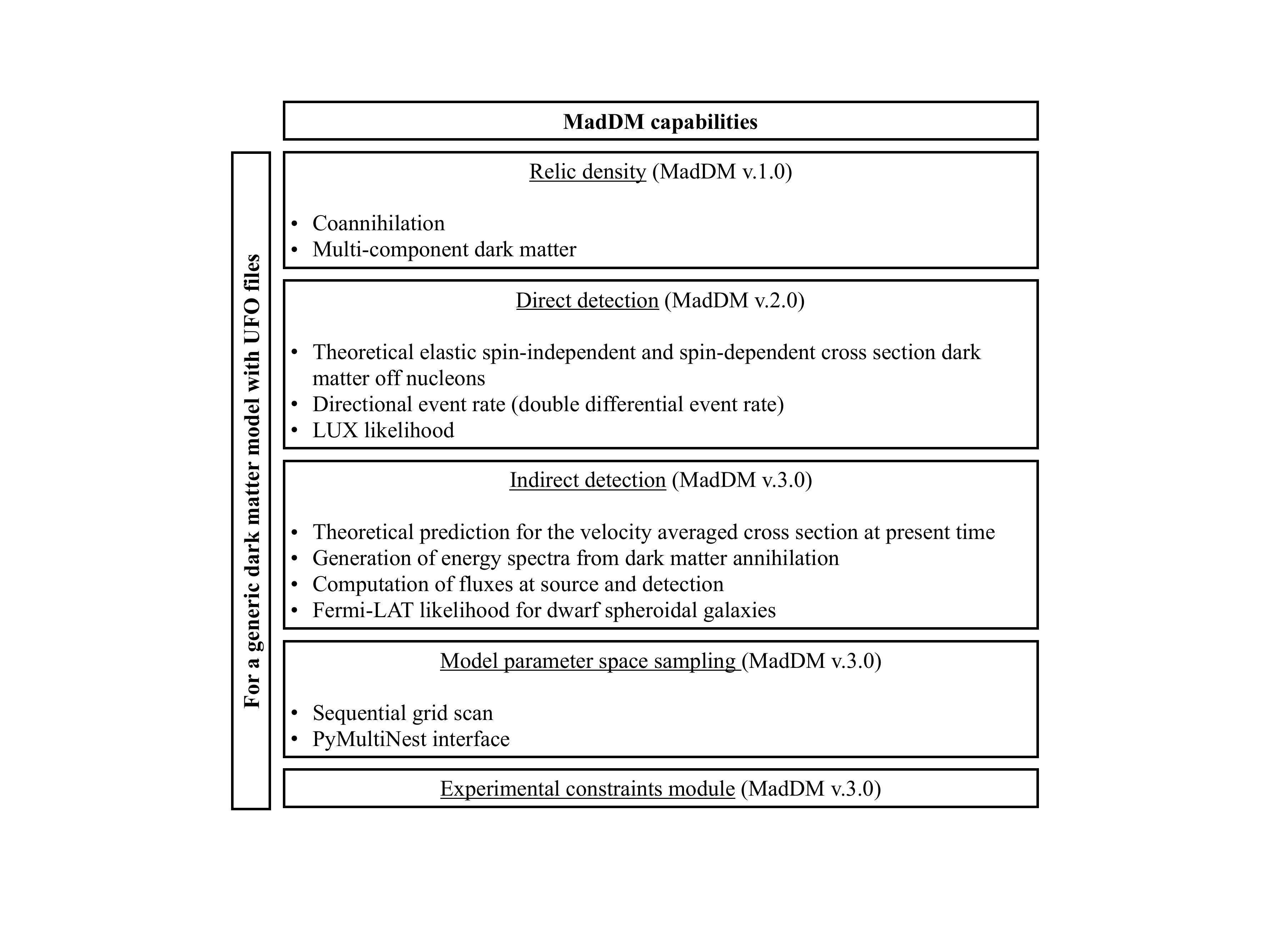}
\caption{Overview of the full capabilities of \maddm.}\label{fig:table1}
\end{figure*}
With similar motivations and goals in mind, \maddm contributes to the effort of providing the necessary tools to the dark matter community. Since its first release in 2013~\cite{Backovic:2013dpa}, the ambition of the project has been not only to provide specific tools for the computation of dark matter observables but to build a fully-fledged and flexible platform together with \texttt{MadGraph5\_aMC@NLO} (\madgraph)~\cite{Alwall:2011uj,Alwall:2014hca} to enable both theorists and experimentalists to perform global fits of generic dark matter models and, in particular, directly exploit all the available technology developed for generic searches of physics beyond the standard model at colliders. The first \maddm version~\cite{Backovic:2013dpa} provided a numerical tool to compute relic density signals together with collider observables for any dark matter model in the \texttt{UFO} (Universal FeynRules Output~\cite{Alloul:2013bka}) format. The second version~\cite{Backovic:2015cra} added the ability to perform the computation of dark matter-nucleon cross sections, dark matter double differential event rates of nuclear recoils for a generic experiment, as well as the LUX experimental likelihood to compare with data. The new version presented here extends the functionality of \maddm to indirect detection, parameter scanning features and integration of experimental limits, hence advancing the original goal of creating a platform for comprehensive dark matter studies. A schematic of the current capabilities of \maddm is given in Fig.~\ref{fig:table1}. 

\begin{figure*}[t!]
\centering
\includegraphics[width=0.9\textwidth,trim=15mm 80mm 15mm 27mm, clip]{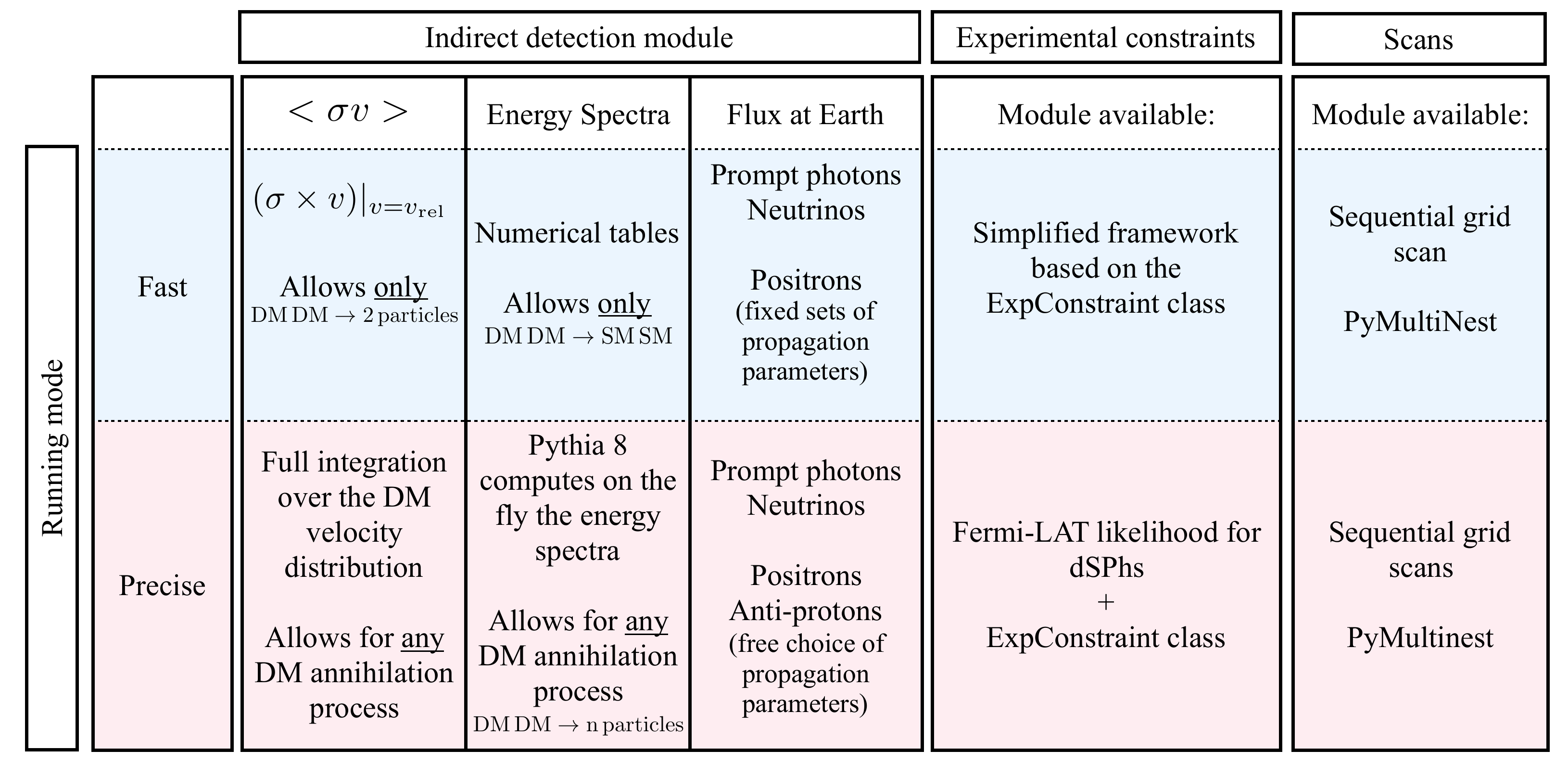}
\caption{\maddm v.3.0: Schematic overview of the new modules with their main features and their link to the `fast' and `precise' running modes. }\label{fig:table}
\end{figure*}

The new features of v.3.0 can be summarised as follows:
\begin{enumerate}
\item  {\it Inheritance from the \madgraph platform}. \maddm v.3.0 is now a \madgraph plugin and therefore automatically integrates all  features of the \madgraph architecture.
\item  {\it Ability to perform calculations of dark matter observables relevant for indirect detection.} The necessary steps to predict the flux of gamma rays, neutrinos and charged particles at the Earth coming from the annihilation of dark matter particles in the Milky Way or in neighboring galaxies are available. The velocity averaged dark matter annihilation cross section ($\sigmav$) can be computed choosing among three different algorithms. The first two methods are equivalent to the simple approximation where $\sigmav$ is evaluated at the fixed velocity characteristic for the respective astrophysical environment, while the third method takes into account the actual velocity distribution of the initial state dark matter particles. Energy spectrum of gamma rays, positrons, anti-protons and neutrinos coming from dark matter annihilation at the production point can be determined via a parton level event generation and then passing the events through  \pythia (\texttt{Pythia\_8.2}~\cite{Sjostrand:2014zea}) for showering and hadronisation. By using the matrix element generator of \madgraph it is also possible to automatically compute the energy spectra for dark matter annihilating into generic $n$ final state particles, which can also belong to theories beyond the SM (BSM final states). To the best of our knowledge \maddm v.3.0 is the only public code with this unique capability. Finally, the code provides the expected flux at the Earth position for gamma rays and for neutrinos coming from distant sources. The cosmic-ray propagation is performed with the numerical code \dragon~\cite{Evoli:2008dv}, to which \maddm provides a user friendly interface.
\item  {\it Inclusion of indirect detection and further direct detection experimental constraints into the code.} More specifically, a simple functionality is provided for testing model points against direct and indirect experimental constraints, whereby the user can choose the upper bounds on elastic scattering cross sections or the upper bound on $\sigmav$ for which the model is consistent with the experiments. As a result, the code automatically compares the output of the calculation to the constraints specified by the user  and determines whether the specific model point is allowed or not. For the case of gamma-ray signals, besides the simplified framework,  the Fermi-LAT likelihoods of Dwarf Spheroidal Galaxies (dSphs)~\cite{Fermi-LAT:2016uux} are provided to precisely confront the model with the flux observed in those Milky Way satellites.
\item {\it Inclusion of advanced, multi-dimensional parameter sampling abilities.} A user friendly interface with the Python version of the nested sampling public algorithm \MN (\texttt{MultiNest}~\cite{Feroz:2007kg,Feroz:2008xx} is provided together with \texttt{PyMultiNest}~\cite{Buchner:2014nha}) that allows the user to perform sampling of the model parameter space in an efficient way, even for large dimensionality of the parameter space. To this end the user interface includes the experimental likelihoods to constrain the dark matter model. In case of a reduced number of parameters, a grid scanning mode is also available, which is directly inherited from the \madgraph platform. 
\item {\it Flexibility}. \maddm allows the user to run the indirect detection module in two ways, depending on the desired output.  It features a `fast' running mode particularly suitable for large sampling of the parameter space and a `precise' mode, where each step of the indirect detection predictions is computed with greatest accuracy for studies of benchmark model points. Notice that the `fast' running mode is available only if the dark matter annihilates directly into a pair of Standard Model (SM) particles. For processes other than $2 \to 2$ only the `precise' running mode is available.  It is also possible to define settings that mix both `fast' and `precise' modes. 
\end{enumerate}
We summarise the main features of the new release of \maddm in Fig.~\ref{fig:table}, together with their interconnections.

The remainder of the paper presents the new features of \maddm and is organised as follows. The next section describes the main ingredients necessary to compute dark matter indirect detection signals and how they have been implemented in the numerical code. Section~\ref{sec:lnlk} summarizes how experimental constraints are taken into account within \maddm,  while Sec.~\ref{sec:parscan} discusses the two ways to perform model parameter sampling within \maddm. Section~\ref{sec:valid} provides applications to physically motivated dark matter models. In particular, we demonstrate the new capabilities of \maddm. Conclusions and future prospects are presented in Sec.~\ref{sec:concl}. All technical details on how to install and run the code in the various modes presented in the paper are described in the Appendices. More specifically~\ref{sec:app2} describes the most important inherited features from the \madgraph platform while~\ref{sec:app1} explains how to install and run the code. Details on the indirect detection capabilities are given in~\ref{sec:app3}, while~\ref{sec:app4} provides details on the Python module for experimental constraints. The commands to perform sequential grid scans or \PyMN runs are described in~\ref{sec:app5}.

\section{Indirect Detection of annihilating Dark Matter}\label{sec:ID}

Indirect detection looks for products of dark matter annihilation in astrophysical environment where the dark matter is denser. For instance, typical benchmarks for gamma-ray searches are the dSphs~\cite{Fermi-LAT:2016uux} or the Galactic Center~\cite{Ackermann:2015lka,Abdalla:2016olq}. For a review on dark matter indirect detection we refer the reader to~\cite{Cirelli:2012tf,Gaskins:2016cha}.

There are three main ingredients necessary to compute predictions for dark matter models and to compare with data: (i) The annihilation cross section $\sigmav$ computed for the environment where the annihilation takes place; this element is discussed in the next section. (ii) The energy spectra ${\rmd} N/{\rmd} E$ of prompt photons, positrons, anti-protons and neutrinos generated at source by the dark matter annihilation products; this will be described in Sec.~\ref{sec:pythia}. (iii) The $J$ factor which depends on the dark matter distribution and is defined as the integral along the line-of-sight (los) of the dark matter density profile squared in a specific sky direction. This quantity will be defined in general for gamma rays and neutrinos. We will further describe how \maddm computes the flux near the Earth for each type of propagated final state particles: prompt photons are detailed in Sec.~\ref{sec:gammas}, charged cosmic rays are discussed in Sec.~\ref{sec:pbep} and neutrinos are provided in Sec.~\ref{sec:nus}.  For the rest of the paper we assume a generic dark matter particle called $\chi$ with mass $m_
\chi$, unless stated otherwise.

\subsection{Computation of $\sigmav$ in astrophysical environments}\label{sec:sigmav}

The computation of the full $\sigmav$ at present time, which might contain $p$-wave terms, calls for the inclusion of the dark matter distribution. In this case $\sigmav$ results in this velocity-weighted expression:
\begin{equation}
\sigmav =   \int\!\!{\rmd}^3 \bvec v_1 {\rmd}^3\bvec v_2P_{\!\bvec r}(\bvec v_1)P_{\!\bvec r}(\bvec v_2) \,\sigma v_{\rm rel}\,,
\end{equation}
where $\bvec v_i$ are the velocities of the two incoming dark matter particles, and $P_{\!\bvec r}(\bvec v_i)$ is the velocity distribution function of the dark matter
at a position $\bvec r$. This can be rearranged as~\cite{Robertson:2009bh,Ferrer:2013cla}:
\begin{eqnarray}\label{eq:sigmav}
\sigmav   = \int\!\! {\rmd} v_{\rm rel} \, \tilde P_{\!\bvec{r},\rm{rel}}( v_{\rm rel})\, \sigma v_{\rm rel}\,,
\end{eqnarray}
with
\begin{equation}\label{eq:vrel}
\tilde P_{\!\bvec{r},\rm{rel}}(v_{\rm rel})\equiv 4\pi v_{\rm rel}^2 \int\!\! {\rmd}^3v_{\rm CM} 
P_{\!\bvec r}(\bvec v_{\rm CM}+\bvec v_{\rm rel}/2)P_{\!\bvec r}(\bvec v_{\rm CM}-\bvec v_{\rm rel}/2) \,,
\end{equation}
where $\bvec v_{\rm CM} \equiv(\bvec v_1+
 \bvec v_2)/2 $ is the velocity in the center-of-mass frame and $\bvec v_{\rm rel} \equiv\bvec v_1-\bvec v_2$ is the relative velocity. 
For a Maxwell-Boltzmann distribution 
$P_{\!\bvec{r}}( v)= \pi^{-3/2}v_0^{-3}\exp(-v^2/v_0^2)$ with most probable velocity $v_0$, the relative velocity
also follows a Maxwell-Boltzmann distribution with most probable velocity $\sqrt{2}v_0$:
\begin{equation}\label{eq:mbdist}
 \tilde P_{\!\bvec{r},\rm{rel}}(v_{\rm rel})=\sqrt{\frac{2}{\pi}}\frac{v_{\rm rel}^2}{v_0^3}\exp\left(-\frac{v_{\rm rel}^2}{2v_0^2}\right)\,.
 \end{equation} 
For a cross section dominated by $p$-wave annihilation, $\sigma v_{\rm rel}\sim b v_{\rm rel}^2$, where $b$ is a
constant, $\sigmav = 3 b v_0^2$. Hence, for cross sections that can be well approximated by the sum of their $s$- and $p$-wave contribution, the velocity averaging is equivalent to the evaluation of $\sigma v_{\rm rel}$ at $v_{\rm rel}=\sqrt{3} v_0$.
In \maddm we consider only the case of Maxwell-Boltzmann velocity distribution. 
 
The velocity averaged annihilation cross section can be computed with two main methods in \maddm: `fast' and `precise', which are described in the following. Further details can be found in~\ref{sec:appsigmav}.
 
 \paragraph{`Fast' running mode} This method has the advantage of being very fast, with an accuracy of about $(10-20)\%$ with respect to the full integral in Eq.~\eqref{eq:sigmav} and to the `precise' method. It consists of computing the leading order $2 \to 2$ matrix elements for the annihilation process(es) and integrate them over the angle between the two final states. The resulting cross section is furthermore evaluated at the required velocity, which is described by a $\delta$ distribution function centred on that specific velocity. This simple evaluation makes this mode a good default choice for extensive model parameter sampling.

The approximated integration over the final state phase-space is allowed only for two initial dark matter particles annihilating into two final state particles. There is an additional caveat if the user wants to compute the predicted flux of for instance photons with this method: this option does not produce events for the annihilation process, hence the computation of the energy spectra can proceed only via the `fast' option, described in the next section, Sec.~\ref{sec:pythia} and in~\ref{sec:appes}, which is available only for final state particles belonging to the SM. 

\paragraph{`Precise' running mode}  This mode incorporates two methods taken from the \madgraph platform: \verb|madevent| and \verb|reshuffling|. Both methods use the event generator \texttt{MadEvent}~\cite{Maltoni:2002qb}. Given the annihilation process(es), \madgraph identifies all the relevant subprocesses, generates both the amplitudes and the mappings needed for an efficient integration over the full phase-space, and passes them to \texttt{MadEvent}. As a result, a process-specific, stand-alone code is produced that allows the user to calculate $\sigmav$ and generate unweighted events in the standard output format (LHE file).
In the method \verb|madevent| annihilation processes are computed at the center-of-mass energy given by 
$\sqrt{s} = 2 m_\chi\big(1 + 1/8\,  v_{\rm rel}^2\big)$ where $v_{\rm rel}=\sqrt{3} v_0$ as discussed above. 

The \verb|reshuffling| option works similarly to the \verb|madevent| method. Once the events have been generated following the $\delta$ distribution for the velocity, the algorithm applies a reshuffling~\cite{Kleiss:1985gy} of the kinematic and of the weight of each event to map a Maxwell-Boltzmann velocity distribution. Additionally it also applies re-weighting~\cite{Mattelaer:2016gcx} of the matrix elements in order to check if those have still the same weight or it has changed. For instance, due to the improved kinematics, an annihilation channel that might have been below threshold, hence suppressed, may now be above threshold and be largely enhanced. This has the consequence of changing the weight of each single amplitude. We have checked that this method is an accurate approximation for the integration over the relative velocity of Eq.~\eqref{eq:sigmav}, which is numerically less stable and slower. 

The \verb|madevent| method gives the same result as the \verb|reshuffling| option, excepts in the case of very light dark matter particles, for which the small velocity dispersion might play a role, or in case of thresholds effects. For instance, if the dark matter mass is very close in mass to the SM final state to which it is annihilating into, high velocity particles belonging to the maxwellian tail can enhance the cross section. In those cases the reshuffling method is more accurate. Therefore the reshuffling method is set by default. The user can switch to \verb|madevent|, which is faster, being aware of the caveats explained before. Both methods have been tested for velocities as low as $v \sim 10^{-6}$ and provide reliable results, whereas we do not guarantee the code to be accurate enough for smaller velocities (\ie~at CMB epoch, $v \sim 10^{-7}$). At present, to the best of our knowledge, such precise computation of $\sigmav$ is a unique feature of \maddm v.3.0.

This method works to compute automatically any possible leading order (LO) final annihilation state in a given dark matter model (ideally $\chi \chi \to n$ particles if kinematically possible). The \madgraph platform is able to perform automatic next to leading order (NLO) calculation: this feature should be inherited automatically by \maddm, however its full testing is kept for a future release. The ability to automatically compute loop induced processes would be a great addition to \maddm, as it will allow the user to easy evaluate dark matter annihilation into a pair of photons or $\gamma Z$ or $\gamma h$ (the so-called smoking gun signatures for dark matter), for which the Fermi-LAT satellite is setting strong exclusion bounds~\cite{Ackermann:2015lka}. At the moment such loop induced processes need to be evaluated analytically within the specific dark matter model, while there are attempts to analytically provide systematic calculations valid for the most popular dark matter candidates~\cite{Garcia-Cely:2016hsk}. A fully automatised numerical procedure for any dark matter model is yet a missing block within the dark matter tools world.

\subsection{Energy spectra from dark matter annihilation}\label{sec:pythia}

Dark matter particles can annihilate into all possible SM final states that are kinematically open. The specific final states of course depend on the detailed properties of the dark matter model. To introduce our implementation, we start with illustrating the standard implementation available in several public tools~\cite{Belanger:2004yn,Gondolo:2004sc,Cirelli:2010xx}, which is the annihilation into pairs of SM particles. This is described by a $2 \to 2$ process: 
\begin{eqnarray}
 \chi \chi & \to & gg,q\bar{q}, c\bar{c}, b\bar{b}, t\bar{t}, e^+e^-, \mu^+\mu^-, \tau^+\tau^-, \nu_e \bar{\nu}_e, \nu_\mu \bar{\nu}_\mu, \nu_\tau \bar{\nu}_\tau, ZZ, W^+ W^-, hh\,,
 \end{eqnarray}\label{eq:smfs}
where $q$ designs collectively the $u,d$ and $s$ quarks and a branching ratio of 100\% into one particle species is assumed.

The standard procedure to compute the energy spectrum of stable particles $i = \gamma, e^+, \bar{p}, \nu_e, \nu_\mu, \nu_\tau$~\footnote{It is a common choice to give the energy spectrum of positrons and anti-protons instead of electrons or protons (even though they are equal unless the dark matter model has a weird symmetry) because the former are subject to a lower background in astrophysical environments.} (and anti-neutrinos) at the production point, ${\rmd }N / {\rmd} \log x_i$ (with $x_i \equiv E_i / m_\chi$ and $E_i$ is the energy of species $i$), is obtained by making decay, shower and hadronise the SM particles via Monte Carlo simulation tools. The annihilation process occurs in the galactic halo or in nearby galaxies, where typically the velocity of dark matter is very small ($v_{0} \sim 220$ km/s or lower); hence the annihilation can be considered at rest with a center of mass energy provided by twice the dark matter mass $\sqrt{s} = 2 m_{\chi}$. Typically the energy spectra are produced in a model independent way by defining in the Monte Carlo simulation tool the decay of a generic resonance $\mathcal{R} \to \rm SM\,  SM$, with $m_{\mathcal{R}} = \sqrt{s}$ and by choosing a specific SM final state among those listed in Eq.~\eqref{eq:smfs} with 100\% branching ratio. For a given choice of SM final state and a set of dark matter masses, high precision tables are produced and stored in the numerical tool. 

For instance \texttt{MicrOMEGAs} computes the energy spectra for a specific dark matter model as follows: for the SM final states allowed by the model, it interpolates among these model independent tables as a function of $m_\chi$ and then rescales each SM final states by the appropriate branching ratio given by the model. The \PPPC~\cite{Cirelli:2010xx} tool on the other hand has released publicly the model independent energy spectra for a variety of SM final states, by providing table files~\cite{pppccode}. 

\maddm has made available both the `fast' and the `precise' running methods (more details are given in~\ref{sec:appes}) to obtain the energy spectra:

\paragraph{`Fast' mode} It first computes $\sigmav$ with the `fast' mode and then it downloads the \PPPC numerical tables with (default) or without weak corrections, depending on the user choice. The energy spectra of the model are interpolated using those tables.  This operation mode is similar to \texttt{MicrOMEGAs} and it is available only if the dark matter annihilates directly into a pair of SM final states. If you use this method please cite the \PPPC reference~\cite{Cirelli:2010xx}. 
\begin{figure*}[t!]
\centering
\includegraphics[width=1.\textwidth,trim=0mm 0mm 0mm 0mm, clip]{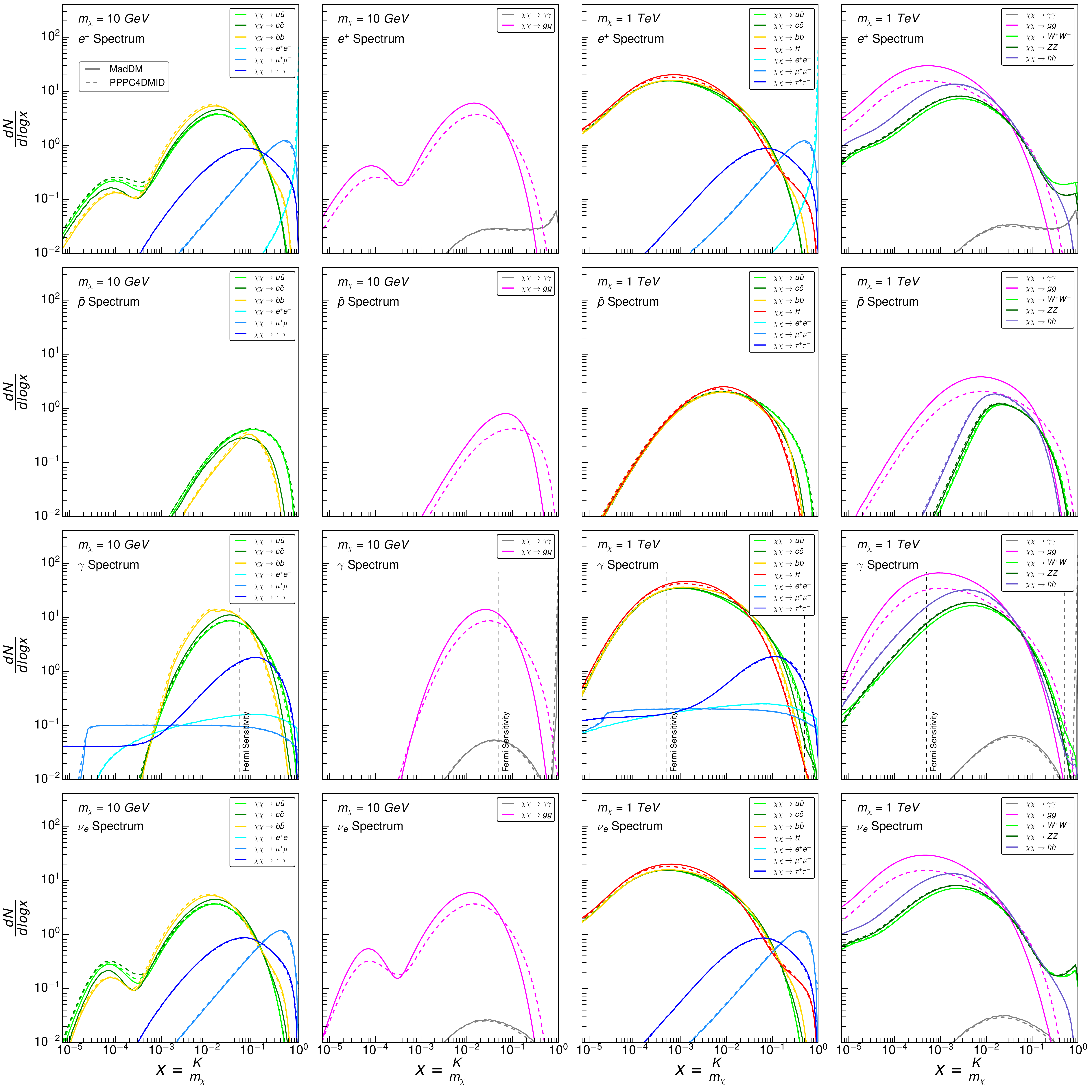}
\caption{Comparison between the energy spectra $\rmd N/\rmd \log x$ at production generated by \maddm and \PPPC, solid and dashed curves as labelled (with $x=K/m_{\rm DM}$ and $K$ being the kinetic energy of the final state stable particle). The dark matter annihilation channels are labelled by the different colours in the panels. The first and second columns show fermionic and bosonic annihilation channels for a mass of the dark matter $m_{\rm DM}$ = 10 GeV, while the third and fourth columns are for a 1 TeV dark matter mass, again fermionic and bosonic annihilation channel respectively.  Positron energy spectra are provided in the first row, while the second, third and fourth rows depict anti-protons, prompt photons and neutrinos (only electron flavour is shown) respectively.  The spectra do not include EW corrections.}\label{fig:pythiaspec}
\end{figure*}

\paragraph{`Precise` mode} Thanks to the embedding of \maddm into the \madgraph platform, it is easy to generate events for the annihilation process the user is interested in and pass it to a Monte Carlo simulation tool to get the energy spectra desired. There are many Monte Carlo simulation tools for decaying, showering and hadronisation; for the purposes of \maddm we have implemented an interface with \pythia. For a discussion on differences on the energy spectra generated with different Monte Carlo simulation tools we refer the reader to~\cite{Cirelli:2010xx,Cembranos:2013cfa}. The energy spectra for gamma rays, positrons, anti-protons and neutrinos are computed by \pythia from the event file generated by the \verb|madevent| or \verb|reshuffling| methods, which make use of the exact matrix element for a given process and for the specific model point in the parameter space. 
\begin{figure*}[t!]
\centering
\includegraphics[width=1.\textwidth,trim=0mm 0mm 0mm 0mm, clip]{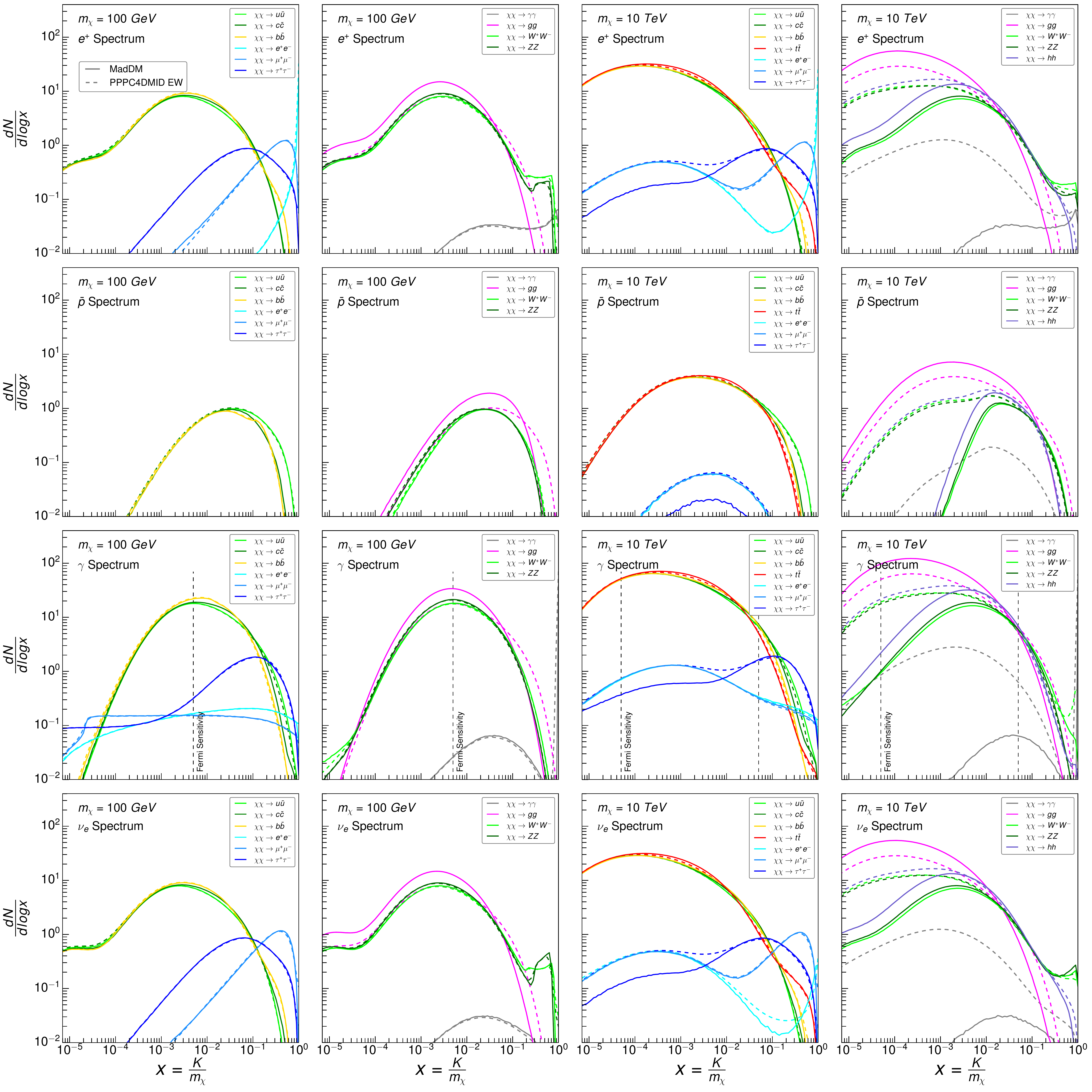}
\caption{Comparison between the energy spectra $\rmd N/\rmd \log x$ at production generated by \maddm and \PPPC including weak corrections. The first and second columns show fermionic and bosonic annihilation channels for a mass of the dark matter $m_{\rm DM}$ = 100 GeV, while the third and fourth columns are for a 10 TeV dark matter mass. The labelling of the plot is the same as Fig.~\ref{fig:pythiaspec}.}
\label{fig:pythiaspecew}
\end{figure*}

A comparison of the spectra generated with \maddm using \pythia with the ones released with the \PPPC code is provided in Fig.~\ref{fig:pythiaspec}. For the purpose of comparison we assume a branching ratio of 100\% into one particle species. We present the results for a dark matter mass of 10 GeV (first two columns) and 1 TeV (last two columns). In order to guarantee sufficiently smooth spectra even in the tails we generated between $5\times10^5$ and $5\times10^6$ events. The first row shows the energy spectrum for $e^+$, while the second, third and fourth depict the energy spectra for $\bar{p}$, $\gamma$ and $\nu_e$, respectively. We obtain perfect agreement with the \PPPC energy spectra (dashed lines) for all SM species, except $c\bar{c}$ and $gg$ final states. These energy spectra deviates slightly for all species of stable particles. At the time of writing we are not able to resolve this discrepancy by a variation of the \pythia settings. Note that this discrepancy has also been found in~\cite{Cuoco:2016jqt}. In the following, when referring to $c\bar{c}$ and $gg$ final states it means that we have generated our own set of model independent energy spectra to be consistent with our `precise' method. In the photon energy spectrum the vertical dashed line indicates the sensitivity window of the Fermi-LAT telescope (500 MeV to 500 GeV energy window): for a 10 GeV dark matter only the upper part of the spectrum matters at large $x$; for 1 TeV dark matter the Fermi-LAT energy window accesses all the energy spectrum, from $x\sim 10^{-3}$ to $x \sim 1$, while a further increase of the dark matter mass shifts the sensitivity range towards smaller $x$ values. 

Coming to the electroweak (EW) corrected energy spectra, \pythia has a partial implementation of the weak corrections~\cite{Christiansen:2014kba}, namely it takes into account the radiation of weak gauge bosons from the fermionic final states only. Once we allow for weak showering in \pythia our energy spectra match those of \PPPC for the case of fermionic final states, as shown in Fig.~\ref{fig:pythiaspecew} (the first and last two columns correspond to a dark matter mass of 100 GeV and 10 TeV, respectively). Energy spectra originating from quarks or gluons are basically unaffected by weak corrections. We are unable to match the energy spectra for weak corrections originating from $W^+W^-, ZZ$ and $hh$ final states, as those are not implemented into \pythia.  It is known that these corrections are large and moreover they open new channels that would be otherwise forbidden: for instance if the annihilation process is $\chi \chi \to e^+ e^-$ in principle there should be no anti-protons as a final results. By including the weak corrections the latter have a non negligible energy spectra, since they arise from hadronisation of the final state quarks originated by the weak bosons.

As \maddm is based on the \madgraph architecture it can easily handle not only the standard $\chi \chi \to 2$ annihilation processes but also dark matter annihilation into more than two particles in the final state, \ie,  $\chi \chi \to n$ processes. Note that \pythia will also automatically produce the energy spectra into $\gamma \gamma, e^+, \bar{p}, \nu_e, \nu_\mu, \nu_\tau$ in this case. We will provide two examples in Sec.~\ref{sec:valid}: the first  is based on $2 \to 3$ processes for indirect detection, where the third particle emitted is a gauge boson. For instance, Majorana or scalar dark matter annihilation into light fermions is $p$- or $d$-wave suppressed, whereas the additional emission of a $\gamma, Z$ or $W$ boson uplift the helicity suppression and lead to a $s$-wave annihilation cross section that can be constrained by present data. The second example is based on a $2 \to 4$ annihilation process  inspired by models of secluded dark matter~\cite{Pospelov:2007mp}. 

The ability to handle $2\to n$ processes is also relevant in the context of weak showering corrections, as the user can test the effect of a single weak boson emission on the standard dark matter annihilation into SM particles. For instance he/she can study for instance the consequence of a single weak boson correction to the $W^+W^-$ final state by considering the following annihilation processes $\chi \chi \to W^+ W^- Z$ and $\chi \chi \to W^+ W^- h$. 

Notice that some of  the energy associated with charged particle final states is redirected into photons, due to inverse Compton scattering of for instance CMB photons, synchrotron emission due to propagation in the magnetic field, and interaction with the interstellar gas producing both bremsstrahlung and neutral pions that further decay into photons. These processes modify the energy spectra of charged particles and of prompt photons (for details see~\cite{Cirelli:2010xx,Boudaud:2014qra,Buch:2015iya}). For this latter the energy spectrum can range from radio to gamma-ray energies. \maddm does not consider however the multi-wave length spectrum originating from these processes, as those depend on the details of the environment, but only the prompt gamma-ray energy spectrum from direct dark matter annihilation. For the former, the loss of energy and the conversion into photons is taken into account together with the propagation in the astrophysical environment, as will be explained in~\ref{sec:pbep}.

\subsection{Gamma-ray flux}\label{sec:gammas}

The study of prompt gamma-ray flux from dark matter annihilation is generically the simplest among the $\gamma, e^+, \bar{p}, \nu_i$ final states, as photons travel straight from the production to the detection points and typically trace the source. 

Let us consider a generic dark matter model that annihilates into the SM particle $i$ with branching ratio $B_i$. The expected gamma-ray flux from dark matter annihilation from a direction $\psi$ in the sky, averaged over an opening angle $\Delta \psi$ is:
\begin{eqnarray}
\frac{{\rmd}\Phi }{\rmd E_\gamma} (E_\gamma, \psi)  =   \frac{\sigmav}{2 m_\chi^2}\,  \sum_{i} B_i \frac{{\rmd}N^i_\gamma}{{\rmd}E_\gamma}\,    \frac{1}{4 \pi} \int_\psi \frac{{\rmd} \Omega}{\Delta \psi}\int_\text{los} \rho^2(\psi,l)\,  {\rmd}l \,.
\label{eq:difflux}
\end{eqnarray}
The second row of the equation defines the $J$ factor $\Big( J \equiv \int_\psi {\rmd} \Omega / \Delta \psi \int_\text{los} \rho^2(\psi,l)\,  {\rmd}l \,\Big )$. For dark matter candidates with distinct particle and antiparticle Eq.~\eqref{eq:difflux} is multiplied by an additional factor of 1/2.\footnote{This factor $1/2$ for non self-conjugate dark matter is automatically computed by \maddm by inferring this information from the UFO model, which stores the particle properties including the label self-conjugate or not.} \maddm provides both the differential flux in Eq.~\eqref{eq:difflux} as well as the total integrated flux, up to the $J$ factor, which should be provided by the user. Details are given in~\ref{sec:appfluxgn}.

\subsection{Anti-proton and positron fluxes}\label{sec:pbep}

Charged cosmic rays (CR), while they travel from the production point to the detection point, suffer of energy loss and diffusion. The energy loss are primarily due to synchrotron radiation, Inverse Compton scattering on CMB photons and on optical and infrared galactic starlight. The diffusion is an effect of the transport through the turbulent magnetic fields. 

The CR propagation is deeply studied by several groups, with basically two approaches: semi-analytical diffusion models, which rely on simplified assumptions of the interstellar gas and sources distributions to solve the cosmic ray transport equations, and fully numerical diffusion models, see \eg~\cite{Donato:2003xg,Delahaye:2010ji,Ciafaloni:2010ti,Boudaud:2014qra,Buch:2015iya} and~\cite{Strong:1998pw,Strong:2004de,galprop,Evoli:2008dv} respectively. Due to the complexity of the problem and the tools already available, we do not attempt to provide within \maddm a CR propagation code, and we rely on two separated codes, depending on the \maddm running mode. If \maddm is run in the `fast' mode as default mode we provide the flux of positrons at Earth only, extracted from the \PPPC numerical tables.\footnote{If you use this method, please cite the appropriate Refs.~\cite{Buch:2015iya}.} If \maddm is set on the `precise' running mode both positrons and anti-protons fluxes are computed via the CR propagation code~\dragon~\cite{Evoli:2008dv}. 

Another distinctive smoking gun for dark matter annihilation in the galactic halo would be the detection of anti-deuteron~\cite{Donato:1999gy}. \maddm has not considered this channel in its standard output. We notice however that the user is provided with all the necessary ingredients to run the \dragon code and get also the expected flux of anti-deuteron for the dark matter model he/she is testing.

\subsection{Neutrino fluxes}\label{sec:nus}

Similarly to prompt photons, neutrinos propagate straight from their source to the detection point. However, their flux at detection must include the effects of vacuum oscillation they experienced on their way to Earth.  
The energy spectrum at Earth\footnote{Notice that neutrinos having traveled through the Earth will experience additional effects due to oscillation into matter, which we do not take into account here.} for the electron flavour is then given by:
\begin{eqnarray}
\hspace*{-1cm}\left.\frac{{\rmd}N_{\nu_e}}{{\rmd} \log x} \right|_{\rm Earth}  =   \frac{{\rmd}N_{\nu_e}}{{\rmd} \log x} \left[1 - P(\nu_e \to \nu_\mu)- P(\nu_e \to \nu_\tau)\right] +  \frac{{\rmd}N_{\nu_\mu}}{{\rmd} \log x} P(\nu_\mu \to \nu_e) + \frac{{\rmd}N_{\nu_\tau}}{{\rmd} \log x} P(\nu_\tau \to \nu_e)\,,
\end{eqnarray}
where $P(\nu_\alpha \to \nu_\beta)$ is the probability for oscillation from flavour $\alpha$ to flavour $\beta$. Similar expressions hold for $\nu_\mu$ and $\nu_\tau$ flavours. The oscillation probabilities are obtained from the three flavour leptonic mixing matrix, assuming $U$ being a unitary matrix~\cite{Esteban:2016qun,nufit}:
\begin{equation}
U_{\alpha,i} = 
\begin{pmatrix}
0.82 & 0.55 & 0.15\\
0.37 & 0.57 & 0.71\\
0.40  & 0.59 & 0.68
\end{pmatrix}\,,
\end{equation}
and are defined as:
\begin{eqnarray}
\label{eq:nuprob}
P(\nu_\alpha \to \nu_\beta)  =  \sum_i \Big| U_{\beta i}\Big|^2  \Big | U_{\alpha i} \Big |^2 + 
\sum_{i < j} 2 \Re \left[ U_{\beta j} U_{\alpha j}^{\ast}U_{\beta i}^{\ast} U_{\alpha i} \,\mathrm{e}^{\mathrm{i} (m_j^2 - m_i^2) L/(2 E_\nu)} \right]\,,
\end{eqnarray}
where the greek letters refer to flavour indices and the latin letters to mass eigenstates (see e.g.~\cite{Lucente:2016vru} containing a recent review). Furthermore, $L$ is the travel distance and $E_{\nu}$ the neutrino energy. For neutrinos having travelled from distant astrophysical sources for which $L\gg E_{\nu}/(m_j^2 - m_i^2)$ the second term in Eq.~\eqref{eq:nuprob} generates very rapid oscillations and can hence be neglected.
In this case it is a good approximation to take the long baseline oscillation limit yielding an energy spectrum with a flavour ratio $\nu_e : \nu_\mu : \nu_\tau = 1: 1: 1$.

Equation~\eqref{eq:difflux}, which is given for prompt photons, has been easily adapted to neutrinos by taking into account vacuum oscillation effects on the energy spectra at detection for each neutrino flavour. Hence \maddm also provide the differential and total flux for neutrinos (same for anti-neutrinos) up to the $J$ factor. Neutrinos from the Milky Way halo or from nearby galaxies are actively searched for by the IceCube telescope~\cite{Aartsen:2016pfc,Aartsen:2013dxa}.

In the \maddm v.3.0 release we do not include the case of neutrinos coming from annihilation of dark matter in the center of the Sun~\cite{,Cirelli:2005gh,Blennow:2007tw,Baratella:2013fya}, which experience oscillation effects due to the interaction with matter. Similarly we do not include oscillation effect due to matter when neutrinos travel through the Earth.\footnote{Dark matter capture and annihilation in the Sun has been implemented in \maddm~for the study in Ref.~\cite{Arina:2017sng} taking into account SI scatterings. However, a public version incorporating also SD scatterings and self interactions is left for future work.}

\section{Experimental Constraints and Likelihoods for Dark Matter}\label{sec:lnlk}

A major part of \maddm v.3.0 is the ability to incorporate experimental constraints. For this purpose, we developed two approaches: (i) a simplified functionality for testing model points against direct and indirect experimental constraints to investigate whether the model point is excluded or not by present observations; (ii) provide likelihood functions of direct and indirect detection experiments. In particular in \maddm v.3.0 we have implemented the Fermi-LAT likelihood for diffuse gamma rays from dSPhs~\cite{Fermi-LAT:2016uux,fermilike}, whose functionalities will be presented in Sec.~\ref{sec:fermilkn}. We do not consider here the LUX likelihood, which has been implemented in \maddm v.2.0~\cite{Backovic:2015cra}. 

Mode (i) basically allows for a fast comparison between the theoretical predictions and the experimental exclusion limits as a function of the dark matter mass. This is done using the \verb|ExpConstraints| class, which is described in details in~\ref{sec:appexpc}. For direct detection it takes the predicted spin-independent (SI) and/or spin-dependent (SD) elastic cross section off nucleon as a function of $m_\chi$ and compares it with the corresponding exclusion limit (\ie XENON1T for SI scattering~\cite{Aprile:2017iyp}, LUX for SD on neutron~\cite{Akerib:2017kat} and Pico60 for SD on proton~\cite{Amole:2017dex}). Indirect detection experimental constraints are provided for prompt photons and gamma-ray lines generated by dark matter annihilation in dSPhs and in the Galactic center respectively. We have chosen to use gamma-ray exclusion limits because of their constraining power and robustness, as compared to neutrino and charged cosmic ray searches, respectively. 
These Fermi-LAT bounds constrain the prompt photon emission or the gamma-ray line emission originating from dark matter annihilation in the plane $\sigmav$ and dark matter mass. Indirect detection constraints are computed assuming a 100\% branching ratio into one SM annihilation channel. To be conservative we rescale the bound correspondingly by the branching ratio of the theoretical model. With this method we test the dark matter model only against SM annihilation channels, which are computed as described in Sec.~\ref{sec:fermilkn}. Of course in order for the comparison to be meaningful the assumptions of the experimental exclusion limits should be the same as for the theoretical predictions, such as the choice of consistent the $J$ factors.

The \verb|ExpConstraints| class is very generic and allows easily the implementation of new limits and/or measurements in case of a discovery. For instance, the implementation of positron and anti-proton constraints from the AMS-02~\cite{PhysRevLett.117.091103} is left for future work.

\subsection{Fermi-LAT likelihoods for dSphs}\label{sec:fermilkn}
\begin{figure*}[t]
\begin{minipage}[t]{0.5\textwidth}
\centering
\includegraphics[width=1.\columnwidth]{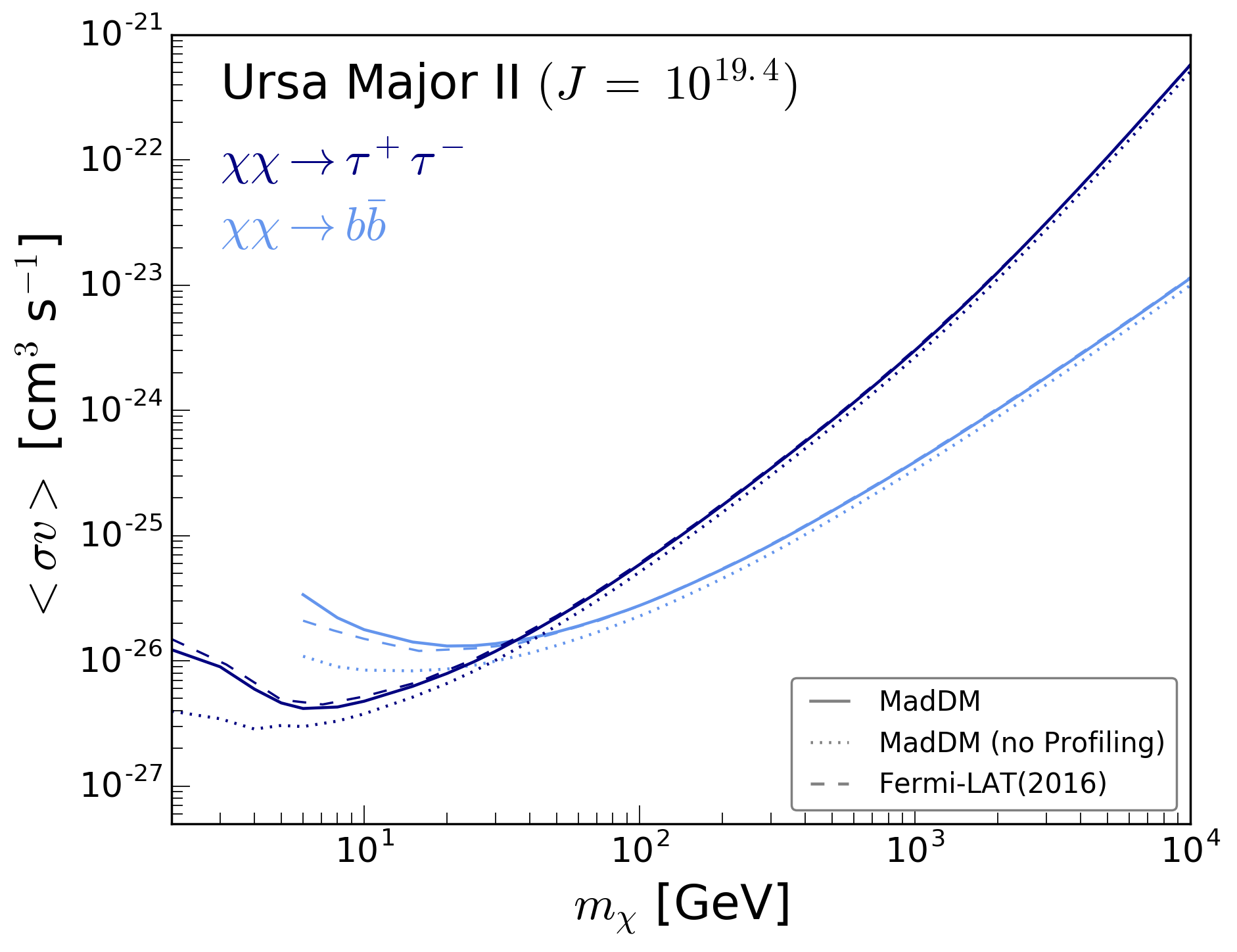}
\end{minipage}
\begin{minipage}[t]{0.5\textwidth}
\centering
\includegraphics[width=1.\columnwidth]{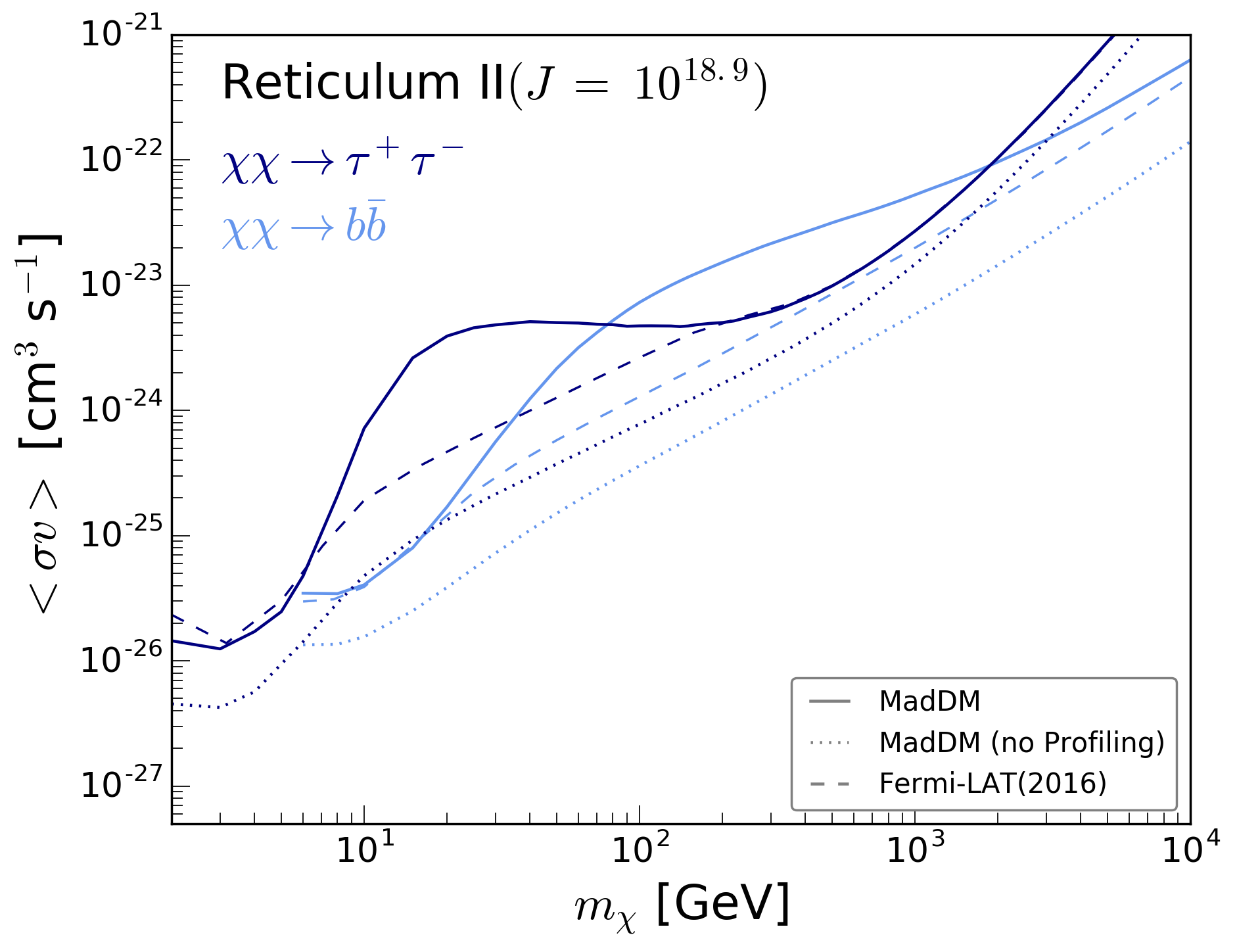}
\end{minipage}
\caption{Comparison between the \maddm and publicly released Fermi-LAT exclusion bounds in the $\{\sigmav - m_\chi\}$-plane for two reference dSphs: Ursa Major II in the left panel and Reticulum II in the right panel. The solid curves are for the \maddm limits profiling on the $J$ factor, the dotted lines do not include profiling, while the dashed lines stand for the Fermi-LAT limits. The light blue curve is for $\chi \chi \to b \bar{b}$, while the dark blue line is for $\chi \chi \to \tau^- \tau^+$.}
\label{fig:dSphcomp}
\end{figure*}
\begin{figure}[t]
\centering
\includegraphics[width=0.5\columnwidth,trim=0mm 0mm 0mm 0mm, clip]{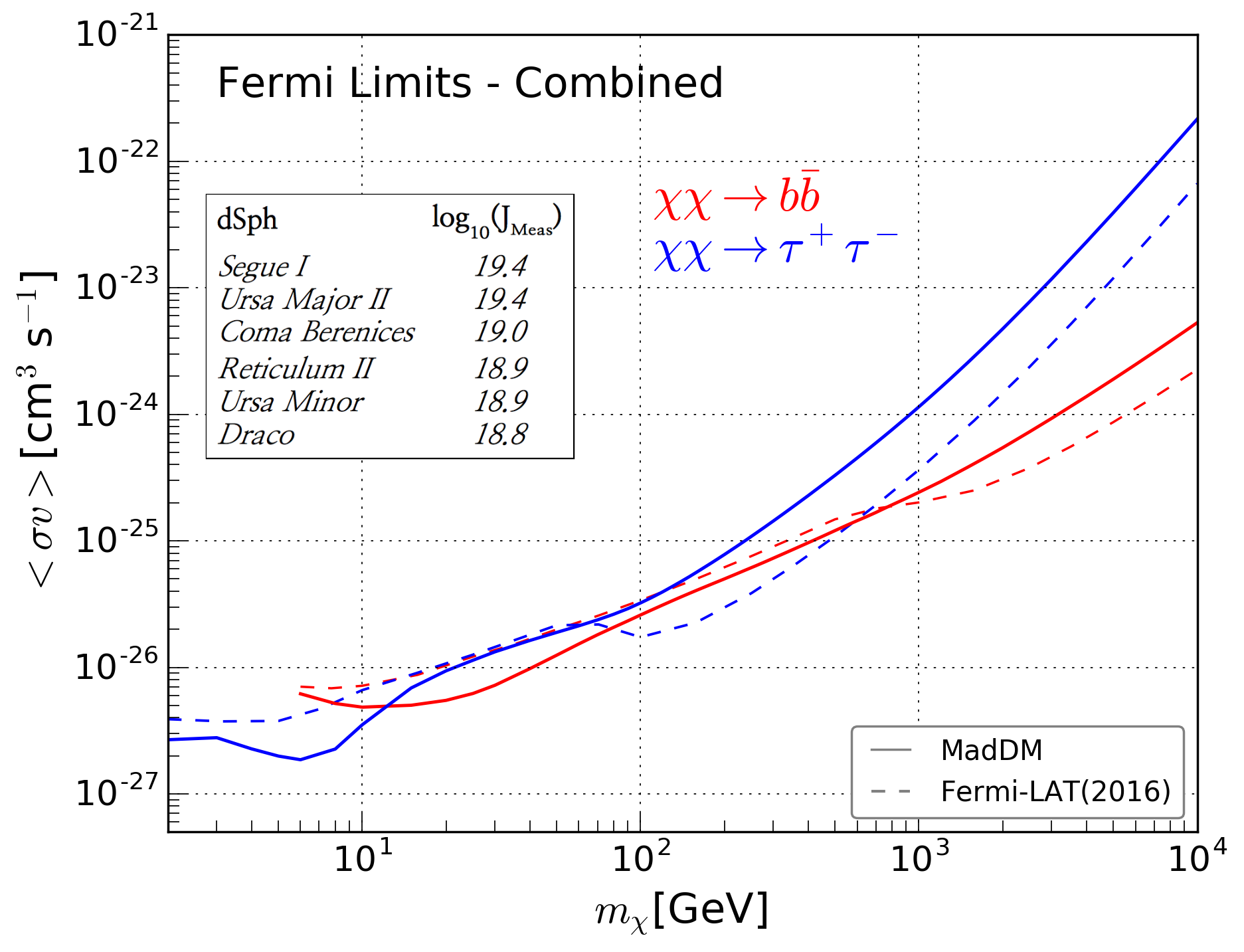}
\caption{Comparison between the \maddm and publicly released Fermi-LAT exclusion bounds in the $\{\sigmav - m_\chi\}$-plane for the combined set of dSphs as labelled. The solid curves are for the \maddm limits profiling on the $J$ factor, while the dashed lines stand for the Fermi-LAT limits. The red curve is for $\chi \chi \to b \bar{b}$, while the blue line is for $\chi \chi \to \tau^- \tau^+$.}
\label{fig:combcomp}
\end{figure}
\begin{figure*}[t]
\begin{minipage}[t]{0.5\textwidth}
\centering
\includegraphics[width=1.\columnwidth]{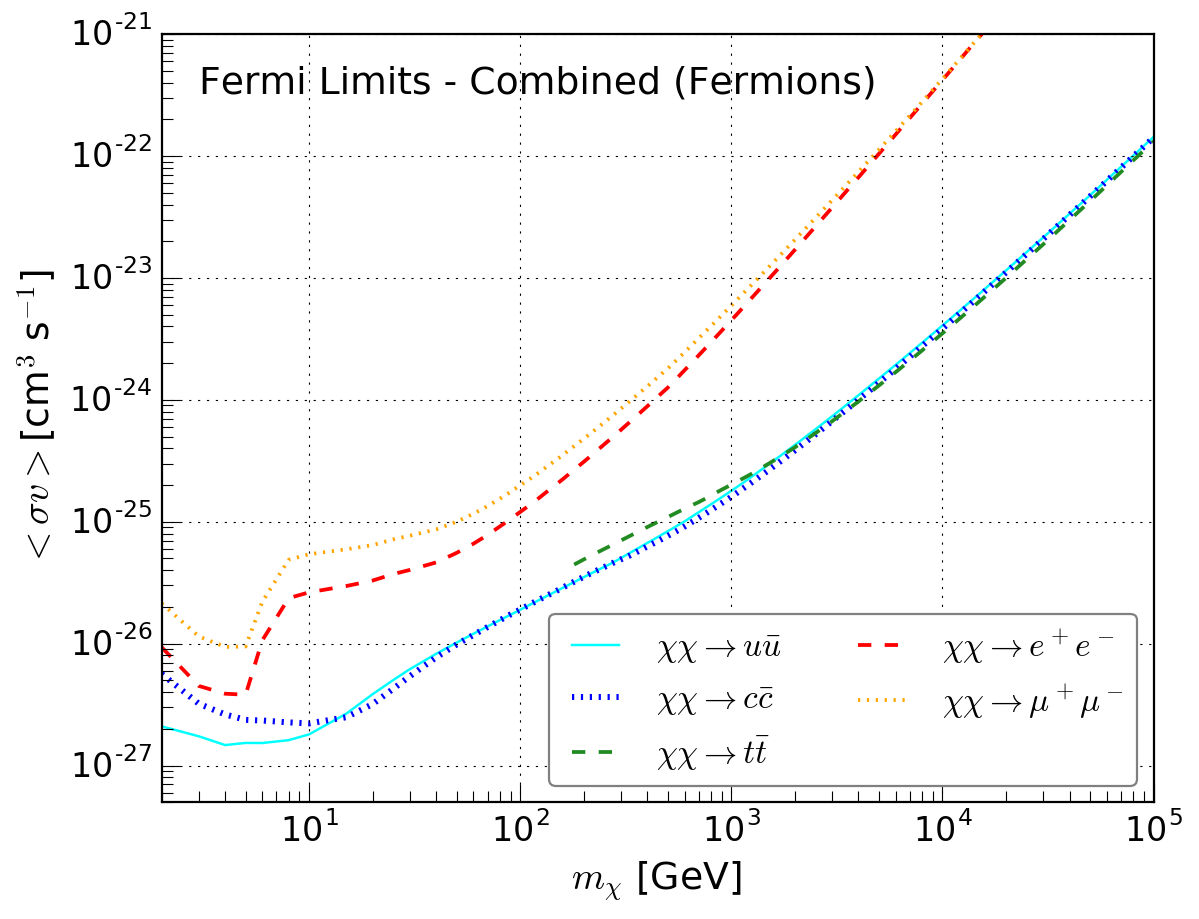}
\end{minipage}
\begin{minipage}[t]{0.5\textwidth}
\centering
\includegraphics[width=1.\columnwidth]{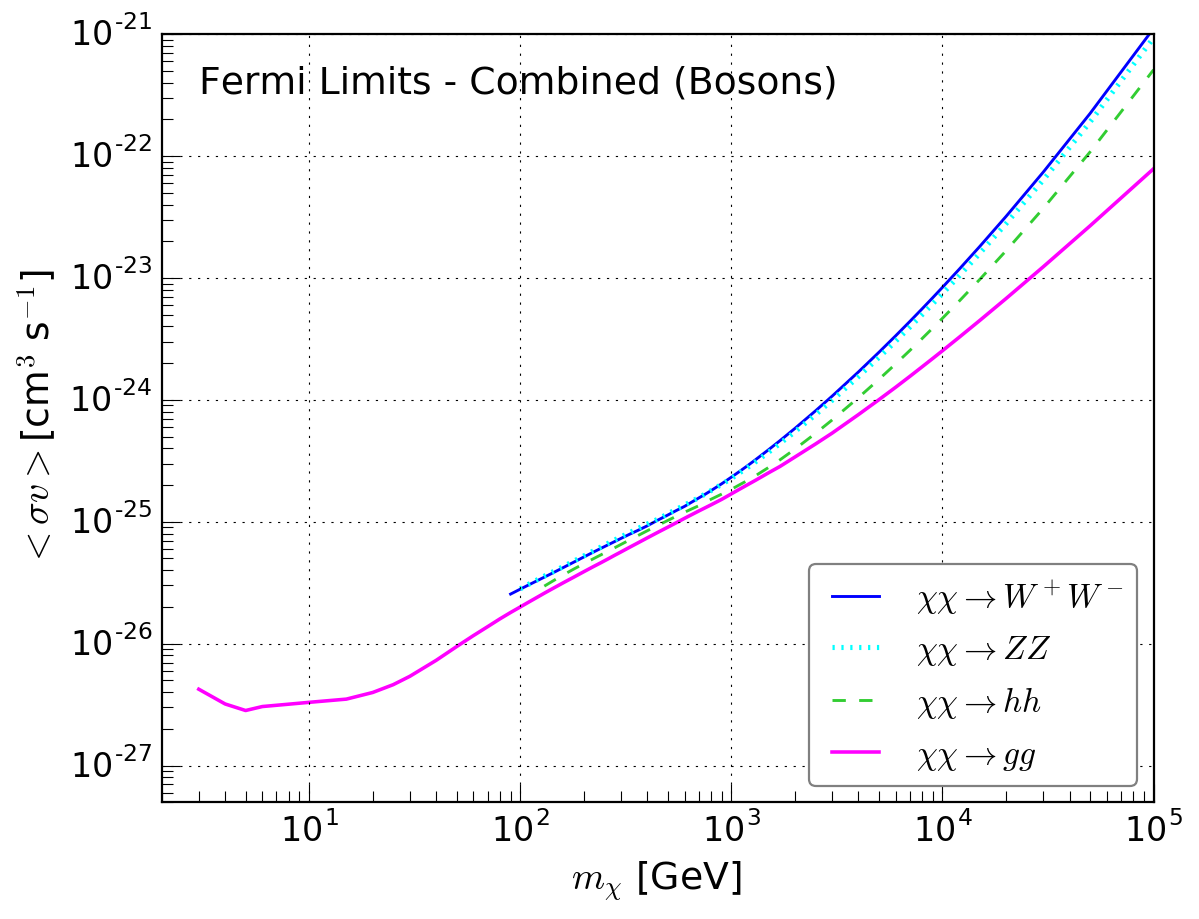}
\end{minipage}
\caption{Exclusion limits for the combined set of dSphs, computed with \maddm for all SM final states, as labelled. Fermions and bosons are presented separately in the left and right panels respectively. These bounds include profiling over the $J$ factor.}
\label{fig:misschann}
\end{figure*}

Dwarf spheroidal galaxies (dSPhs) are optimal targets to test the hypothesis of annihilating dark matter in the galactic halo. These Milky Way satellite galaxies are known to be dark matter dominated objects, with a reduced baryonic content in terms of recent stellar formation activity and interstellar gas medium~\cite{Mateo:1998wg,Weisz:2011gp,Brown:2012uq,Courteau:2013cjm}. In the past few years there has been a refurbished activity in this field, with the discovery of many low brightness dSphs (also called ultra-faint dSphs) by optical wide-field imaging surveys such as the Sloan Digital Sky Survey~\cite{Yanny:2009kg}  or dark energy surveys such as DES~\cite{Abbott:2005bi} that sum up to the 9 luminous classical dSphs known so far and largely increase the number of confirmed dSphs. This reinvigorated activity makes these satellite galaxies attractive targets for dark matter searches and are worth to be included with their full power as experimental constraints in the \maddm code.

The non-observation of a gamma-ray emission from dSphs galaxies sets very strong constraints on the prompt photon flux originating from a given dark matter model. The analysis implemented in \maddm is based on the public data released by the Fermi-LAT collaboration~\cite{Fermi-LAT:2016uux,fermilike}. The Fermi-LAT satellite has analysed the 6 years data (Pass 8) to search for an excess in the gamma-ray emission coming from 45 confirmed and candidate dSphs, finding no significant excess in the analysis of the combined data. Hence exclusion limits are set on the prompt photon flux generated by dark matter annihilation and consequently on the annihilation cross section. There are four dSphs recently discovered by DES, which, taken individually, show a slight excess over the background of the order of 2$\sigma$; this set contains the Reticulum II dSph, for which also other analyses~\cite{Geringer-Sameth:2015lua,Hooper:2015ula} have pointed out a possible excess over the background. As a consequence, the combined exclusion limits of Fermi-LAT in the $\sigmav$ and dark matter mass plane are weaker than the exclusion limits of Fermi-LAT 2015~\cite{Ackermann:2015zua}, where only 15 confirmed dSphs (without Reticulum II) were considered. For instance, the exclusion limit for the channel $\chi \chi \to \tau^+ \tau^-$ gets weakened below $m_\chi=100$ GeV by roughly a factor of 2, excluding $\sigmav\sim 5 \times 10^{-27} \rm cm^3/s$ for a dark matter mass of 10 GeV and passing the thermal freeze-out cross section at about $m_\chi = 90$ GeV.

For each of the 45 dSphs the Fermi-LAT collaboration has made available likelihood profiles for all energy bins that allow for the derivation of gamma-ray flux upper limits. By using these likelihood profile functions, we perform a likelihood analysis to constrain the model point the user is testing. To this end we bin the predicted flux accordingly and sum up the likelihood for each bin for a given dSph. We use the $J$ factors provided by the collaboration (adopted from~\cite{Geringer-Sameth:2014yza}) which are based on 
spectroscopic observations when possible and distance scaling relationships otherwise. We take into account the uncertainties on the $J$ factors by including them as nuisance parameters and profiling over them, according to Ref.~\cite{Drlica-Wagner:2015xua}. The total likelihoods of all dSphs included are then summed up and interpreted
as a test statistic in order to derive the $p$-value of the model. The default set of dSphs contains Segue I, Ursa Major II, Coma Berenices, Reticulum II, Ursa Minor and Draco which are the dSphs with the six largest $J$ factors. Based on this method \maddm also finds the corresponding 95\% confidence level (CL) cross-section upper limit for the model by demanding a $p$-value of 0.05. 
The likelihood method to compare theoretical predictions with the Fermi-LAT data is very generic and can constrain any dark matter model, no matter what are the annihilation final state. This method is the default procedure if the user selects the `precise' running mode for \maddm. 

Figure~\ref{fig:dSphcomp} shows the comparison between the Fermi-LAT (dashed line) and the \maddm (solid line) exclusion limits in the $\sigmav$ and dark matter mass plane for two sample dSph galaxies: Ursa Major II  (left panel) and Reticulum II (right panel). For Ursa Major II \maddm reproduces the Fermi-LAT limits over the whole dark matter mass range for both the $b\bar{b}$ final state (dark blue) and the $\tau^-\tau^+$ final state (light blue).
The same degree of agreement  
is found for all considered dSphs, with the exception of Reticulum II.\@ In the mass range that fits the slight excess we obtain a significantly stronger limit, revealing the limitations of the use of the public likelihood with respect to the full Fermi-LAT limits.
Here, we find deviations up to almost an order of magnitude. 
Nevertheless, by default we include Reticulum II in the computation of the combined limit (see below) in order to stay as close as possible to the combined Fermi-LAT analysis.

In Fig.~\ref{fig:combcomp} we present the \maddm combined limits in comparison with the Fermi-LAT bounds, which have been released publicly only for the $b\bar{b}$ and $\tau^+\tau^-$ channels. In order to speed up our numerical routine we include the 6 dSphs, as labelled in the plot, which are the ones with the largest $J$ factors. 
We have verified that stacking additional dSphs does not affect significantly our exclusion limits. Hence we choose this set of 6 dSphs to be our reference set of satellite galaxies from which to compute the Fermi-LAT exclusion limits. We find that 
the \maddm bounds are approximately in agreement with the public exclusion limits. 
Deviation of up to a factor of 3 arise, again, due to the limitations of the use of public likelihood with respect to the full Fermi-LAT limits in the course of the combination of dSphs. While in the mass range that fits the access seen in Reticulum II (below roughly 100\,GeV and 1\,TeV for annihilation into $\tau^+\tau^-$ and $b\bar b$, respectively) the limits from \maddm slightly overestimate the constraints, above this region the full Fermi-LAT analysis provides somewhat stronger bounds. The dip-like shape in the curve for the $\tau^+\tau^-$ channel just below 10\,GeV arises from an interplay of the weaker limit from Reticulum II between 10 and 100\,GeV and the over-all weakening of the limits for dark matter masses below a few GeV due to the low-energy bound of 500\,MeV in the experimental analysis.

From this set we additionally compute the exclusion limits for dark matter annihilating with a branching ratio of 100\% into the following SM final states: $e^+e^-,\mu^+\mu^-,gg,q\bar{q},c\bar{c},t\bar{t},hh,ZZ,W^+W^-$ ($q$ includes the light quarks $u, d, s$ for which the prompt photon energy spectrum is the same). Those are presented in Fig.~\ref{fig:misschann}: in the left panels we show the limits for dark matter annihilation into fermions, while on the right panel the limits for annihilation into bosons are depicted. All the exclusion limits presented in Figs.~\ref{fig:combcomp} and~\ref{fig:misschann} are encoded in the \verb|ExpConstraints| class. Note that these precomputed limits are only used for the fast limit settings on individual channels. For the upper limit on the total annihilation cross section we always follow the prescription above utilizing the public likelihood.

\subsection{Rescaling of fluxes}\label{sec:rescale}

Within \maddm we provide likelihoods and upper cross-section limits for two distinct scenarios regarding the composition of dark matter.

\begin{enumerate}
\item  `All DM': In this scenario we assume that the dark matter candidate under consideration makes up all gravitationally interacting dark matter, $(\Omega h^2)_{\rm theo}=(\Omega h^2)_{\rm Planck}$ regardless of the abundance that result from thermal freeze-out, $(\Omega h^2)_{\rm thermal}$, within the model. In particular our assumption concerns the local dark matter densities that enter the fluxes for indirect and direct detection experiments. Accordingly, no rescaling of the fluxes is made. For $(\Omega h^2)_{\rm thermal}<(\Omega h^2)_{\rm Planck}$ this scenario could \eg~be realised by additional non-thermal contributions to dark matter production while $(\Omega h^2)_{\rm thermal}>(\Omega h^2)_{\rm Planck}$ could be accommodated by a non-standard cosmological history. As an example for the former, in supersymmetric scenarios where the dark matter is higgsino, the late gravitino decay is a popular mechanism to augment its relic density and bring it to the Planck measured value~\cite{Allahverdi:2012wb}. 

\item  `Thermal' scenario: In the thermal scenario we assume that the dark matter candidate under consideration is produced solely via the freeze-out mechanism within a standard cosmological history. Hence we assume that its relic density is set by the thermal value $(\Omega h^2)_{\rm theo}=(\Omega h^2)_{\rm thermal}$. We consider
$(\Omega h^2)_{\rm theo}\le(\Omega h^2)_{\rm Planck}$ while the missing fraction of gravitationally interacting dark matter is assumed to originate from other species. We define the fraction of dark matter originating from the candidate under consideration:
\begin{equation}
\xi \equiv (\Omega h^2)_{\rm theo} / (\Omega h^2)_{\rm Planck}\le1\,.
\end{equation}
Assuming that there is no difference in the clustering properties of the different dark matter species, $\xi$ also denotes the corresponding fraction of the local dark matter density. Accordingly we rescale the fluxes for indirect and direct detection with $\xi^2$ and $\xi$, respectively, with respect to the `all DM' scenario.
\end{enumerate}

Note that the two scenarios coincide for $(\Omega h^2)_{\rm thermal}=(\Omega h^2)_{\rm Planck}$. For $(\Omega h^2)_{\rm thermal}>(\Omega h^2)_{\rm Planck}$ only the former scenario is allowed and numbers for the thermal scenario are not provided.

\section{Sampling the Model Parameter Space} \label{sec:parscan}

Another valuable new feature of \maddm v.3.0 is the capability to perform sampling of the dark matter model parameter space. \maddm puts forward two different sampling procedure: (i) a sequential grid scan, whose details are provided in~\ref{sec:appsg}; (ii) a nested sampling method, which is described in~\ref{sec:appmn}.

The sequential sampling method belongs to the inherited features from the \madgraph platform. It scans the model parameter space on a grid, once the range and the amount of variation in the parameter value is specified. This method provides an efficient way to explore low dimensional
parameter spaces, \eg~2-dimensional projections as considered in Sec.~\ref{sec:mnvad}.

\maddm now supports model parameter sampling with \PyMN~\cite{Buchner:2014nha}, a Python implementation of the \linebreak \MN algorithm~\cite{Feroz:2007kg,Feroz:2008xx,2013arXiv1306.2144F}. \MN is a Bayesian inference program based on the nested sampling algorithm. It allows for parameter estimation and model comparison. We are not going to discuss here the model comparison part, and refer the interested reader to this review~\cite{Trotta:2008qt}. What makes \PyMN particularly interesting for \maddm is its capability of model parameter inference. The algorithm is faster and samples more efficiently an high dimensionality model parameter space with respect to a Markov Chain Monte Carlo algorithm, and is particularly suited for multi-modal likelihood functions. For more details on the \MN and on the \PyMN algorithms we refer the interested reader to Refs.~\cite{Feroz:2007kg,Feroz:2008xx,2013arXiv1306.2144F,Buchner:2014nha}. If you use \PyMN in the context of \maddm please cite the previous references.

\section{Applications}\label{sec:valid}

In this section we provide applications to physically motivated dark matter theories. In Sec.~\ref{sec:cval} we first compare the cross-section predictions of \maddm v.3.0 and \texttt{MicrOMEGAs} v.5.0 for benchmarks of the Minimal Supersymmetric Standard Model (MSSM), Minimal Universal Extra Dimensions (MUED) and Higgs portal models, similarly to previous \maddm v.1.0 and \maddm v.2.0 releases.
In subsequent sections we utilize the new capabilities of \maddm studying various aspects of dark matter simplified models.
In particular, in Sec.~\ref{sec:2to4} we derive constraints on secluded dark matter~\cite{Pospelov:2007mp} where dark matter annihilates into 
metastable mediators that subsequently decay into the standard model demonstrating the automatic computation of annihilation into $n$ final state particles. In Sec.~\ref{sec:vib} we study the case of helicity suppression lifting through internal bremsstrahlung. Finally, we demonstrate the model parameter sampling capabilities by comparing the sequential grid and \PyMN scanning in Sec.~\ref{sec:mnvad}.

\subsection{Comparison with \texttt{MicrOMEGAs} for some benchmark scenarios}\label{sec:cval}
\begin{table}[t]
  \centering
  \begin{tabular}{|c|c|c|c|c|} 
  \hline
  & & \multicolumn{2}{c|}{$\sigmav$ [$\rm cm^{3}s^{-1}$]} & \\
   model &  $m_{\rm DM}$ [${\rm GeV}$] & {\maddm{} v3.0}  & \texttt{MicrOMEGAs} v5.0 & difference (\%)\\
  \hline
  MSSM (SPS1a) & 100 & $3.99 \times 10^{-28}$ & $3.92\times10^{-28} $& 1.8\\
  \hline
  MSSM (SPS1b) & 100 & $3.41 \times 10^{-28}$ & $3.42 \times 10^{-28}$ & 0.3\\
  \hline
  MSSM (SPS3) & 100 & $9.20 \times 10^{-29}$ & $9.32 \times 10^{-29}$ & 1.3\\
     \hline
  MUED & 1500  & $5.94 \times 10^{-27}$ & $5.85\times10^{-27} $& 0.38\\
  \hline
  Higgs Portal Vector &  200 & $2.35 \times 10^{-23}$ & $2.35\times10^{-23} $& 0.0\\
  \hline
  \end{tabular}
  \caption{\maddm{} v3.0 and \texttt{MicrOMEGAs} v5.0 comparison for the velocity averaged annihilation cross sections for various benchmark models.}
  \label{sigvtable}
\end{table}

In this section we compare the results of \maddm v.3.0 and \texttt{MicrOMEGAs} v.5.0~\cite{Belanger:2018ccd} for 
the velocity averaged annihilation cross sections, which is used in the indirect detection modules of each simulation package. For illustration, we choose several benchmark models that offer rich dark matter phenomenology and have been studied extensively in the literature. 

We first compare the $\sigmav$ output for the MSSM\@. This model introduces a large parameter space, requiring
dedicated studies to fully explore their complexity. 
For the purposes of illustration we present here three Snowmass Points and Slopes (SPS) for the MSSM model, each of which have different physical interpretation. The SPS convention is a method of relating the high energy MSSM input parameters to the low energy parameters of the theory. 
As examples, we show three benchmark scenarios for the MSSM: SPS (1a), SPS (1b) and SPS (3) points. The first two points represent an mSUGRA region with intermediate values of $\tan \beta$, with the dark matter candidate being the lightest neutralino that is bino-like \cite{Hooper:2003ka}. The latter point represents a coannihilation region in the mSUGRA space, in which the dark matter candidate, a bino-like neutralino, undergoes rapid coannihilation with sleptons \cite{Allanach:2002nj, Hooper:2003ka,Backovic:2013dpa}. 
For the MSSM models we see a very good agreement between \maddm{} and \texttt{MicrOMEGAs}, as it is illustrated in Tab.~\ref{sigvtable}, with a percentage difference on the order of $\sim(1\!-\!2)\%$. We have assumed a neutralino mass of 100 GeV for each SPS model.
Another class of models we test are models with Universal Extra Dimensions. These models provide a viable dark matter candidate with rich phenomenology. We focus on the MUED model. In this scenario, the space-time dimensions of the standard model (SM) are extended by one extra dimension, resulting in a tower of Kaluza-Klein (KK) partners of the SM fields. The lightest of these KK fields is the level 1 hypercharge gauge boson ($B^{(1)}$) and is the dark matter candidate \cite{Kong:2005hn, Servant:2002aq}. 
In the last row of Tab.~\ref{sigvtable} we show the comparison between \maddm{} and \texttt{MicrOMEGAs} for this model, assuming a Benchmark dark matter mass of 1.5 TeV. 
For this model point we see a percentage difference of roughly $\sim 1\%$ between the two simulation packages. 

We furthermore consider the Higgs portal vector dark matter model, which along with its variations has been studied extensively in the literature. In this paper, we present an example model similar to the one studied in Ref.~\cite{Lebedev:2011iq}, the only difference being that the dark matter can couple directly to the $W$, $Z$ and Higgs bosons and thus requires no new dark sector scalar which would mix with the Higgs. 
We choose a dark matter mass point of 200 GeV and see very good agreement with the \texttt{MicrOMEGAs} output as is shown in Tab.~\ref{sigvtable}. We further compare \maddm{} against \texttt{MicrOMEGAs} for $\sigmav$ by scanning over the dark matter mass, as seen in Fig.~\ref{fig:sigvcomp}. For the default settings of \maddm we observe a $\sim 30\%$ difference in the outputs of the two packages for dark matter masses below the $WW$ threshold, $m_\text{DM}\le 80$\,GeV. This difference is due 
to the missing $2\to3$ annihilation processes via an off-shell $W$-boson, $\chi\chi\to W W^\star$ (and similarly but less importantly for $ZZ^\star$) which contributes $\sim 30\%$ to $\sigmav$ (this is well compatible with the findings in Ref.\cite{Cline:2013gha}). This process is considered automatically within \texttt{MicrOMEGAs}, but so far not by default in \maddm. The respective result is superimposed in Fig.~\ref{fig:sigvcomp} and is in good agreement with the result of 
\texttt{MicrOMEGAs}. The increase of the annihilation cross section at around $80$, $90$ and $125\,$GeV 
correspond to the $W$, $Z$ and Higgs thresholds, respectively. Note that below the Higgs threshold only Higgs mediated $s$-channel processes contribute while above the threshold various diagrams interfere. In the plot we choose a Higgs portal coupling of $\lambda_\text{HP}=1$.

The off-shell decays can easily be included upon specification of the $2\to3$ process:
\begin{verbatim}
define ferm = ve vm vt ve~ vm~ vt~ u c t d s b u~ c~ t~ d~ s~ b~ e- mu- ta- e+ mu+ ta+
define wz = w+ w- z
add indirect_detection wz ferm ferm
\end{verbatim}
\begin{figure}[t]
\centering
\includegraphics[width=0.5\columnwidth,trim=0mm 0mm 0mm 0mm, clip]{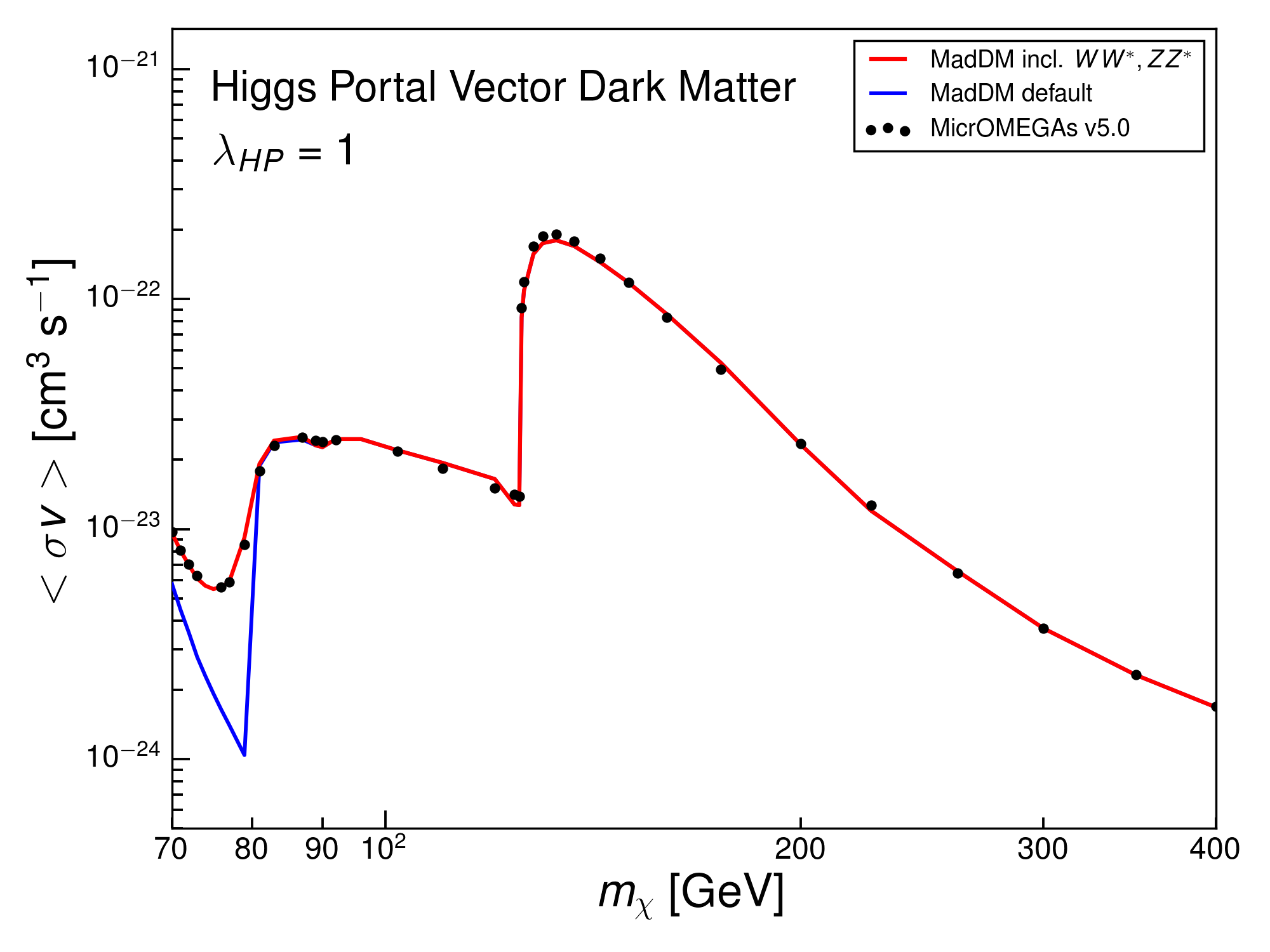}
\caption{Comparison between \maddm{} and \texttt{MicrOMEGAs} in the computation of the velocity averaged annihilation cross-section $\sigmav$. The comparison is done in a benchmark Higgs Portal vector dark matter model where $M_{DM}$ is the mass of the vector dark matter candidate of the model.}
\label{fig:sigvcomp}
\end{figure}

\subsection{Annihilation into the dark sector}\label{sec:2to4}
\begin{figure*}[t]
\begin{minipage}[t]{0.5\textwidth}
\centering
\includegraphics[width=1.\columnwidth,trim=10mm 0mm 10mm 0mm, clip]{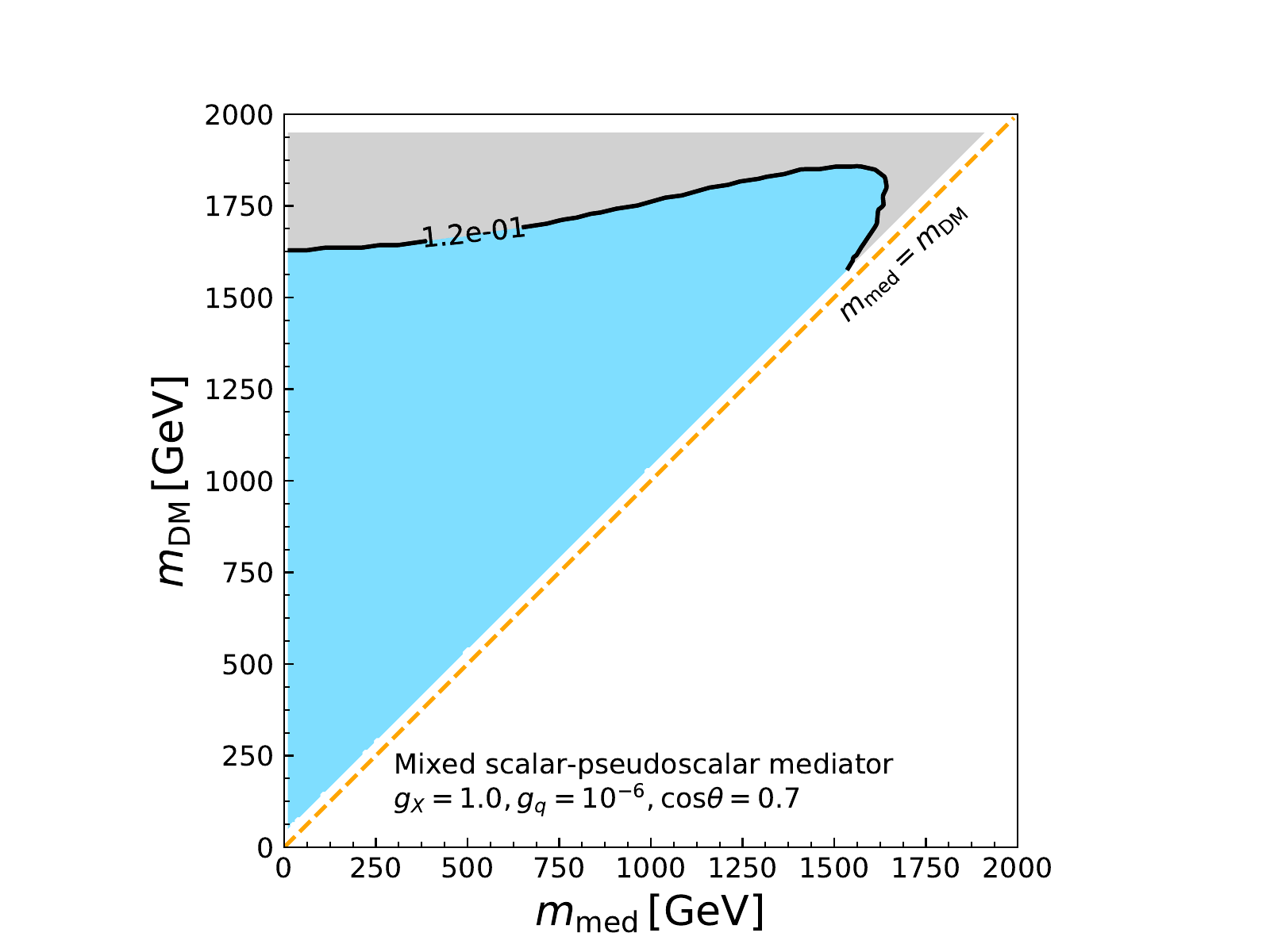}
\end{minipage}
\begin{minipage}[t]{0.5\textwidth}
\centering
\includegraphics[width=1.07\columnwidth,trim=0mm 0mm 10mm 0mm, clip]{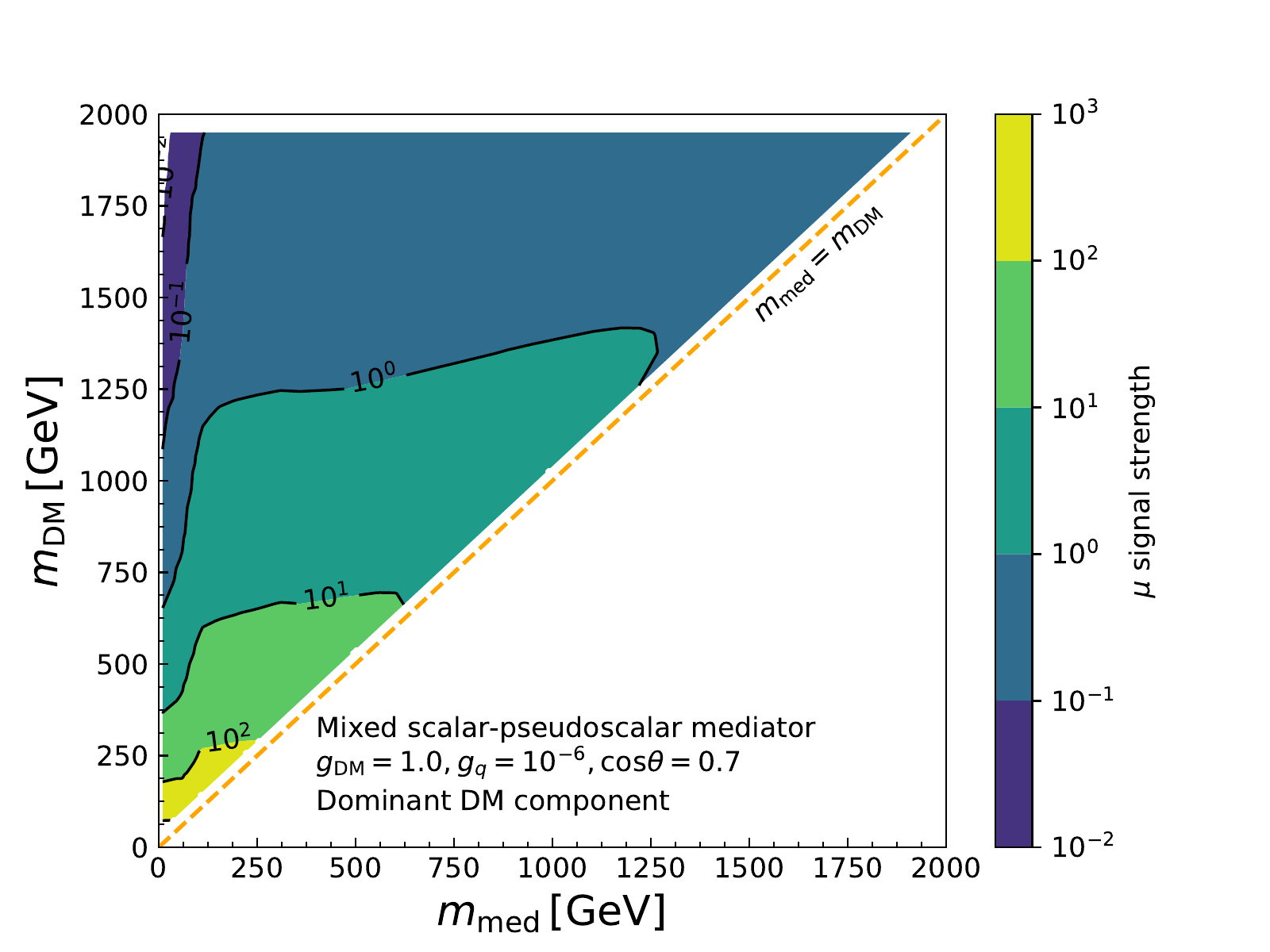}
\end{minipage}
\caption{Results for sequential grid scan for a mixed scalar pseudoscalar mediator model with Dirac dark matter. We study the $X_d X_d \to Y_0 Y_0 \to 4 t$ process. In the left panel we show the behaviour of the relic density: the grey region denotes over-abundant dark matter, while the blue region is for under-abundant dark matter. The black solid line denotes the correct relic density value. In the right panel we show the exclusion by the Fermi-LAT dSph likelihood: this is signaled by the region with signal strength $\mu \equiv \sigmav_{\rm pred}/\sigmav_{\rm ul} \ge 1$. In both panels, the lower triangle designated by $m_{\rm DM} < m_{Y_0}$ has not been considered in the sequential scans as it leads to over-abundant dark matter. The model benchmark is as labelled in the plot.}
\label{fig:yy4t}
\end{figure*}

The model we use in this section is an extension of the dark matter simplified model with spin-0 mediator in the $s$-channel. The interaction Lagrangian is given by:
\begin{eqnarray}
	\mathcal{L}  =  g_q y_q \, \bar{q} \left[ {\rm cos} \,\theta +  i\, {\rm sin}\, \theta\, \gamma_5 \right] q \,Y_0 
			   +   g_X \, \bar{X}_d \left[ {\rm cos} \,\theta+ i \, {\rm sin}\, \theta\, \gamma_5 \right] X_d\, Y_0\,, \label{eq:lagrmixed}
\end{eqnarray}
that allows for a mixed scalar-pseudoscalar mediator $Y_0$. The amount of scalar against pseudoscalar component is quantified by the $\theta$ mixing angle: for instance for  $\cos\theta=1$ we retrieve the usual scalar mediator model. In Eq.~\eqref{eq:lagrmixed} $y_q\equiv \sqrt{2} m_q / v_h$ is the quark Yukawa coupling with $v_h=246 \GeV$ and $m_q$ is the quark mass. 

This model presents a rich dark matter phenomenology, depending for instance on the $Y_0$ mass scale and on the size of the coupling with the SM. The case of light and long-lived mediators has been studied in Ref.~\cite{Arina:2017sng}, showing that solar gamma rays are a powerful tool to constrain such scenario. The same model has been considered in the context of self-interacting dark matter
in~\cite{Kahlhoefer:2017umn}. In this work we focus on a very different region of the model parameter space, along the line of secluded dark matter~\cite{Pospelov:2007mp}. 

The main idea of secluded dark matter consists in having a dark matter candidate that is a thermal relic, however the freeze-out is achieved through a combination of dark matter annihilations to a metastable mediator, which subsequently decays to SM particles. Provided the dark matter mass ($m_{\rm DM}$) is greater than that of the mediator ($m_{\rm med}$), it can be secluded from the SM by setting an extremely small coupling $g_q$ with the quarks. We show in Fig.~\ref{fig:yy4t} (left panel) the behaviour of the relic density for the benchmark point with $g_q=10^{-6}$ and maximal mixing. Even though the dark matter can annihilate via $s$-channel $Y_0$ mediation into quarks directly, the $g_q$ coupling is so small that such interactions are negligible at freeze-out and lead to over-abundant dark matter (lower triangle, for $m_{\rm DM} < m_{\rm med}$, not considered in the sequential scan). The viable region for achieving the correct relic density in denoted by the upper coloured triangle and  is dominated by the $t$-channel process $X_d X_d \to Y_0 Y_0$. The blue region indicates under-abundant dark matter, while the grey region is for over-abundant dark matter, while the black line show were the relic density comes out just right. 

The region with $m_{\rm DM} > m_{\rm med}$ is a difficult region to probe at direct detection because the smallness of $g_q$ has the effect of leading to very small elastic scattering cross-section on nuclei, far below the XENON1T exclusion limit. It is also difficult to test at collider, for two reasons. First, the standard mono-$X$ plus missing energy searches for dark matter are sensitive to the region where $m_{\rm med} \ge m_{\rm DM}$. The $m_{\rm DM} > m_{\rm med}$ can be probed by direct production of the mediator, which subsequently decays back into SM quarks, however, again the smallness of $g_q$ depletes enormously the rate for direct $Y_0$ production (see \ie Ref.~\cite{Arina:2016cqj}). 

In contrast with direct detection and collider searches, indirect signatures from dark matter annihilation are potentially consistent with a weak scale cross-section.  Dark matter annihilation at present time in dSphs is given by the same process that sets the relic density: $t$-channel annihilation into a pair of mediators. For a Dirac dark matter candidate, if the mediator is a pure scalar or a pure pseudoscalar mediator, this process is $p$-wave suppressed, hence it is useful to set the relic density but cannot be detected in gamma rays at present time. In the case of a mixed scalar and pseudoscalar mediator the $p$-wave suppression is lifted and becomes an $s$-wave (see Ref.~\cite{Arina:2017sng} for the analytic behaviour of the $t$-channel process, for which we have checked that \maddm v.3.0 reproduces the analytic computation). Hence dark matter annihilation do provide observable gamma-ray signals and are already challenged by the Fermi-LAT exclusion limits from dSphs. This is clearly illustrated in the right panel of Fig.~\ref{fig:yy4t}: the Fermi-LAT exclusion limits already constraints this model for dark matter and mediator masses up to around 600\,GeV. The excluded region is denoted by the line with signal strength $\mu =1$, where we have defined as usual the signal strength as $\mu  \equiv \sigmav_{\rm pred}/ \sigmav_{\rm ul}$ with $\sigmav_{\rm pred}$ being the theoretical annihilation cross-section and $\sigmav_{\rm ul}$ being the experimental excluded cross-section at 95\% CL. Here we assume that $X_d$ constitutes 100\% of the cold dark matter content of the universe, regardless of the thermal abundance (`all DM' scenario). 
Note that in this scenario $\mu\propto g_X^4$, so that the constraints for other choices of the coupling can easily be inferred. On the contrary, for the `thermal' scenario (see Sec.~\ref{sec:rescale}) $\sigmav$ has to be rescaling by $\xi^2$ and the slice in the parameter space shown in the right panel of Fig.~\ref{fig:yy4t} is entirely unconstraint. In this scenario, Fermi-LAT only provides sensitivity to smaller couplings, $g_X\lesssim1/4$, and masses, $m_{\rm DM}\lesssim100\,$GeV.

\noindent
This study is easily reproducible with \maddm v.3.0 by typing the following commands:\footnote{The UFO files for this model can be found in~\cite{dmsimp}.}
\begin{verbatim}
import model DMsimp_s_spin0_mixed_MD
generate relic density
define q= u d s c b t
define qbar=  u~ d~ s~ c~ b~ t~
add indirect detection y0 y0, y0 > q qbar
output secluded_dm_gammarays
\end{verbatim}
This first option asks the decay of on-shell mediators directly within \madgraph. A second option is:
\begin{verbatim}
import model DMsimp_s_spin0_mixed_MD
generate relic density
define q= u d s c b t
define qbar=  u~ d~ s~ c~ b~ t~
add indirect detection y0 y0
output secluded_dm_gammarays
\end{verbatim}
where the user asks only for the annihilation channel and \pythia takes care of the $Y_0$ decay subsequently. We have checked that the two methods are strictly equivalent for on-shell mediators. After having chosen the preferred option, upon the launch command the scan can be performed as:
\begin{verbatim}
launch secluded_dm_gammarays
set WY0 auto
set nevents 50000
set gq 1e-6
set theta 0.7
set MXd scan1:[10*x for x in range(1,200) for y in range(1,200) if x < y]
set MY0 scan1:[10*y for x in range(1,200) for y in range(1,200) if x < y]
set save_output all
\end{verbatim}
The last set saves all the energy spectra generated by Pythia for each sampled point in the parameter space, while the default option erases them.

\subsection{Final state radiation in dark matter annihilation: the case of internal bremsstrahlung}\label{sec:vib}
\begin{figure*}[t]
	\begin{minipage}[t]{0.5\textwidth}
		\centering
		\includegraphics[width=1.\columnwidth]{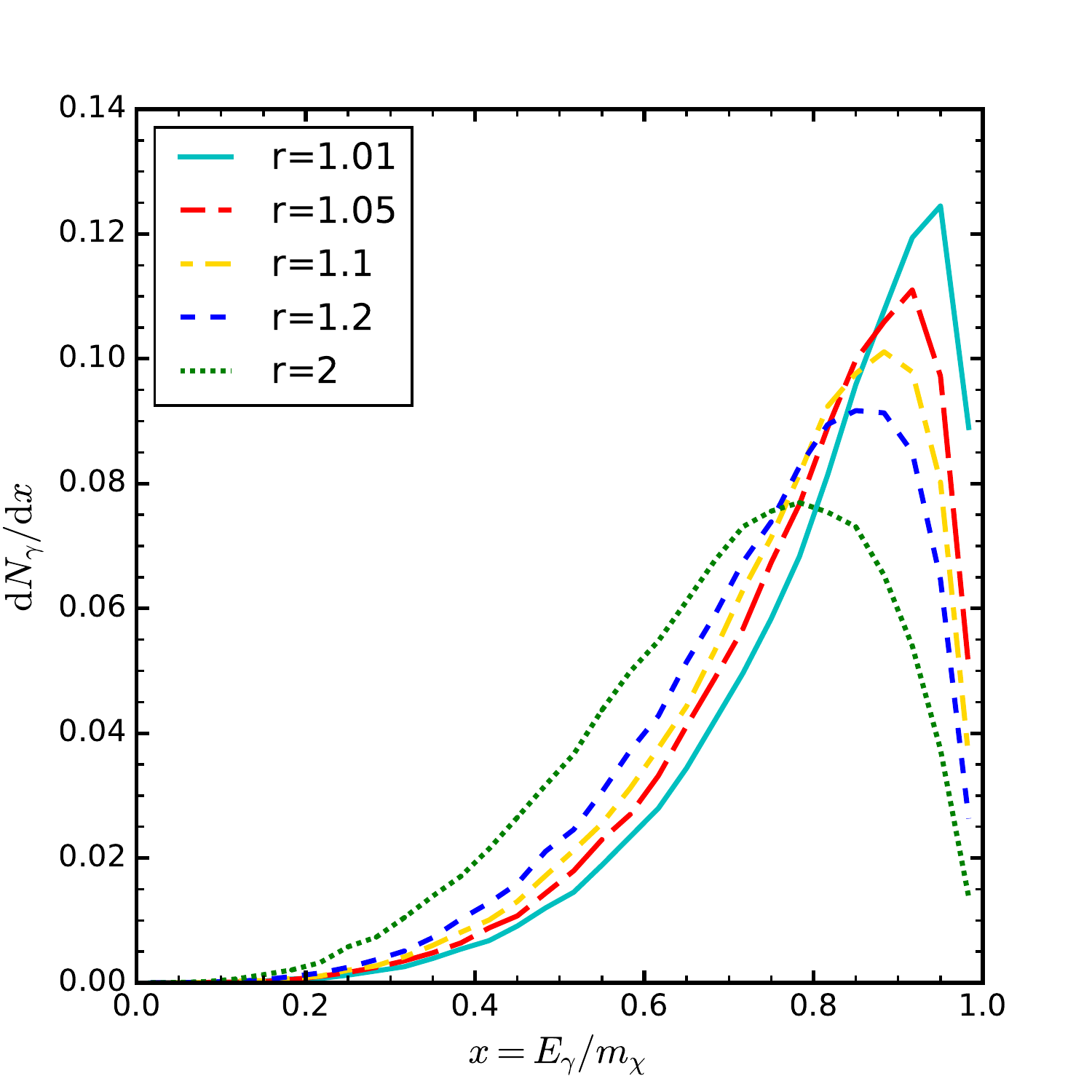}
	\end{minipage}
	\begin{minipage}[t]{0.5\textwidth}
		\centering
		\includegraphics[width=1.\columnwidth]{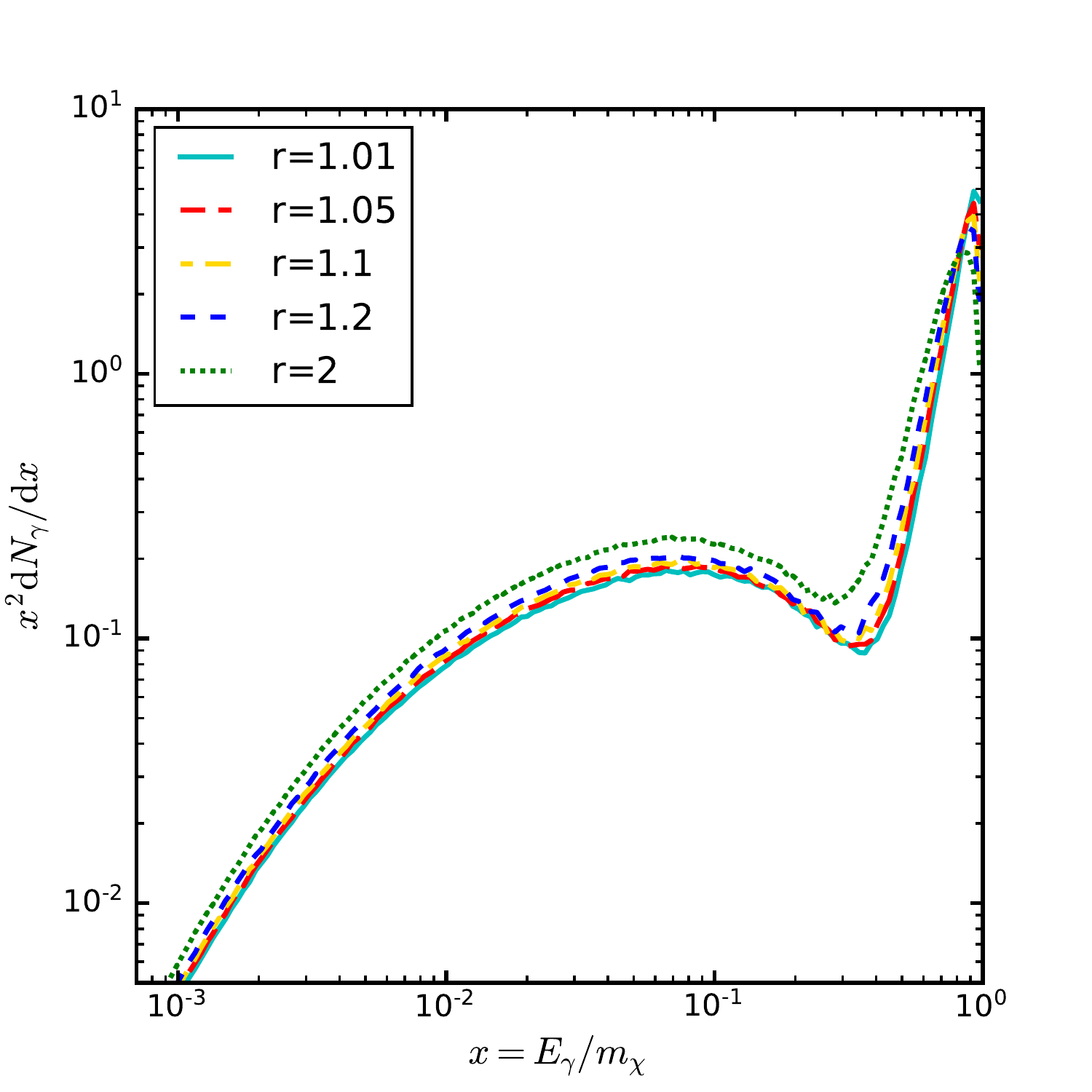}
	\end{minipage}
	\caption{Left:  energy spectrum of the photon for $\chi_r \chi_r \to q \bar{q} \gamma$ annihilation. Different values of $r=M_\Psi/M_\chi$ have been considered, keeping the dark matter mass fixed at 100 \GeV. Right:  gamma-ray distribution produced after parton showering and hadronisation of the final state with \pythia, for different values of $r$. The sharp increase at high $x$ values corresponds to internal bremsstrahlung.}
	\label{gamma_distro}
\end{figure*}

In this section we show how \maddm can easily reproduce the internal bremsstrahlung phenomenon in $2 \to 3$ annihilation processes~\cite{Giacchino:2014moa,Giacchino:2015hvk,Bringmann:2007nk,Ibarra:2014qma,Bringmann:2012vr,Kachelriess:2007aj,Toma:2013bka}. Let us consider a simplified $t$-channel mediator model where dark matter is a real scalar gauge singlet $\chi_r$ that couples to the fermions of the SM (either quarks or leptons) and a heavy fermion  mediator $\Psi$. Interactions with quarks (leptons) imply the mediator $\Psi$ to be a colour triplet (singlet). Being  the two cases analogous (the results differ only by an overall factor due to the gauge charges) we focus on the interaction with quarks. The Lagrangian describing the Yukawa interaction  is
\begin{equation}
	\mathcal{L} \supset y_R \, \chi_r \, \bar{\Psi}_R \, q_R + y_L \, \chi_r \, \bar{\Psi}_L \, Q_L + h.c. \, ,
\end{equation}
where $q_R$ and $Q_L$ are the SU(2) singlet right-handed quark and the SU(2) doublet left-handed quarks, respectively, while $y_R$ and $y_L$ are the coupling parameters. In order to preserve SU(2) gauge symmetry, $\Psi_R$ is a singlet, while $\Psi_L$ is a doublet. For our purposes, it is enough to restrict to a model involving only the interaction with right-handed quarks.

This model (and simple variations thereof) exhibits interesting features when considering dark matter annihilation at threshold.  In the case of annihilation to massless fermions $\chi_r \chi_r  \to f \bar f$,  the $2 \to 2$  amplitude is  $d$-wave suppressed ($\sigma v \propto v^4 $). As a result, despite formally being higher-order in $\alpha$, the $2 \to 3$ annihilation process with bremsstrahlung of a photon gives the dominant contribution. The 3-body final state cross section is no longer suppressed at threshold and  therefore becomes insensitive to the non-relativistic relative velocity ($v \sim 10^{-3}$ in the galactic centre for example). In addition to that, depending on the available phase space, the internal bremsstrahlung can give a rather sharp spectral signal resembling a monochromatic line, see Fig.~\ref{gamma_distro}. The results shown there for different values of the mass of the mediator  have been obtained by \maddm and found to be in perfect agreement with the analytical computation of Ref.~\cite{Giacchino:2013bta}. In fact, such behaviour is not specific to the model considered here, but it is a common feature of scenarios in which the produced fermions are massless or very light, as the $s$-wave contribution in the $2 \to 2$ annihilation is proportional to the mass of the fermion. Had we considered Majorana dark matter interacting through a scalar or a vector mediator, we would have observed the same features (see Refs.~\cite{Barger:2011jg,Garny:2011ii,Bell:2008ey,Bringmann:2017sko}).

\begin{figure*}[t]
	\begin{minipage}[t]{0.32\textwidth}
		\centering
		\includegraphics[width=1.\columnwidth]{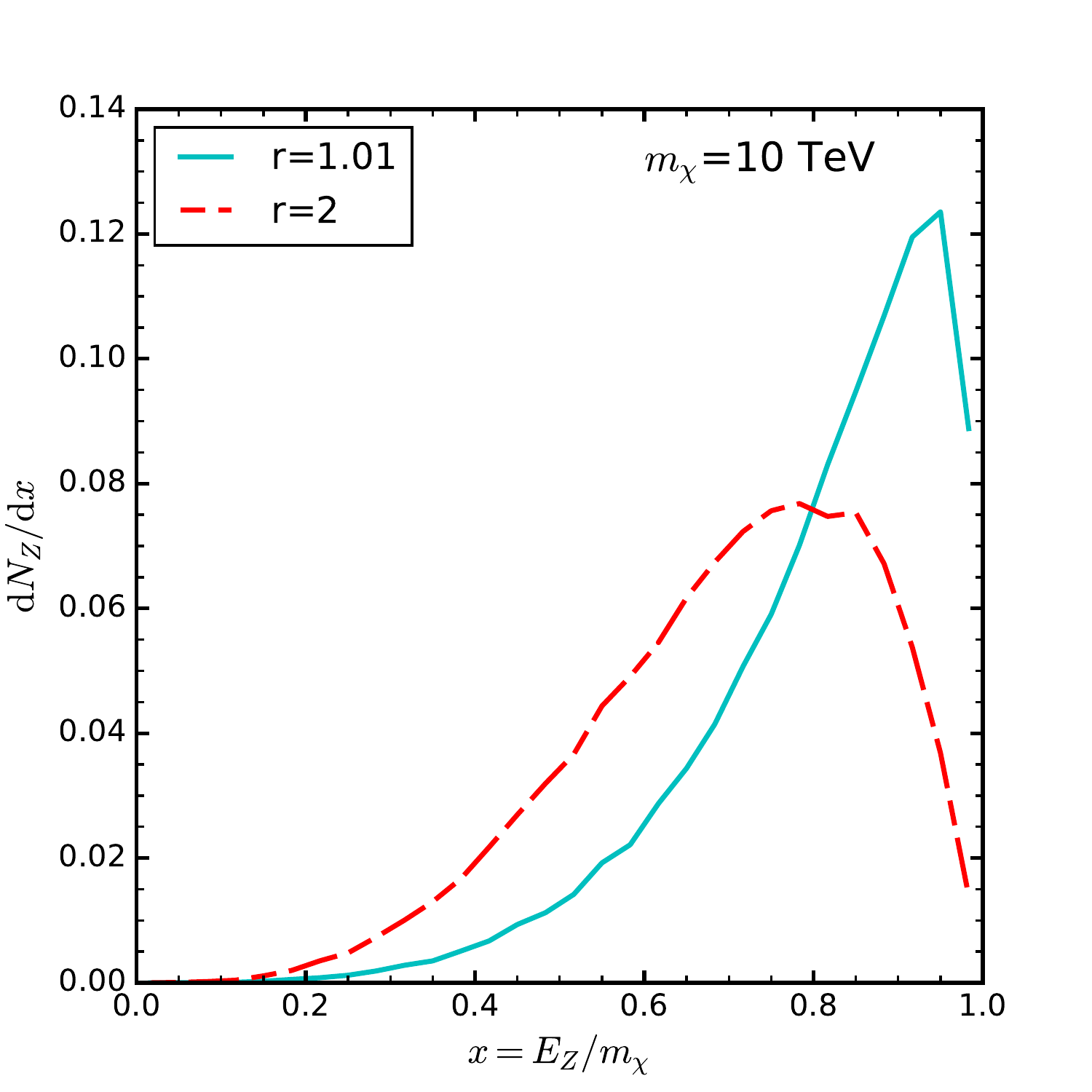}
	\end{minipage}
	\begin{minipage}[t]{0.32\textwidth}
		\centering
		\includegraphics[width=1.\columnwidth]{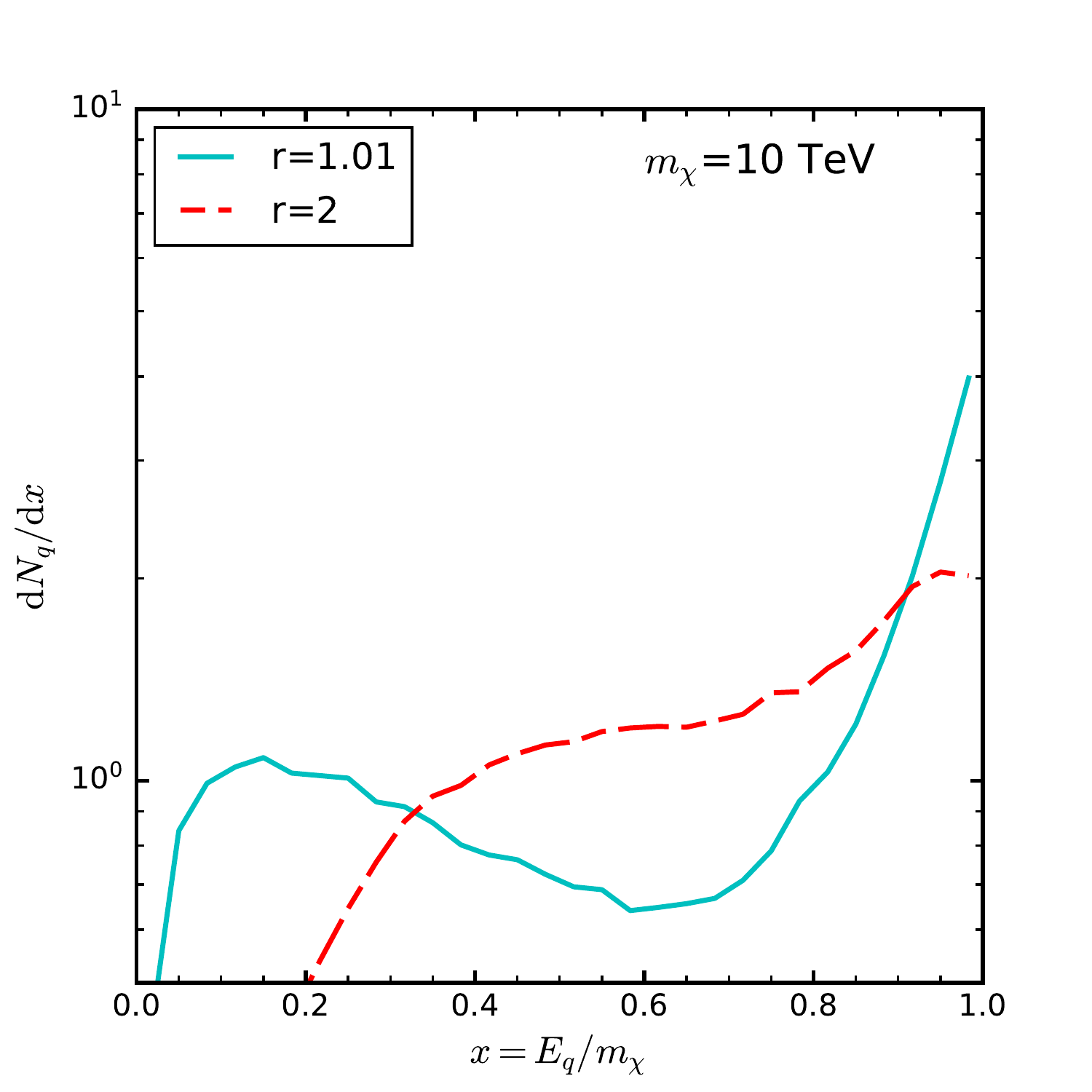}
	\end{minipage}
	\begin{minipage}[t]{0.32\textwidth}
		\centering
		\includegraphics[width=1.\columnwidth]{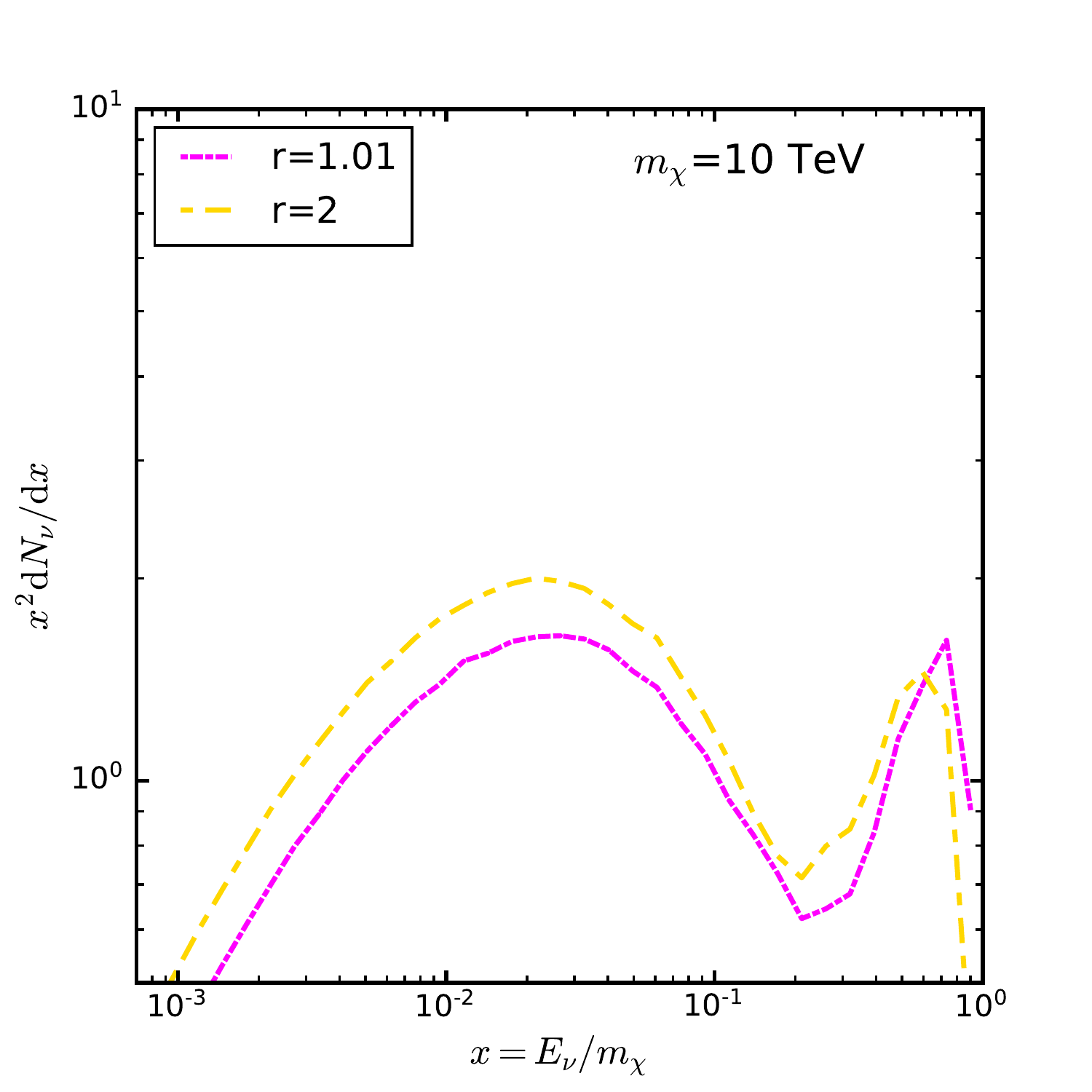}
	\end{minipage}
	\caption{Left:  $Z$ boson energy distribution and quark spectrum, respectively, for two different values of $r=M_\Psi/M_\chi$, for $\chi_r \chi_r \to q \bar{q} Z$ annihilation. The mass of dark matter has been fixed to 10 \TeV. Centre: quark distributions. Right: neutrino distribution after parton shower and hadronisation has been performed by \pythia, for two different values of $r$. The bump in high-energy fraction region is still visible, yet not as pronounced as in the case of photon radiation.}
	\label{z_distro}
\end{figure*}

By employing \texttt{MadAnalysis5}~\cite{Conte:2012fm}, we analyse the events at the parton as well as at the hadron level, after parton shower and hadronisation as obtained by \pythia.   At the parton level, we find that the peak of the spectrum moves to lower energies for large values of $r=M_\Psi/m_\chi$ and that the distribution is also more spread.
At the hadron-level,  the distribution of gamma rays gets sizable contributions at low $x$ from hadrons decaying to photons (such as $\pi^{0} \to \gamma \gamma$), see Fig.~\ref{gamma_distro}.  Yet, the sharp profile given by the photons produced by internal radiation is clearly visible at high $x$, see Ref.~\cite{Bringmann:2013oja} for a similar plot obtained with a Majorana dark matter.

Having reproduced the photon radiation pattern, we then consider the spectrum of neutrinos coming from $Z$ bremsstrahlung (see Ref.~\cite{Bell:2010ei,Bell:2011if}). In this case we expect that when the dark matter particle is heavy compared to the $Z$ boson and the mediator mass is nearly degenerate with the dark matter mass, a similar behaviour to the photon emission should be found. To check this explicitly, we have considered a dark matter particle with a mass of 10 \TeV. We have found  that the spectrum of the radiated $Z$ boson is identical to the photon one, see  Fig.~\ref{z_distro}. Decaying the $Z$ into neutrinos and accounting for the corresponding branching ratio, gives a less prominent bump compared to photon emission (see Fig.~\ref{z_distro}), as expected. 
\\
\\
\noindent The series of commands necessary to obtain the results shown above are:

\begin{verbatim}
import model DMsimp_t_f3
define darkmatter xr
generate indirect detection u u~ a
output test_uuxa
launch test_uuxa
set sigmav_method madevent
set indirect_flux_source pythia8
set vave_indirect 1e-3
set nevents 100000
set Mxr 100
set MYur1 scan:[101,105,110,120,200]
set save_output all
\end{verbatim}

\subsection{Model parameter sampling}\label{sec:mnvad}

In this section we provide an example on how to run a sequential grid scan and a \PyMN sampling over two free parameters of the model. We use this example to validate the results obtained with \maddm v.3.0 with respect to the past versions of \maddm and with respect to the official results presented by the LHCDM working group~\cite{Albert:2017onk}.                       
                        
We consider two simplified dark matter models, an $s$-channel scalar mediator and a $s$-channel spin-1 axial-vector mediator and Dirac dark matter (see~\eg~\cite{Boveia:2016mrp}). The interaction Lagrangian of the $Y_0$ mediator with a Dirac dark matter candidate $X_d$ is given by:
\begin{equation}
 {\cal L} = \bar X_d \left(g^{S}_{X_d}+ig^{P}_{X_d}\gamma_5\right)X_d\, Y_0 + \sum_{i,j} \left[\bar d_i \frac{y_{ij}^d}{\sqrt{2}}
       \left(g^{S}_{d_{ij}}+ig^{P}_{d_{ij}}\gamma_5\right)d_j + \bar u_i \frac{y_{ij}^u}{\sqrt{2}}
       \left(g^{S}_{u_{ij}}+ig^{P}_{u_{ij}}\gamma_5\right)u_j\right] Y_0 \,,
\end{equation}
where $g^{S/P}$ are the scalar/pseudo-scalar couplings of the dark matter and of the quarks.
Similarly, the interaction Lagrangian of the  spin-1 mediator $Y_1$ with $X_d$ and with the quarks is given by:
\begin{equation}\label{eq:vector_mediator}
 {\cal L} = \bar X_d \gamma_{\mu} \left(g^{V}_{X_D}+g^{A}_{X_d}\gamma_5\right)X_d\,Y_1^{\mu} + \sum_{i,j} \left[\bar d_i\gamma_{\mu}
    \left(g^{V}_{d_{ij}}+g^{A}_{d_{ij}}\gamma_5\right)d_j  +\bar u_i\gamma_{\mu} \left(g^{V}_{u_{ij}}+g^{A}_{u_{ij}}\gamma_5\right)u_j\right] Y_1^{\mu} \,,
\end{equation}
where $d$ and $u$ denote down- and up-type quarks, respectively,  ($i,j$=1,2,3) are flavour indices, and $g^{V/A}$ are the vector/axial-vector couplings of the dark matter and the quarks.

\begin{figure}[t!]
\begin{minipage}[t]{0.5\textwidth}
\centering
\includegraphics[width=1.\columnwidth]{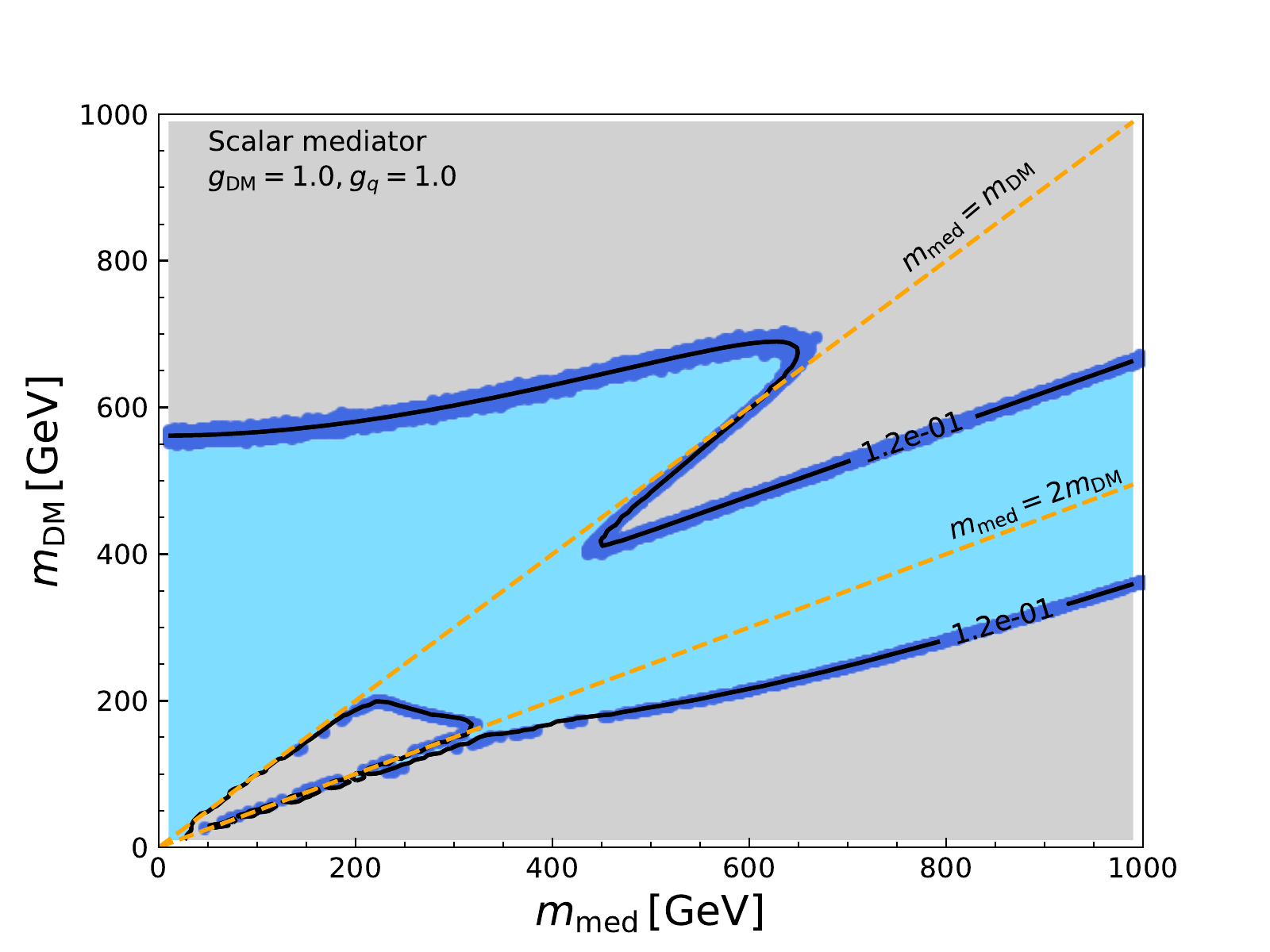}
\end{minipage}
\begin{minipage}[t]{0.5\textwidth}
\centering
\includegraphics[width=1.\columnwidth]{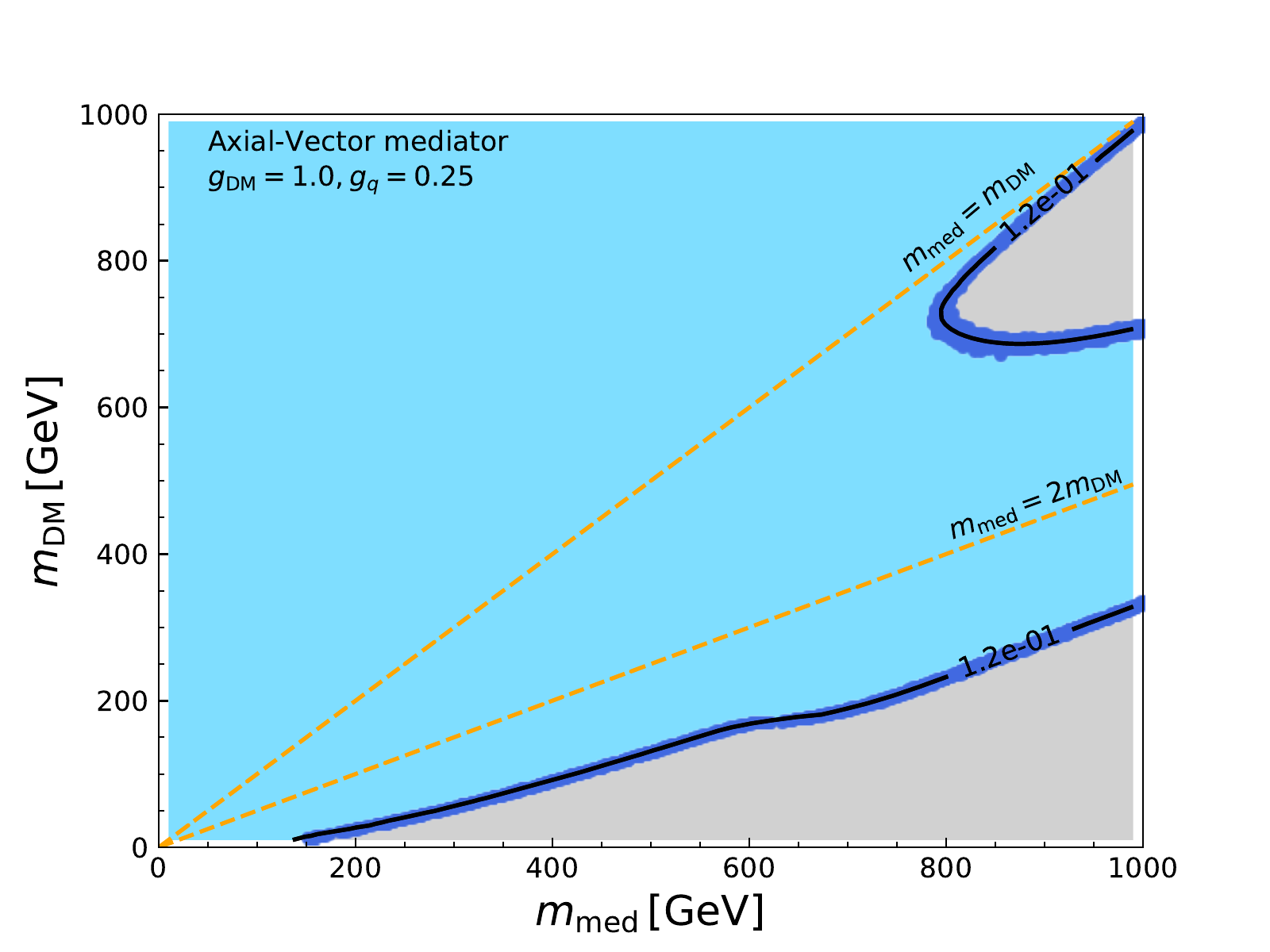}
\end{minipage}
\caption{Validation plots for sequential grid scan and \PyMN sampling using two dark matter simplified models. The left panel is for a scalar mediator, while the right panels denotes the case of spin 1 mediator, in both cases the dark matter is a Dirac fermion. The gray region denotes over-abundant dark matter, while the blue region is for under-abundant dark matter. The black solid line denotes the correct relic density value obtained with the sequential grid scan. The dark blue points are the results of the \PyMN parameter sampling, asking for 100\% dark matter component. The orange dashed lines stand for mediator mass equal to the dark matter mass and twice the dark matter mass, as labelled. In both panels the couplings are fixed at the values labelled in the plots.}
\label{fig:validgridmn}
\end{figure}

Figure~\ref{fig:validgridmn} shows the regions where the dark matter $X_d$ fulfills the relic density constraints in both models. First, we comment on the scalar mediator model shown in the left panel. The results obtained with the sequential grid scan are represented by the black contour which traces $\Omega h^2=0.12$ and the shaded regions: the blue region denotes under-abundant dark matter while the grey region denotes over-abundant dark matter. The \PyMN scan was guided by the requirement of fitting the thermal abundance to the measured value. Accordingly,
the \PyMN scan (given by the dark blue dots) properly samples the required parameter space, falling on top of the black line. If we had chosen under-abundant dark matter as well, the \PyMN would have filled in all the blue region. The results of both sampling methods are well in agreement with the previous version of \maddm v.2.0 and with analytic estimations, which has been used by the LHCDM working group~\cite{Albert:2017onk}. The main improvement in \maddm v.3.0 is the automatisation of the sampling capability, which can be performed simply with few lines of code.

The right panel of Fig.~\ref{fig:validgridmn} shows the results for the analogous scans performed for the case of a spin-1 mediator. Again, these results are in perfect agreement with the ones obtained with \texttt{MicrOMEGAs} and the analytic estimations presented in~\cite{Albert:2017onk}.\footnote{The respective result from \maddm v.2.0 reported in~\cite{Albert:2017onk} appears to be inconsistent with the others. However, we could not reproduce this discrepancy with neither \maddm v.2.0 nor v.3.0.}

This study is easily reproducible by the following commands.
Let us suppose we want to sample the parameter space over the mediator mass and the dark matter for fixed couplings for the scalar mediator model, to reproduce Fig.~\ref{fig:validgridmn}. We can first generate the output directory:\footnote{The UFO files for these models can be found in~\cite{dmsimp}.}
\begin{verbatim}
   import model DMsimp_s_spin0_MD 
   define darkmatter ~xd
   generate relic_density
   output sampling_s0_mxd_my0 
\end{verbatim}
Then we can run the sequential grid scan by performing the following commands upon the \verb|launch| command:
\begin{verbatim}
   launch sampling_s0_mxd_my0
   set MXd scan:[10*x in range(0,100)]
   set MY0 scan:[10*x in range(0,100)]
   set gSXd 1
   set gSu11 1
   set gSu22 1
   set gSu33 1
   set gSd11  1
   set gSd22 1
   set gSd33 1
   set WY0 AUTO
\end{verbatim}
Instead the \PyMN run is launched by doing:
\begin{verbatim}
   launch sampling_s0_mxd_my0
   nestscan = 0N
   \end{verbatim}
The switch \verb|nestscan=ON| turns on the use of \PyMN. Notice that direct detection and indirect detection should be turned \verb|OFF| because we asked only for the relic density computation. In the \PyMN card the user can set up the parameters for the nested sampling run, such as choose the number of live points, the parameters over which the user wish to scan with their range and the type of likelihood for the observable computed (to open the \verb|multinest_card.dat| file the user can either type 7 in the prompt shell upon the \verb|launch| command). The requirement of having the  dark matter to be always 100\% of the cold dark matter content of the universe implies a gaussian likelihood choice whereas under-abundant dark matter corresponds to the half gaussian choice. There are informations for the other likelihoods as well, which however are not read by \PyMN if direct detection and indirect detection are set to \verb|OFF|. More informations on how to run \PyMN are provided in~\ref{sec:appmn}.
The sequential grid scan and the \PyMN sampling for the axial-vector spin 1 mediator model are achieved by following exactly the same procedure as above for the model \verb|DMsimp_s_spin1_MD|.

\section{Conclusions and future prospects}\label{sec:concl}

The need of efficiently combine results from various dark matter searches and globally test them against theoretical models has triggered an increasing demand for more complete dark matter numerical tools. \maddm is an ongoing effort to connect dark matter collider phenomenology with astroparticle physics and cosmology, with the ultimate goal of providing a comprehensive dark matter package that can be easily incorporated into the interpretation of current and future dark matter searches at the LHC\@. 

In its first release, \maddm v.1.0 performed only relic density computations. In \maddm v.2.0 the possibility of computing the cross sections relevant for dark matter direct detection was added. The latest version, \maddm v.3.0,  completes the original goals of the project and provides dark matter indirect detection tools for generic new physics models. The code computes the velocity averaged cross section at present time at a given velocity,  produces the energy spectra of gamma rays, positrons, anti-protons and neutrinos originating from dark matter annihilation and evaluates the flux at Earth for prompt photons and neutrinos. The propagation of cosmic rays from production to detection can be obtained with the \dragon propagation code, for which an interface is provided in \maddm\ .   Predictions can be computed in two different modes: the `fast' mode, suitable for quick scans via $2 \to 2$ processes with final state composed by SM particles only and the `precise' mode where arbitrary $2 \to n$ annihilation process can be considered, also including BSM final states and which can take into account the velocity distribution of dark matter particles in the halos.

Besides the theoretical predictions for dark matter indirect detection, \maddm v.3.0 now provides a platform to test  generic dark matter models against a broad set of experimental data. Using the available features of the \madgraph platform (including the library of LHC analyses in \texttt{MadAnalysis5} ) new physics models can be tested against collider data on the one hand and cosmological, direct detection and dark matter annihilation on the other.  In particular,  gamma-ray exclusion limits from dwarf spheroidal galaxies released by the Fermi-LAT satellite, which are at present among the most constraining and robust bounds from indirect detection, are now accessible in the platform. Moreover, an interface to perform either sequential grid scans or nested sampling of the parameter space is available. 

For illustration purposes, we have tested and run \maddm v.3.0 on several scenarios. First, we have derived 
constraints on a secluded dark matter model where annihilation proceeds via a $2\to4$ process. We have found that for small couplings of the mediator to the SM, indirect detection constitutes the only promising search strategy. This allowed us to set limits for dark matter masses up to around 600\,GeV for a coupling of the mediator to dark matter of order one. Furthermore, we have computed the annihilation spectra for $2\to3$ processes including internal bremsstrahlung which lifts the helicity suppression present in certain classes of $t$-channel mediator models. Finally, we have compared the two scan modes of \maddm v.3.0 in simplified dark matter models with a scalar and axial-vector mediator, respectively.

In the current release \maddm  includes all basic elements necessary to make it a tool for global tests of dark matter models. There are, however, several features whose implementation would be certainly welcome in the near or mid-term future, and would allow to fully explore the phenomenology of generic dark matter models. For the next \maddm releases, we envisage to focus both on  theoretical improvements and on the comparison between experiments and theoretical predictions. 

On the experimental side, constraints can easily be reinforced by adding further experimental likelihoods encompassing direct and indirect detection experiments. On the theoretical side, several improvements/extensions could be considered. First, the range of mechanisms leading to the measured relic density of dark matter has been extended, recently.  For example, one could consider coannihilation in the absence of relative chemical equilibrium (see \eg~\cite{Garny:2017rxs}), and the freeze-in mechanism, which has become popular with the advent of feebly interacting massive particle dark matter models (see \eg~\cite{Hall:2009bx}). Second, the dark matter indirect detection module could be expanded in several directions ranging from high energy physics effects to pure astrophysics. The inclusion of non-perturbative corrections to the annihilation cross-section belongs to the first category. Examples of such effects are provided by the Sommerfeld enhancement~\cite{sommerfeld,Hisano:2006nn} and bound state formation~\cite{MarchRussell:2008tu,vonHarling:2014kha}, relevant for dark matter models with light mediators (as compared to the dark matter mass) at low velocities such as the one of indirect detection or recombination epoch. The advent of future precision radio telescopes such as the Square Kilometer Array (SKA)~\cite{Acero:2017vei} calls for a precise determination of the low energy photon spectrum produced by dark matter annihilation, taken into account by considering, for instance, synchrotron radiation. Third, cashing on the results achieved in collider physics where one-loop and NLO computations in QCD and EW interactions can be automatically performed in the standard model~\cite{Alwall:2014hca,Hirschi:2015iia,Frixione:2015zaa,Frederix:2016ost} and beyond, see \eg~\cite{Degrande:2014sta,Degrande:2015vaa,Backovic:2015soa,Mattelaer:2015haa,Mattelaer:2016ynf,Arina:2016cqj,Das:2016pbk},  loop-induced and/or NLO accurate computations for several dark matter relevant observables could be now achieved. In this respect, the full integration of \maddm into the \madgraph platform is the first step in this direction as it gives the opportunity to seamlessly access the one-loop and NLO modules. For example, the possibility of automatically computing relic density and indirect detection signals originating from loop-induced processes is of phenomenological relevance and it is currently under study.

\section*{Acknowledgements}
We would like to thank Andrea De Simone and Thomas Jacques for discussions and support in the initial stages of this project. CA would like to thank Andrea Vittino for very helpful discussions on the \dragon code. We are grateful to Marco Cirelli and Mario Kadastik for providing feedback on the annihilation spectra and for the consent of using the spectra released with the \PPPC package. We also would like to thank Kyoungchul Kong, Michele Lucente and Nishita Desai for useful discussions on the code development as well as Alessandro Cuoco and Mattia Di Mauro for helpful discussion on the Fermi-LAT analysis. FA has received fundings from the European Union's Horizon 2020 research and innovation programme as part of the Marie Sk\l{}odowska-Curie Innovative Training Network MCnetITN3 (grant agreement no. 722104). CA is supported by the grant Attract Brains for Brussels 2015 of Innoviris. JH acknowledges support by the German Research Foundation (DFG) through the research unit ``New physics at the LHC''. GM is supported by the United States Department of Energy under Grant Contract de-sc0012704. Computational resources have been provided by the supercomputing facilities of the Universit\'e catholique de Louvain (CISM/UCL) and the Consortium des \'Equipements de Calcul Intensif en F\'ed\'eration Wallonie Bruxelles (C\'ECI) funded by the Fond de la Recherche Scientifique de Belgique (F.R.S.-FNRS) under convention 2.5020.11. This work was partly supported by  F.R.S.-FNRS under the `Excellence of Science` EOS be.h project n. 30820817.

\appendix

\section{\maddm \ {\tt v.3.0} as \madgraph plugin}\label{sec:app2}

The main reason for recasting the \maddm code into a \madgraph plugin is inheritance of \madgraph features. To better explain what this means, we are going to provide a few examples below.

\begin{enumerate}
\item {\it Automatic Width Computation.} The automatic resonance width calculation has been a \madgraph feature for some time. The user can simply set the value of any particle width to \verb|AUTO| in the model parameter card, and \madgraph  will compute the width automatically, together with all the branching ratios. \maddm now uses the same feature inherited from \madgraph. 
\item {\it Set parameter to value.} Similarly to (1.), the user can now directly edit the model parameter card in the prompt shell interface of \maddm to set the value of a model parameter, by typing \verb|set PARAM VALUE|. For instance, let us assume that the user wants to modify the dark matter mass \verb|mx| value appearing in the parameter model card and set it to 100 GeV. This can be done now by typing:
\begin{spverbatim}
set mx 100
\end{spverbatim}
in the prompt shell interface of \maddm, besides editing directly the model parameter card.
\item {\it Sequential scans.} Now \maddm will carry out sequential grid scan by simply using the same scanning syntax allowed already in \madgraph. More precisely, the user can now scan over any number of parameters in a sequential manner, by replacing the value of the parameter in the parameter card with 
\begin{verbatim}
scan:[array of values]
\end{verbatim}
For instance, let us assume that the user wants to perform a scan over a two dimensional grid representing the values of dark matter mass (\verb|mx|) from 100 to 500 GeV in steps of 100 GeV and the coupling of dark matter to the SM particles (\verb|gx|), from $10^{-5}$ to $10^{-1}$ in steps of a factor of 10. This can be achieved by the following parameter card changes:
\begin{spverbatim}
<id> scan:[100, 200 ,300, 400, 500] # mx
<id> scan:[1e-5, 1e-4, 1e-3, 1e-2, 1e-1] # gx
\end{spverbatim}
Here \verb|<id>| is the parameter id which appears in the parameter card.
     
The \verb|scan| syntax will accept any Python command. For instance, the above scan can also be implemented as
\begin{spverbatim}
<id> scan:[100*x for x in range(1, 6)] # mx
<id> scan:[10**x for x in range(-5, 0)] # gx
\end{spverbatim}

Sequential scans can also be performed without a grid. For instance, if the user wanted to perform the above scan over the dark matter mass and couplings, but in a way which only uses \verb|mx| = 100 GeV with \verb|gx| = $10^{-5}$, \verb|mx| = 200 GeV with \verb|gx| = $10^{-4}$ and so on, this can easily be done by replacing \verb|scan| with \verb|scan0|. We will give more details about the output in case of sequential grid scans in Sec~\ref{sec:appsg}.

\item  {\it Simple \maddm parameter card.} Following the structure of \madgraph, the \verb|maddm_card.dat| has been refined and by default it appears in a reduced form. It is divided into four main blocks: (i) Relic Density, (i) Direct Detection, (iii) Indirect Detection and (iv) Global Parameters. Each block contains the most relevant switches that can be edited by the user.  If the user wants to modify more technical parameters, such as for instance the precision of the integrator routine computing $\sigmav$ for the relic density, he/she can do so by switching to the full \verb|maddm_card_full.dat|. This is achieved by typing in the user interface the command:
\begin{verbatim}
update to_full
\end{verbatim}

\end{enumerate}

\section{Installing and Running \maddm}\label{sec:app1}

\subsection{Installation}

Installing \maddm is easy and can be done from the \madgraph command line, by executing\footnote{For command line installation \madgraph version 2.6.2 or higher is required.}
\begin{verbatim}
install maddm
\end{verbatim}

The command will fetch the \maddm code from the repository and create \verb|maddm.py| executable in the \verb|bin| folder of \madgraph. \maddm will run on most Linux and Mac OS X operating systems. At the moment, Windows is not supported.

\subsection{Running MadDM v.3.0}\label{sec:running}

\maddm can be used by executing \verb|maddm.py| from the command line in the \verb|bin| folder of \madgraph. The code will bring up the \maddm command line, equivalent to the command line of \madgraph 5. The user can compute various dark matter observables by executing commands analogous to the standard \madgraph syntax. For instance, if the user has a UFO model called \verb|MyDMmodel|, with the dark matter candidate \verb|chi|, the user can compute all the available observables by typing the following commands:

\begin{verbatim}
     import model MyDMmodel
     define darkmatter chi     
     generate relic_density
     add direct_detection
     add indirect_detection 
     output MyDMproject
     launch
\end{verbatim}

The \verb|define| command can be omitted, in favour of MadDM automatically finding the dark matter candidate. Upon the execution of \verb|launch|, the code will prompt the user to select which of the available calculations should be executed and to change parameters of the computation if needed. The prompt takes the familiar \madgraph format:
 \begin{figure}[h]
\centering
\includegraphics[width=1\columnwidth,trim=0mm 0mm 0mm 0mm, clip]{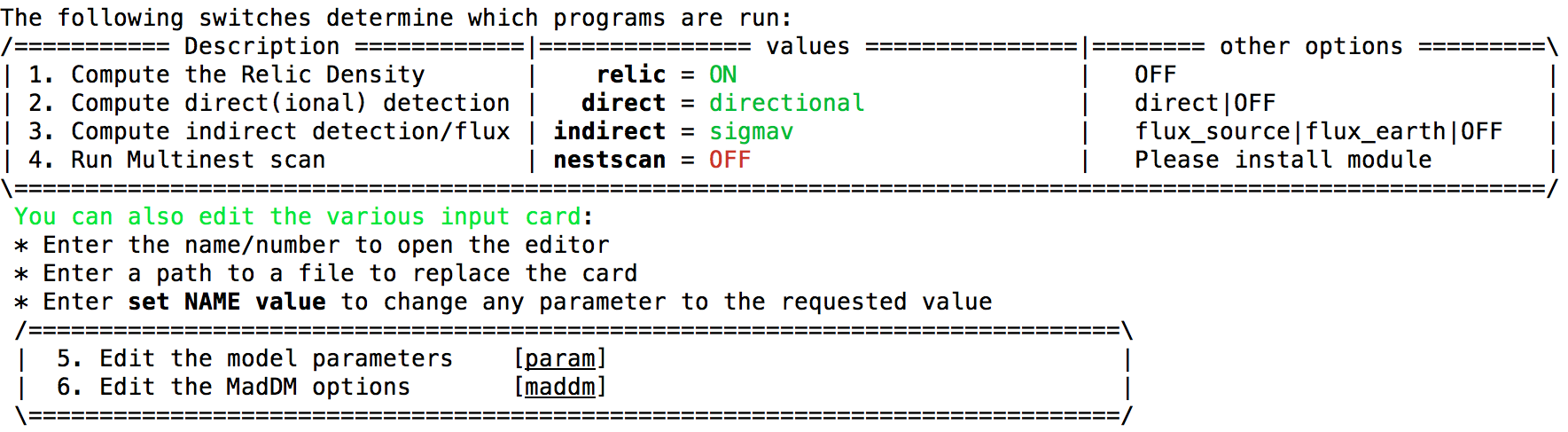}
\end{figure}

Switches 1-3 turn on/off the calculations of various dark matter observables, as labelled. The main features of these switches are presented in the maddm tutorial, that can accessed by typing the \verb|tutorial| command in the user interface.

By setting the \verb|nestscan| switch no.4 to ON, the parameter scan is performed by using the \PyMN algorithm (see Sec.~\ref{sec:parscan} for more details). Switches 5 and 6 allow the user to edit the computation parameters using the \verb|vi| text editor, or the user can also change parameters using the \verb|set| function, as in \madgraph, as it will be explained below in more details (this is the command used throughout the rest of the paper). 

In the next section we will describe thoroughly the case of indirect detection of dark matter (switch 3). For the rest of the paper we will not consider anymore the computation of dark matter relic density or direct detection and we refer the interested user to \maddm v.1.0~\cite{Backovic:2013dpa} and \maddm v.2.0~\cite{Backovic:2015cra} respectively. We briefly mention that the direct detection layout has been changed in this \maddm version: instead of having two separated switches for direct and directional detection, switch no.2 sets now direct detection as `direct', `directional' or `OFF'. If the `direct' option is chosen it computes elastic cross sections between dark matter and nucleons. To have directional information (default if the user chooses \verb|add direct_detection|) the user has to type 2 once more.

\section{The indirect detection module}\label{sec:app3}

Below we provide details on how to run the indirect detection module, described in Sec.~\ref{sec:ID}. 
Table~\ref{tab:summary} schematically provides a brief summary of the capabilities of the two main running modes, `fast' and `precise', together with their interconnection with the new \maddm modules. There are two short cuts available to run either in the fully `fast' mode or 'precision' mode. Those are obtained by typing upon the \verb|launch| command:
\begin{verbatim}
set fast
\end{verbatim}
or
\begin{verbatim}
set precise
\end{verbatim}
The default settings for these two modes are provided in Tab.~\ref{tab:summary}, whereas the details for each command and switch of indirect detection module are described in the following.
\begin{table}[h!]
\caption{Summary of the \maddm indirect detection functionalities upon the execution of the launch command and description of the default settings of the two shortcuts \texttt{set\,fast} and \texttt{set\,precise}.}
\label{tab:summary}
\begin{center}
\begin{tabular}{|c|l|l|}
\hline
& `fast' mode  & `precise' mode \\
& \verb|set fast| & \verb|set precise| \\
& ONLY for SM final states $(2 \to 2)$  & ALL possible final states $(2\to n)$ \\
\hline
& & \verb|sigmav_method = reshuffling| (default) \\
& \verb|sigmav_method = inclusive|  & or can be changed to\\
$\sigmav$ & & \verb|sigmav_method = madevent| \\
\verb|Indirect = sigmav| & & \\
& NO EVENTS generated & EVENTS generated (LHE file) \\
& & \\ 
& output: $\sigmav$ for each $2\to 2$ & $\sigmav$ for ANY\\
& annihilation process (SM and BSM) & annihilation process\\
\hline
& \verb|indirect_flux_source_| & \\
& \verb|method = PPPC4DMID_ew| (default) & \\
Spectra at source & or can be changed into & \verb|indirect_flux_source_|\\
\verb|Indirect = flux_source| & \verb|indirect_flux_source_| & \verb|method = pythia8|  (default)\\
& \verb|method = PPPC4DMID| & \\
& & \\
& computes $\sigmav$ with \verb|inclusive| & computes $\sigmav$ with \verb|reshuffling|\\
& & \\
& output: energy spectra   & output: energy spectra \\
& coming ONLY from SM final states & coming from ANY final state \\
\hline
&  \verb|indirect_flux_earth_| & \verb|indirect_flux_earth_|  \\
& \verb|method = PPPC4DMID_ep| (default) &  \verb|method = dragon| (default)\\
Flux at Earth & & \\
\verb|Indirect = flux_earth| & computes $\sigmav$ with \verb|inclusive|  & computes $\sigmav$ with \verb|reshuffling| \\
& &  \\
& output: $\gamma$ and $\nu_i$ using & output: $\gamma$ and $\nu_i$ using \\ 
&  \PPPC tables at production (ew case) &  \pythia for the prompt energy spectra  \\
& & \\
& $e^+$ flux at Earth  & $e^+$ and $\bar{p}$ fluxes at Earth\\
& using \PPPC tables & using \dragon \\
& or can be changed into &  \\
& $e^+$ and $\bar{p}$ fluxes at Earth & \\
& using \dragon & \\
\hline
Experimental constraints & \verb|ExpConstraints| class & \verb|ExpConstraints| class\\
& + full  Fermi-LAT likelihood for dSphs & + full  Fermi-LAT likelihood for dSphs \\
& ONLY for SM final states &  \\
\hline
\end{tabular}
\end{center}
\end{table}

As a general remark, all computed indirect detection quantities are stored in the directory \verb|MyDMproject/output/| if a single model point is run, while for a sequential grid scan the relevant files are stored in \verb|MyDMproject/output/|\\\verb|output_indirect|. This directory is structured as a typical \madgraph folder that contains the subfolders \verb|run_01|, \verb|run_02|, etc... for each run the user has done. Output files produced by the \dragon code will be stored in the \verb|output| folder of  \dragon and not in \verb|MyDMProject/output|.

\subsection{Velocity averaged cross section}\label{sec:appsigmav}

Assuming the user is only interested in indirect detection, the computation of the velocity-weighted annihilation cross section at present time is selected by typing:
\begin{verbatim}
     import model MyDMmodel
     define darkmatter chi     
     generate indirect_detection 
     output MyDMproject
     launch
\end{verbatim}
The prompt already selects the correct switch:
\begin{verbatim}
 | 3. Compute indirect detection/flux  |  indirect = sigmav  |  flux_source|flux_earth|OFF  |
\end{verbatim}
If nothing is specified together with the command line \verb|generate indirect_detection|, \maddm will compute all possible annihilation channels for the dark matter. If the user is interested in a specific final state, for instance $\chi \chi \to b \bar{b}$, he/she can do instead:
\begin{spverbatim}
generate indirect_detection b b~; output MyDMproject; launch
\end{spverbatim}
We provide the  two following running modes for the $\sigmav$ computation. The selection of the mode is always done by editing the \verb|maddm_card.dat| file as described below.

\paragraph {\it `Fast' Method}

This option corresponds to the switch \verb|inclusive| in the \verb|maddm_card.dat| file. To edit the \maddm card, type in the user interface:
\begin{spverbatim}
set sigmav_method inclusive
\end{spverbatim} 
We remind the user that the \verb|set| command is an inherited feature of \madgraph and allows to quickly edit the parameter or \maddm card without directly opening them.

The evaluation of $\sigmav$ is handled by the fortran side of \maddm. It consists in computing the leading order 2-2 matrix elements for the annihilation process(es) and integrate them over the angle between the two final states. The resulting cross section is furthermore evaluate at the required velocity, which is described by a $\delta$ function distribution centred at $\sqrt{3}v_0$ (see Sec.~\ref{sec:sigmav} for further details). The user can set the desired velocity $v_0$ by editing the \verb|maddm_card.dat|:
\begin{spverbatim}
set vave_indirect 1e-03
\end{spverbatim}
for instance if he/she is interested in the Milky Way halo. The velocity should be provided in natural units. This method allows to compute $\sigmav$ for any $2 \to 2$ process.

The output is provided in \verb|MyDMproject/output| within the file \verb|MadDM_results.txt|. If only relic density and indirect detection modules are selected by the user, the output contains the value of $\Omega h^2$, $\xi$, of $\sigmav$ for each available annihilation channel, together with the corresponding $\sigmav_{\rm ul}$ 95\% CL upper limit from Fermi-LAT dSphs for each SM annihilation final state available in the model. Finally it also provides the total theoretical $\sigmav_{\rm tot}$ together with the total experimental 95\% $\sigmav_{\rm ul}$. We have illustrated its content with an explicit example in Sec.~\ref{sec:valid}. 

\paragraph{\it `Precise' Method} 

This method, as a matter of fact, contains two ways of computing $\sigmav$, \verb|madevent| and \verb|reshuffling| (default choice), the latter being more sophisticated than the former one, but basically all relying on the generation of events via \madgraph. Both options are obtained by setting:
\begin{spverbatim}
set sigmav_method madevent
\end{spverbatim}
or
\begin{spverbatim}
set sigmav_method reshuffling 
\end{spverbatim}
The functioning and the physics of both methods has been described in Sec.~\ref{sec:sigmav}.

Similarly to the `fast' method, the output file \verb|MadDM_results.txt| is created. The generated events are saved inside the output directory  or in \eg \verb|MyDMproject/output/output_indirect/run_01/unweighted_.lhe.gz|, if the grid scan mode is selected. In the scan mode, the user can choose to erase the LHE file only or to delete all files including the energy spectra by setting the parameter \verb|save_output| to \verb|spectra| or \verb|off| respectively in the \verb|maddm_card.dat|. The default value is \verb|off = save_output|, whereas if \verb|all = save_output| all files will be saved.

\subsection{Energy spectra}\label{sec:appes}

The approach of \maddm to compute energy spectra is twofold, namely we made available two operation modes (similarly as for the $\sigmav$ evaluation). Both modes are obtained by typing in the user interface:
\begin{verbatim}
     import model MyDMmodel
     define darkmatter chi     
     generate indirect_detection 
     output MyDMproject
     launch
\end{verbatim}
By typing the switch no. 3, the prompt should select:
\begin{verbatim}
 | 3. Compute indirect detection/flux  |  indirect =  flux_source |  |flux_earth|OFF|sigmav  |
\end{verbatim}

\paragraph {\it `Fast' Method}

This method is selected by editing the \verb|maddm_card.dat|. This is achieved by typing:
\begin{verbatim}
set indirect_flux_source_method PPPC4DMID 
\end{verbatim}
It automatically computes $\sigmav$ with the `fast' method (\verb|inclusive|) without event generation. Then it uses the \PPPC tables\footnote{At the first usage the user will be asked to download a processed version of the \PPPC numerical tables. The download is performed by typing the command \texttt{install} \texttt{PPPC4DMID} within the user interface and is done once for all.} to get the energy spectra of the model point by interpolating among them as a function of $m_\chi$. The tables are available for a dark matter mass ranging from 5 GeV to 100 TeV and for SM final states only.

If the user types:
\begin{spverbatim}
set indirect_flux_source_method PPPC4DMID_ew
\end{spverbatim}

\noindent \maddm will use the tabulated energy spectra that include EW corrections~\cite{Ciafaloni:2010ti}\footnote{Please cite this paper as well when using those tables.}. The weak corrections account for the radiation of weak gauge bosons from the final state SM particles, when the dark matter mass $m_\chi$ is much larger than the EW scale.

For each model point tested by the user \maddm produces the corresponding energy spectra. If the user is testing one single model point at a time the events are saved in  \verb|MyDMproject/output| and are overwritten each time the user executes the \verb|launch| command. If the user is in grid scan mode, the energy spectra are saved in, \eg, \verb|MyDMproject/output/output_indirect/run_01/| and labelled as: \verb|positrons_spectrum_pppc4dmid.dat| 
\linebreak
(positron energy spectrum), \verb|antiprotons_spectrum_pppc4dmid.dat| (anti-proton energy spectrum), 
\linebreak
\verb|gammas_spectrum_pppc4dmid.dat| (photon energy spectrum) and \verb|neutrinos_e_pppc4dmid.dat| (for the electron neutrino energy spectrum; similar names are given for the neutrino $\mu$ and $\tau$). We provide the results for the energy spectra in terms of:
\begin{equation}
\log x = \log\left(\frac{K}{m_\chi}\right) \, \, \, {\rm and}\, \, \, \frac{{\rmd}N_i}{{\rmd} \log x}\equiv \sum_{\rm SM} B_{\rm SM} \frac{{\rmd}N^{\rm SM}_{i}}{{\rmd} \log x}\,,
\end{equation}
where $K$ is the kinetic energy of the particle $i=\gamma, e^+, \bar{p}, \nu_e, \nu_\mu, \nu_\tau$, and the sum of the branching ratio is understood over all possible SM final states (given in Eq.~\eqref{eq:smfs}). The user can choose to erase the energy spectra files by setting the parameter \verb|save_output| to \verb|OFF| in the \verb|maddm_card.dat|, if he/she runs in scan mode.

\paragraph{\it `Precise' Method} 

This method is selected by typing: 
\begin{spverbatim}
set indirect_flux_source_method pythia8 
\end{spverbatim}
It computes $\sigmav$ with the default \verb|reshufflingt| method (which can be changed by the user into \verb|madevent| by typing \verb|set sigmav_method madevent|) and generates the event file {\bfseries \verb|unweighterd_events.lhe.gz|}. This files is subsequently passed to \pythia for decaying, showering and hadronisation. 

We generate the energy spectra with a logarithmic binning to ensure a sufficient number of events in each energy range. The number of events generated by \verb|madevent| can be changed by typing upon the \verb|launch| command:
\begin{verbatim}
set nevents 1000000
\end{verbatim}
The default is \verb|nevents = 10000|, which is quite fast but might be not accurate enough. We suggest the reader to use in between \verb|nevents = 50000| and \verb|nevents = 100000| to obtain a precise energy spectrum. The storing of the spectral files is exactly the same as in the `fast' running mode, see above.  The only difference is the labelling of the files, which now are \verb|species_spectrum_pythia8.dat| (species=gammas, positrons, antiprotons, neutrinos) instead of \verb|species_spectrum_pppc4dmid.dat| to trace the origin of the file. We provide the results for the energy spectra in terms of:
\begin{equation}
\log x = \log\left(\frac{K}{m_\chi}\right) \, \, \, {\rm and}\, \, \, \frac{{\rmd}N_i}{{\rmd} \log x}\equiv \sum_{\rm all} B_{\rm all} \frac{{\rmd}N^{\rm all}_{i}}{{\rmd} \log x}\,,
\end{equation}
where $K$ is the kinetic energy of the particle $i=\gamma, e^+, \bar{p}, \nu_e, \nu_\mu, \nu_\tau$, and the sum of the branching ratio is understood over all possible final states, including new particles beyond the SM (BSM).
\begin{figure*}[t!]
\begin{minipage}[t]{0.5\textwidth}
\centering
\includegraphics[width=1.\columnwidth]{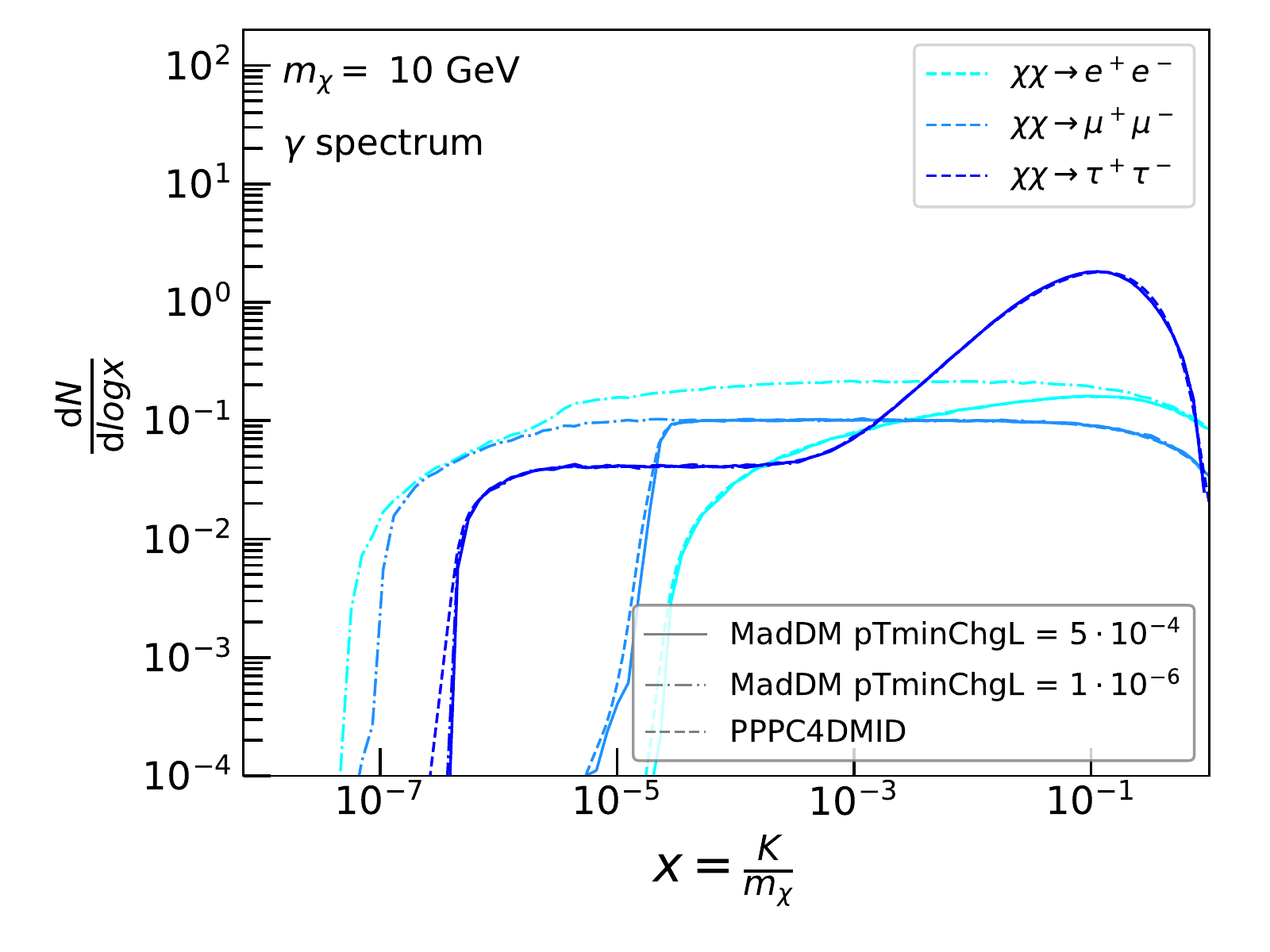}
\end{minipage}
\begin{minipage}[t]{0.5\textwidth}
\centering
\includegraphics[width=1.\columnwidth]{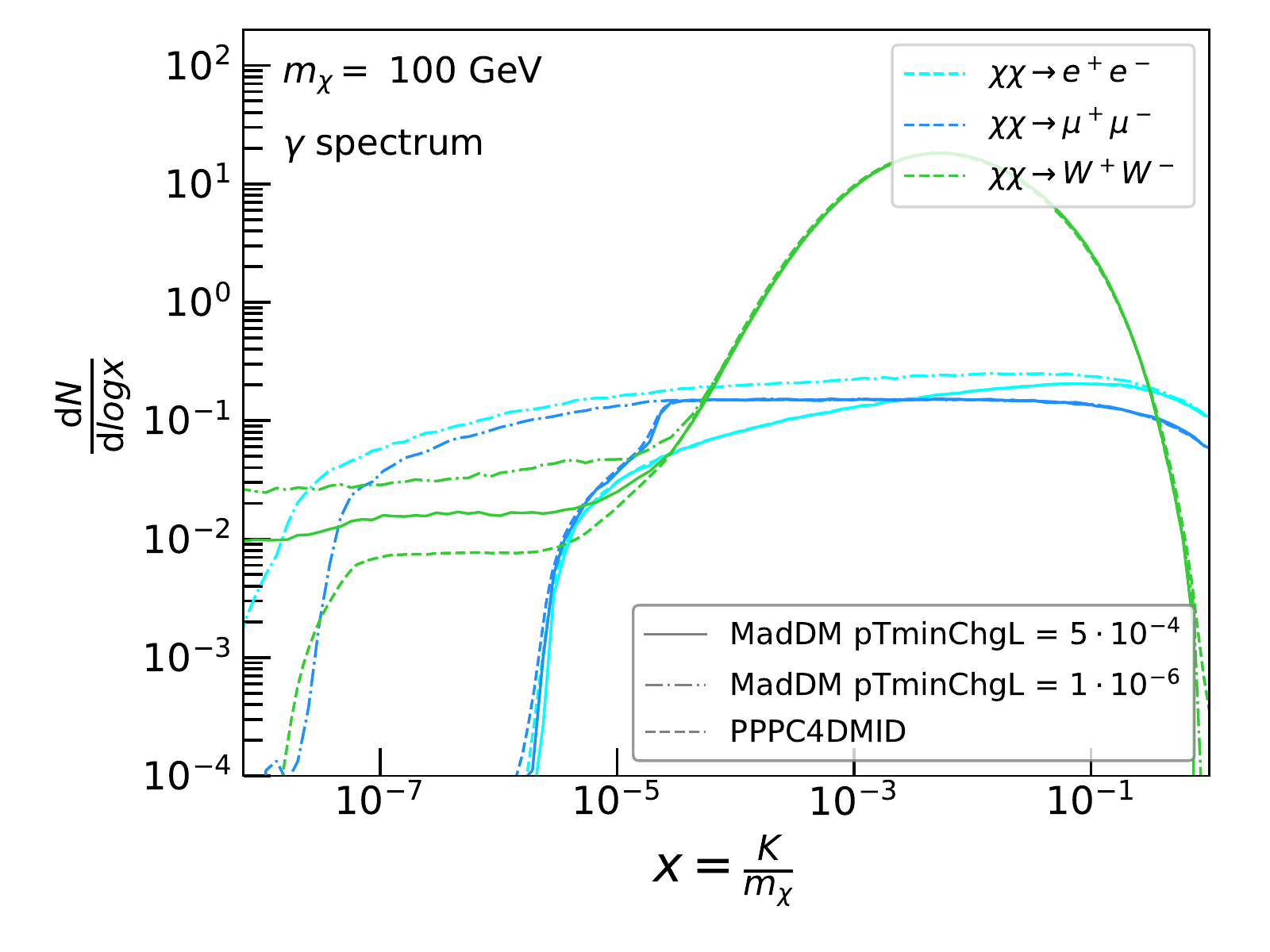}
\end{minipage}
\caption{Dependence on the \pythia\, parameter for  pure QED final state radiation (\texttt{TimeShower}) called \texttt{pTminChgL}. The default value in the most recent \pythia version if \texttt{pTminChgL} = $10^{-6}$, while \PPPC has it fixed at \texttt{pTminChgL} = $5 \times 10^{-4}$. The left panel is for a 10 GeV dark matter mass, while the right panel for a 10 GeV dark matter mass.}
\label{fig:ptmin}
\end{figure*}

We have encountered a small difference with respect to the \PPPC energy spectra, due to the parameter for pure QED final state radiation called \texttt{pTminChgL}. In the last version of \pythia, which we used, it is set by default to the value $10^{-6}$, while in the \pythia version used by \PPPC it is fixed at \texttt{pTminChgL} = $5 \times 10^{-4}$. This parameter sets the cutoff of QED radiation, hence the smaller its value the softer will be the photon energy spectra produced. This is illustrated in Fig.~\ref{fig:ptmin} where we show the photon energy spectrum for selected SM final states, as labelled. The default value of $10^{-6}$ populates the very end of the photon energy spectrum in the case of $\chi \chi \to e^+ e^-, \mu^+ \mu^-$ (left panel) and $\chi \chi \to W^+ W^-$ (right panel), while the other SM channels are unaffected. Such tail at low $x$ has practically no impact for a wide dark matter mass range when predictions or constraints are computed for current experiments such as Fermi-LAT. It can have an impact if the dark matter is very heavy, close to 100 TeV, because in that case $x \simeq 10^{-6}$ falls in the experimental sensitivity range. Besides the differences due the \texttt{pTminChgL} flag, the \maddm spectra of prompt photons originating from $\chi \chi \to W^+ W^-$ is different from the \PPPC one because in the latter \pythia has been modified to allow $W$ boson to radiate photons. 

The user can switch \verb|OFF| weak showering in \pythia (the default is \verb|ON|) by editing the \verb|pythia_card.dat|:
\begin{verbatim}
TimeShower:weakShower = off
\end{verbatim}
Details on the various parameters are provided on the \pythia manual webpage~\cite{pythiaws}.

\subsection{Fluxes of $\gamma$ and $\nu_i$ at Earth}\label{sec:appfluxgn}

The flux at Earth for prompt photons and neutrinos is obtained by the following commands in the user interface:
\begin{verbatim}
     import model MyDMmodel
     define darkmatter chi     
     generate indirect_detection 
     output MyDMproject
     launch
\end{verbatim}
By typing twice the switch no. 3, the prompt should select:
\begin{verbatim}
 | 3. Compute indirect detection/flux  |  indirect =  flux_earth |  OFF|sigmav|flux_source  |
\end{verbatim}
The flux is a \maddm output for both the `fast' and `precise' modes for prompt photons and for all neutrino flavors separately. The difference between the two modes is simply that \maddm will interpolate numerical tables for the prompt photon energy spectrum in the `fast' option, while it will compute the energy spectrum with \texttt{MadEvent} and \pythia for the `precise' mode. We provide the total flux of gamma rays, integrated from 0 to the dark matter mass, and the differential flux, up to the $J$ factor. To be more precise we compute the flux normalised to the Draco dSph galaxy. The user will have to rescale the \maddm predictions by  $J_{\rm user}/J_{\rm Draco}$, with $J_{\rm user}$ being the $J$ factor he/she is interested in. Example values for the $J$ factors are provided in the \maddm folder \verb|Jfactors|: we have stored all the $J$ factors used by the Fermi-LAT collaboration for both gamma-ray line searches towards the Galactic Center~\cite{Ackermann:2015lka} and for dSphs limits~\cite{fermilike}. Examples on how to compute $J$ factors are provided in~\cite{Cirelli:2010xx}. 

The fluxes of gamma rays and neutrinos are provided in the file \verb|MyDMproject/output/MadDM_results.txt|.

\subsection{Fluxes of $e^+$ and $\bar{p}$ at detection}
The flux at Earth of $e^+$ and $\bar{p}$ is obtained by the following commands in the user interface:
\begin{verbatim}
     import model MyDMmodel
     define darkmatter chi     
     generate indirect_detection 
     output MyDMproject
     launch
\end{verbatim}
By typing twice the switch no. 3, the prompt should select:
\begin{verbatim}
 | 3. Compute indirect detection/flux  |  indirect =  flux_earth |  OFF|sigmav|flux_source  |
\end{verbatim}
Additionally the user has to choose the `fast' or `precise' running mode as follows.

\paragraph {\it `Fast' Method}
This method is selected by typing in the prompt:
\begin{spverbatim}
set indirect_flux_earth_method PPPC4DMID_ep
\end{spverbatim}
\maddm computes $\sigmav$ with the `fast' method (\verb|inclusive|) without event generation. Then it calls the \PPPC tables, which are available for positrons only, and interpolates for obtaining the flux at Earth for the dark matter mass $m_\chi$ the user is interested in. The user can choose: (i) the desired dark matter density profile among those four: NFW, Moore~\cite{Ghigna:1999sn}, Isothermal~\cite{Binneybook} and Einasto; (ii) the propagation model (MIN, MED and MAX) and the modeling of the galactic magnetic field (MF1, MF2 and MF3). For details about those settings we refer to~\cite{pppccode} and references therein. The default settings are: Einasto - MED - MF2. These setting can be changed by editing the \verb|maddm_card_full.dat|. 

The differential flux at Earth is given in \verb|MyDMproject/output/run_01/positrons_dphide_pppc4dmid.dat|; 
the first column of this file contains the particle kinetic energy $K$ in units of [$\log_{10}(K/\rm GeV)$] while the second column contains the differential flux ${\rmd} \Phi / {\rmd} \log_{10} K$ in units [GeV$^{-1}$\,cm$^{-2}$\,s$^{-1}$\,sr$^{-1}$].

\paragraph{\it `Precise' Method} 

This method is selected by typing in the prompt:
\begin{spverbatim}
set indirect_flux_earth_method dragon
\end{spverbatim}
\maddm computes $\sigmav$ with the `precise' method, namely \verb|madevent| or \verb|reshuffling|, it generates an output event file in the LHE format that is passed to \pythia for showering and hadronisation. This step is needed to obtain the energy spectra of positrons and anti-protons at production, as a function of their kinetic energy $K$. The propagation of these primary particles produced by dark matter annihilation is then achieved by interfacing with the fully numerical code \dragon~\cite{Evoli:2008dv}. 

The user will have to separately install \dragon on his/her machine to be able to run it.\footnote{Details for the code installation can be found here~\cite{dragon}.} Once the installation is done, the user has to set up the correct path to the \dragon executable within \madgraph. This is achieved by typing in the user interface:
\begin{verbatim}
set dragon_path /the_path_to_the_dragon_executable_folder
\end{verbatim}

The flux at detection of $e^+$ and $\bar{p}$ originating from dark matter annihilation in the Milky Way halo is obtained by running \dragon with the files produced by \maddm. \dragon needs two inputs:
\begin{itemize}
\item energy spectra of $e^+$ and $\bar{p}$ at source: those are provided by \maddm but not stored;
\item an input file called \verb|MyDMproject\Cards\dragon_card.xml|, provided again by \maddm in \linebreak \verb|MyDMproject/Cards|. In this file the CR propagation model is fixed to the default value, and no particles other than $e^+/e^-$ and $p/\bar{p}$ produced by the dark matter are propagated. This is obtained by commenting out the following elements: \verb|<PropLepton />|, \verb|<PropSecAntiProton />| and \verb|<PropExtraComponent />| and setting \verb|<Zmax value="1" />|, \verb|<Zmin value="1" />|. In addition \maddm automatically edits the dark matter related part of the \dragon card, which looks like:
\end{itemize}
\begin{verbatim}
<DarkMatter Model="SelfTable" Profile="NFW" Reaction="Annihilation">
  <PropDMLepton/>
  <PropDMAntiProton/>
  <Mass value="m_\chi"/>
  <SigmaV value="<\sigma v>"/>
  <SSDensity value="0.41"/>
  <LeptonDatafile value="MyDMproject/Indirect/Events/run_01/positrons_dnde.txt"/>
  <AntiprotonDatafile value="MyDMproject/Indirect/Events/run_01/antiprotons_dnde.txt"/>
</DarkMatter>
\end{verbatim}
by inserting the correct value of $\sigmav$, the dark matter mass $m_\chi$ and the path to the CR energy spectra. The dark matter density profile is by default NFW, however the user can easily change to the desired dark matter distribution by editing the \verb|dragon_card.xml| file in \verb|MyDMproject/Cards|, before the \verb|launch| command. The \dragon code allows for the following dark matter density profiles: Isothermal, NFW, Moore, Einasto and Kravtsov~\cite{Bullock:1999he}. 

\dragon provides two output files, which are stored in the \dragon output folder and are called \eg \linebreak \verb|1_DRAGON.fits.gz| and \verb|1_DRAGON.txt|:
\begin{itemize}
\item  \verb|fits| file: it has a series of units (HDU), one for each particle \dragon has propagated (equal particles but produced by different sources, \eg $\bar{p}$ from dark matter annihilation and $\bar{p}$ from CR spallation are saved in different HDU). Each HDU has is own header to classify the particle with some keywords. The most relevant keywords are:
\begin{itemize}
\item Z  = the charge, \eg +1 for protons and -1 for $\bar{p}$;
\item A    = mass number, \eg 0 for leptons, 1 for $p$ and $\bar{p}$;
\item DM = 1 if the particle has been produced by dark matter annihilation, DM = 0 if particle is from astro origin.
\end{itemize}
Inside the HDU the fluxes at the Sun position are saved in units [GeV n$^{-1}$ m$^{-2}$ s$^{-1}$ sr$^{-1}$], while the energy is in [GeV nucleon$^{-1}$].
\item \verb|txt| file with the following column labelling: Kinetic Energy per nucleon [GeV$^{-1}$] (first column) and fluxes at the Sun in [GeV$^{-1}$ m$^{-2}$ s$^{-1}$ sr$^{-1}$] for the various cosmic ray propagated (as labelled in the first raw).
\end{itemize}
The text file has the same degree of informations of the \verb|fits| file only for nucleus with A $>$ 1 if the option \linebreak \verb|<partialstore />| is selected. The two files are however different if in the \dragon card the option \verb|<fullstore />| is selected: the \verb|fits| file saves the flux in all the galaxy in the coordinates $r$ and $z$. 

At the end of its run, \dragon provides the CR fluxes close to Earth, however the solar modulation is not taken into account. This latter can be implemented by the user afterwards by employing either the code \texttt{HelioProp}~\cite{Maccione:2012cu,Vittino:2017fuh,helio} either the force-field approximation~\cite{Boudaud:2014qra,Buch:2015iya}. We do not store a copy of the \dragon card.

\subsection{Output example and explanation for a single run}\label{sec:singlerun}

In this section we consider the spin-0 mediator model introduced in Sec.~\ref{sec:mnvad}.  
Let us assume that the user wants to test the scalar mediator model for the following choice of masses and couplings: $m_{X_d} = 200$ GeV (dark matter mass), $m_{Y_0} = 100$ GeV (mediator mass), $g_{S_X} = 1$ (scalar coupling between $X_d$ and $Y_0$) and $g_{S_q}=1$ (scalar coupling of $Y_0$ to quarks). This can be achieved by typing:\footnote{The coupling and mass names match the one used in the UFO model files for spin 0 mediator in the $s$-channel.}
\begin{verbatim}
   import model DMsimp_s_spin0_MD
   define darkmatter ~xd
   generate relic_density
   add direct_detection
   add indirect_detection 
   output test_y0y0_scalar
   launch test_y0y0_scalar
   indirect=flux_earth
   set sigmav_method madevent
   set indirect_flux_source pythia8
   set nevents 100000
   set MXd 200
   set MY0 100
   set gSXd 1
   set gSu11 1
   set gSu22 1
   set gSu33 1
   set gSd11  1
   set gSd22 1
   set gSd33 1
   set WY0 AUTO
\end{verbatim}

\noindent  The above commands first import the model, then define the dark matter candidate and ask for the desired theoretical predictions: relic density, direct detection and indirect detection. The directory for the project is named \verb|test_y0y0_scalar|. Upon the \verb|launch| command, the two commands \verb|3| set \verb|Indirect=Flux_Earth|, to get the gamma-ray and neutrino fluxes. The choice of \verb|madevent| and \verb|pythia8| means that \maddm will run in the `precise' mode and will compute 100000 events for the annihilation processes, which will be showered and hadronised by \pythia to produce the energy spectra of final stable particles. Finally the last 10 \verb|set| commands fix couplings and masses to the desired values and enforce the automatic evaluation of the $Y_0$ width. 

The output is divided into three blocks: (i) relic density, (ii) direct detection and (iii) indirect detection. 
In the relic density block we provide the value of $\Omega h^2$, of $x_f$, of the thermally averaged cross section at decoupling and of $\xi = (\Omega h^2)_{\rm theo} / (\Omega h^2)_{\rm Planck}$ (defined in the fourth line of the output). The specific tested model point has under-abundant dark matter with $\Omega h^2 = 0.0193$ and $\xi=0.16$, if the dark matter is thermally produced. The output on screen related to the relic density block is:

\begin{figure}[h!]
\centering
\includegraphics[width=1\columnwidth,trim=0mm 0mm 0mm 0mm, clip]{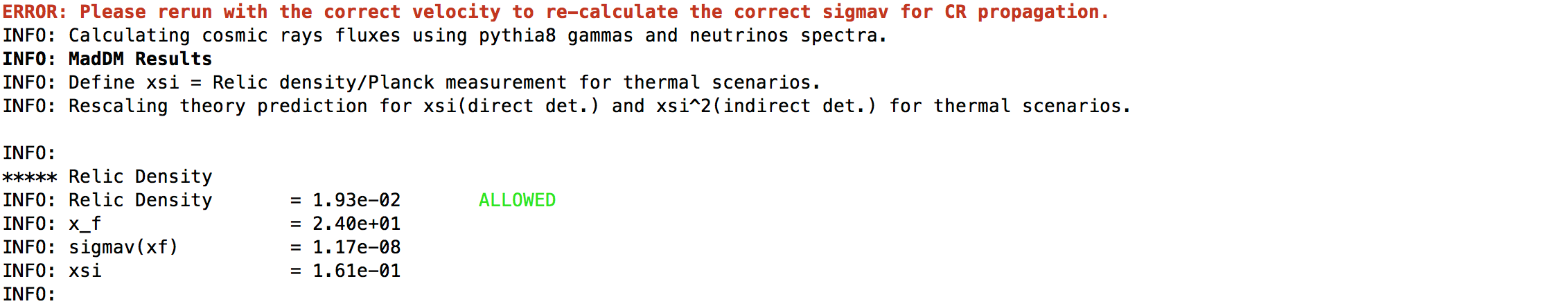}
\end{figure}

The direct detection block provides informations on the elastic cross section for SI scattering of $X_d$ off proton and neutron (given in units of $\rm cm^2$) as labelled. These two values are confronted with the 90\% CL of the XENON1T exclusion limit contained in the \verb|ExpConstraint| class, whose value is explicitly stated in the last column. As defined in Sec.~\ref{sec:rescale} we present the results for two scenarios, the `thermal' scenario and the `all DM' scenario. 

\clearpage
\noindent  The output relative to the direct and indirect detection blocks is:
\begin{figure}[h!]
\centering
\includegraphics[width=1\columnwidth,trim=0mm 0mm 0mm 0mm, clip]{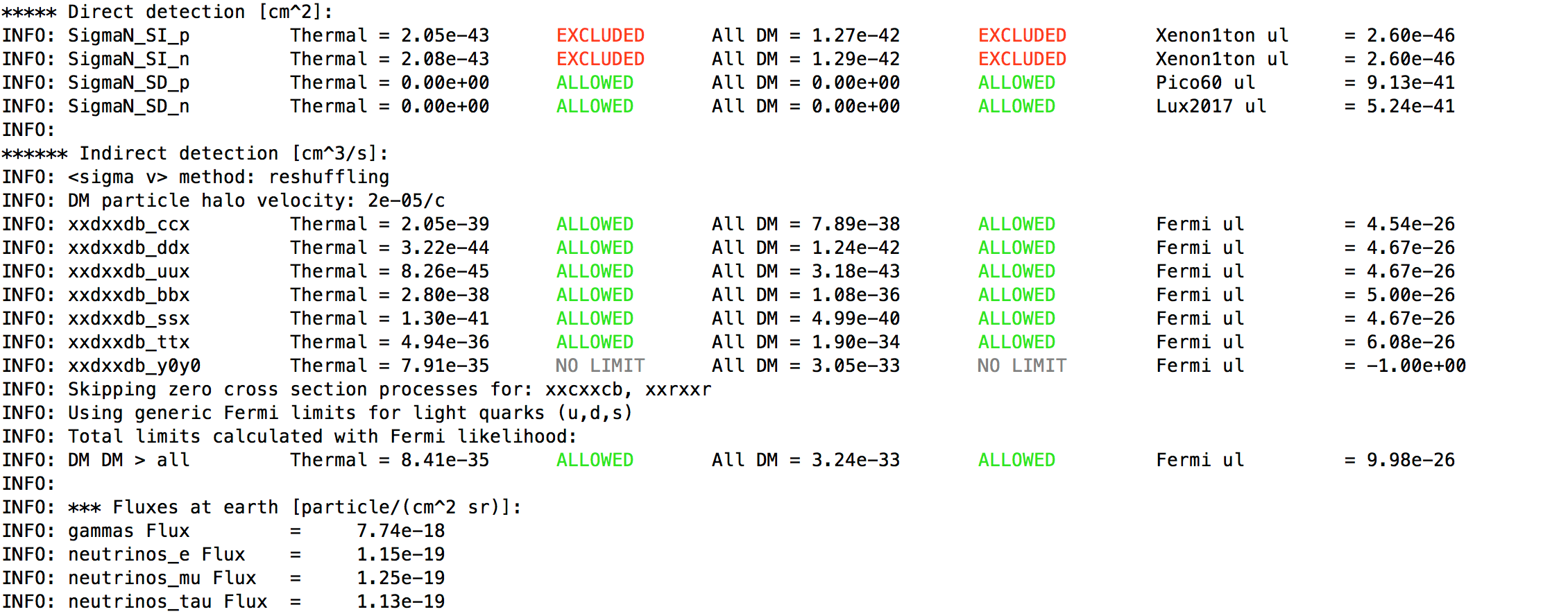}
\end{figure}

\noindent The SD interaction is suppressed by powers of the momentum transferred, hence in this scenarios it is negligible~\cite{Fitzpatrick:2012ix}. In any case, as the user can see, also for SD we do provide the corresponding experimental constraint: the exclusion limit originating from the PICO detector for SD off proton and the upper bound from LUX for SD on neutron.

Concerning the indirect detection block, the output first recaps the method used to compute $\sigmav$ at present time and the relative velocity chosen by the user. In this case the velocity is given for dSPhs.  The output lists the value of all possible partial annihilation cross section computed with $v \equiv v_{\rm rel}$ and the total velocity averaged $\sigmav$. For the `Thermal' scenario, the $\sigmav$ have all been rescaled by $\xi^2$, while the `all DM' scenario implies no rescaling. Since the annihilation of Dirac dark matter into fermions via $s$-channel scalar mediator, or the $t$-channel annihilation into a pair of mediators $Y_0$, are both $p$-wave suppressed, there are no significant bounds from prompt gamma-ray from dSphs, which are displayed  at the end of each line at 95\% CL. Hence this model point is allowed as far as it concerns the Fermi-LAT exclusion limits. Note that for the individual channels pre-computed upper cross sections are displayed (only present for $2\to2$ processes with SM final states) while the limit on the total annihilation cross section is derived via the generated spectrum (containing all annihilation channels) performing the Fermi-LAT likelihood analysis as described in Sec.~\ref{sec:fermilkn}. Finally the last lines give the value of the total integrated flux for prompt photons and of neutrinos normalised to the Draco dwarf spheroidal galaxy $J$ factor.

Notice the error reported in the second and third lines of the output. The predictions for cosmic rays should be done within the Milky Way, hence for $v_{\rm rel} = 220 \rm km/s \simeq 10^{-3}$, which is incompatible with the velocity chosen by the user. Hence the code stops and prevents to run \dragon, unless the user does not provide the correct relative velocity. The value of the fluxes reported at the end should be understood as gamma-ray and neutrino fluxes originating from dSphs and not from the Milky Way halo or the Galactic Center. Conversely, when the user runs the model with the correct velocity for the Milky Way, he/she will be able to run \dragon but the Fermi-LAT exclusion limits will not be displayed as the relative velocity value does not match the typical dSph velocity.

The output file \verb|MadDM_results.txt| stored in \verb|test_y0y0_scalar/output/run_01/| is as follow:

{\tiny 
\begin{verbatim}
#############################################
#                MadDM v. 3.0               #
#############################################


#############################################
# Relic Density                             #
#############################################

Omegah2               = 1.93e-02
Omegah_Planck         = 1.20e-01
xsi                   = 1.61e-01 	 # xsi = (Omega/Omega_Planck)
x_f                   = 2.40e+01
sigmav_xf             = 1.17e-08




#############################################
# Direct Detection [cm^2]                   #
#############################################

SigmaN_SI_p           = [1.27e-42,2.60e-46]   # Xenon1ton
SigmaN_SI_n           = [1.29e-42,2.60e-46]   # Xenon1ton
SigmaN_SD_p           = [0.00e+00,9.13e-41]   # Pico60
SigmaN_SD_n           = [0.00e+00,5.24e-41]   # Lux2017

#############################################
# Indirect Detection [cm^3/s]               #
#############################################

# Annihilation cross section computed with the method: reshuffling 
# Global Fermi Limit computed with pythia8 spectra

xxdxxdb_ccx           = [7.89e-38,4.54e-26]   
xxdxxdb_ddx           = [1.24e-42,4.67e-26]   
xxdxxdb_uux           = [3.18e-43,4.67e-26]   
xxdxxdb_bbx           = [1.08e-36,5.00e-26]   
xxdxxdb_xxcxxcb       = [0.00e+00,-1.00e+00]  
xxdxxdb_ssx           = [4.99e-40,4.67e-26]   
xxdxxdb_ttx           = [1.91e-34,6.08e-26]   
xxdxxdb_xxrxxr        = [0.00e+00,-1.00e+00]  
xxdxxdb_y0y0          = [3.05e-33,-1.00e+00]  
TotalSM_xsec          = [-3.46e-05,9.98e-26]  
Fermi_Likelihood      = -1.76e+01             
Fermi_pvalue          = 3.34e-04              
Fermi_Likelihood(Thermal)= -1.76e+01             
Fermi_pvalue(Thermal) = 5.38e-05              
 
##############################################
# CR Flux at Earth [particles/(cm^2 s sr)]   #
##############################################

# Fluxes calculated using the spectra from pythia8

Flux_neutrinos_e      = 1.15e-19              
Flux_neutrinos_mu     = 1.25e-19              
Flux_neutrinos_tau    = 1.13e-19              
Flux_gammas           = 7.74e-18  
\end{verbatim}}

\noindent This file contains actually the same information as the output screen, however in a format that is easier to use for parsing and plotting. In the direct and indirect detection blocks the cross sections are given as array of numbers: the first number is the theoretical prediction for the model point while the second one is the experimental exclusion limit, as labelled. We additionally provide information on the Fermi-LAT likelihood and on the $p$-value of the tested model point for the case of the `all DM' scenario first and for the `Thermal' scenario. We do not provide anymore any indication about the model point being excluded or allowed by experimental constraints as the user possesses all the necessary informations for its own analysis. The prompt energy spectra are stored in the \verb|test_y0y0_scalar/output/run_01/Output_indirect| folder.

\section{Experimental Constraints}\label{sec:app4}

\subsection{Experimental Constraints class}\label{sec:appexpc}

We have developed the \verb|ExpConstraints| class, which holds information on the relic density constraint, various direct detection constraints on SI and SD cross section and the indirect detection cross-section constraints.

The default constraint on dark matter relic density is set as:
\begin{equation}\label{eq:planck}
	\Omega h^2 = 0.1198 \pm 0.0015.
\end{equation}
from the latest Planck data release~\cite{Ade:2015xua}. The constraint can be changed by executing the following command upon \verb|launch|:
\begin{verbatim}
     set relic_limits <Oh^2 value> <uncertainty>
\end{verbatim}
where \verb| <oh^2 value>| is the new value for the relic density and \verb|<uncertainty>| represents the user specified confidence level of the new $\Omega h^2$ value. 

Direct detection constraints are split into SI and SD cases, and are described by data stored in the \verb|ExpData| folder of \maddm. The default constraints for spin independent cross section is set by the XENON1T exclusion limit~\cite{Aprile:2017iyp}, found in the file \verb|ExpData/Xenont1T_data_2017.dat| (also available in the same folder is the latest LUX 2016 exclusion limit~\cite{Akerib:2016vxi}). Similarly, the default limits on the spin dependent cross section on protons and on neutrons are the PICO-60~\cite{Amole:2017dex} and the LUX~\cite{Akerib:2017kat} exclusion bounds respectively (data files (\verb|ExpData/Pico60_sd_proton.dat| and \verb|ExpData/Lux_2017_sd_neutron.dat|). All direct detection limits are functions of dark matter mass in GeV, and are expressed in ${\rm cm}^2$, and are automatically interpolated between the data points in the files.

Similarly to relic density constraint, the direct detection limits can be modified by executing
\begin{verbatim}
     set dd_si_limits <limit_filename>
     set dd_sd_limits p/n <limit_filename>
\end{verbatim}

Here \verb|<limit_filename>| is the path to the new file which contains the limit and \verb|p/n| signals whether the limit applies to protons or neutrons (in case of SD scattering). The user can also set an observed value for direct detection cross section (in case a signal is ever observed) by executing
\begin{verbatim}
    set dd_si_limits <sigma> <uncertainty>
    set dd_sd_limits p/n <sigma> <uncertainty>
\end{verbatim}
The value of \verb|<sigma>| is assumed to be in ${\rm cm}^2$, and \verb|<uncertainty>| represents the confidence interval of the observed cross section (also in ${\rm cm}^2$). 

The indirect detection experimental constraints are provided for prompt photons and gamma-ray lines generated by dark matter annihilation in dSPhs and in the Galactic center respectively. 
Concerning prompt photons, the default limits on the diffuse components are the Fermi-LAT pass 8 exclusions limits from dSphs galaxies~\cite{Fermi-LAT:2016uux} \linebreak after  6 years  of data taking. These data files are stored in the \verb|ExpData| folder of \maddm and are called \eg \linebreak \verb|ExpData/MadDM_FermiLim_bb.dat|. We provide limits for all SM final states, $gg,\,q\bar{q},\,c\bar{c},\,b\bar{b},\,t\bar{t},\,e^+e^-,\,\mu^+\mu^-,\,\tau^+\tau^-,$ $hh,\,ZZ,\,W^+W^-$; these limits are computed with the Fermi-LAT likelihood as explained in the next section. For dark matter annihilation into a pair of photons (\ie~gamma-ray lines) we provide the Fermi-LAT~\cite{Ackermann:2015lka} and HESS~\cite{Abdalla:2016olq} exclusion limits. The limits strongly depend on the chosen dark matter density profile: we set as default constraint \verb|ExpData/Fermi_lines_2015_Einasto_R16.dat|, which is computed in a region of $16^\circ$ around the Galactic center and is optimised for an Einasto density profile. The HESS bound is given for the same region. Fermi-LAT constraints on other three different regions are also available:
\begin{enumerate}
\item \verb|ExpData/Fermi_lines_2015_NFWcontracted_R3.dat| for a $3^\circ$ region around the Galactic center for the NFW contracted dark matter density profile;
\item \verb|ExpData/Fermi_lines_2015_NFW_R41.dat| for a $41^\circ$ region around the Galactic center for the NFW dark matter density profile;
\item \verb|ExpData/Fermi_lines_2015_Isothermal_R90.dat| for a $90^\circ$ region around the Galactic center for an iso\-thermal dark matter density profile.
\end{enumerate}

Limits on indirect detection can be modified/added using a syntax similar to the one described for direct detection exclusion limits. This is done by the command
\begin{verbatim}
     set id_limits <ave. velocity> <ann. channel> <file path> 
\end{verbatim}  
 in order to set the upper limits. Alternatively, the user can set the observed cross section and the uncertainty using
\begin{verbatim}
     set id_limits <ave. velocity> <ann. channel> <obs. cross section> <obs. uncertainty>
\end{verbatim} 
Units of cm$^3$/s are assumed for the cross-section values/limits, while velocities are in $v/c$ units. \verb|<ave. velocity>| is the average velocity of dark matter in the astrophysical object whose observation yielded the limit (\eg $v\approx 10^{-5}$ for dSph galaxies, $10^{-3}$ for the Galactic center). 

\subsection{Using ExpConstraints outside of MadDM}
Users can also use the \verb|ExpConstraints| class in \verb|python| scripts outside of \maddm, for the purpose of custom analyses. The class can be imported as
\begin{verbatim}
     from maddm_run_interface import ExpConstraints
\end{verbatim}

The user can modify directly the relevant data members of the class:
\begin{itemize}
	\item \verb|_oh2_planck, _oh2_planck_width|: relic density central value and uncertainty.
	\item \verb|_dd_si_limit_file|: location of the file containing the upper limit on SI cross section. The code assumes the first column in the file are values are dark matter mass in GeV while the second column is the upper limit on the cross section in ${\rm cm}^2$.
	\item \verb|_dd_sd_<nucleon>_limit_file|: Same as previous item, but for spin dependent cross section. Here \\ \verb|<nucleon> = proton, neutron|.
	\item \verb|_allowed_final_states|: A list of final states allowed for indirect detection. The default is
	\verb|{`qqx', `gg',|\linebreak
	\verb|`bbx', `ttx', `e+e-', `mu+mu-', `ta+ta-', `w+w-', `zz', `hh', `hess2013',|\;\verb|`hess2016',| \linebreak
        \verb|`aaER16',`aaIR90',`aaNFWcR3',`aaNFWR41'}|. The code will only allow for indirect detection cross-section constraints in the channels defined by this list. The last 6 items of the list are for gamma-ray lines final states and represent the different limits by HESS or Fermi-LAT around the Galactic Center, as described in the previous section.	
	\item \verb|_id_limit_file[<channel>]|: A dictionary of file locations containing upper limits on indirect detection cross section for various annihilation channels. \verb|<channel>| is assumed to fall in the \verb|_allowed_final_states|. The code assumes the first column in the file are values are dark matter mass in GeV while the second column is the upper limit on the cross section in ${\rm cm}^3/{\rm s}$.
	\item \verb|_id_limit_vel[<channel>]|: A dictionary containing the average velocity of dark matter in the astro-physical system from which the limit in the above item is derived. For instance, if the limit on the \verb|bbx| channel comes from Dwarf spheroidal galaxies, the corresponding \verb|_id_limit_vel['bbx'] = 2d-5|.
\end{itemize}
In case a signal of particle dark matter is observed, the \verb|ExpConstraints| also allows to set the observed values via the following class members:
\begin{itemize}
	\item \verb|_sigma_<type>|: Direct detection cross section in cm$^2$. \verb|<type> = SI, SDp, SDn| for spin independent, spin dependent (proton) and spin dependent (neutron) cross sections.
	\item \verb|_sigma_<type>_width|: Uncertainty in the cross-section measurement in cm$^2$.  \verb|<type> = SI, SDp, SDn| for spin independent, spin dependent (proton) and spin dependent (neutron) cross sections.
	\item \verb|_sigma_ID[<channel>]|: Same as \verb|_sigma_<type>| but for indirect detection in the final state \verb|<channel>|. It assumes units of cm$^3$/s.
		\item \verb|_sigma_ID_width[<channel>]|: Same as \verb|_sigma_<type>_width| items but for indirect detection in the final state \verb|<channel>|. It assumes units of cm$^3$/s.
\end{itemize}

Once all constraints are set up, it is necessary to run the \verb|load_constraints()| method on the \verb|ExpConstraints| object in order to load all the data. 
Interpolations of various constraints can be accessed by the member functions:
\begin{itemize}
	\item \verb|SI_max(mdm)|: Returns the value of the upper limit on the spin independent cross section (in cm$^2$) for the dark matter mass \verb|mdm| in GeV.
	\item \verb|SD_max(mdm, nucleon)|: Returns the value of the upper limit on the spin dependent cross section (in cm$^2$) for the dark matter mass \verb|mdm| in GeV. \verb|nucleon = 'n', 'p'| stands for neutrons and protons respectively.
	\item \verb|ID_max(mdm, channel)|: Returns the value of the upper limit on the indirect detection cross section (in cm$^3$/s) for the dark matter mass \verb|mdm| in GeV in the \verb|channel| final state. 
\end{itemize}

\section{Parameter space sampling}\label{sec:app5}

\subsection{Sequential grid scans}\label{sec:appsg}

This functionality has been described in~\ref{sec:app2}; here we provide the output informations and examples. 

The output for sequential grid scans is stored in the \verb|MyDMproject/output| folder (where \verb|MyDMproject| is folder created by \maddm, following the example in Sec.~\ref{sec:running}) and consists in a multi-column file. For instance, if the user requires a grid scan over 25 model points, the output file is called \verb|scan_run_01.txt| and contains $N$ columns, defined as follow:
\begin{itemize}
\item 1$^{\rm {st}}$ column: run\_name, namely 1,2,3...;
\item $(1+n)$ columns: the $n$ variables over which the scan is run. In the example below, $n=1$ as the scan runs over the dark matter mass MXd;
\item (2+n) - (5+n) columns, relic density outputs, if it is generated;
\item (6+n) - (15+n) columns, direct detection outputs, if required;
\item (16+n) - $N$, indirect detection outputs, if it has been also selected. These differ between the `fast' and `precise' running modes.
\end{itemize}
Inside the \verb|MyDMproject/output| folder there are several subfolders, one for each point of the scan. For a scan over 25 points there will be 25 subfolders named \verb|run_01_0i|, with $i=1,25$. In these subfolder the user can find all results that \maddm provides for a single run. If dark matter indirect detection is selected, there are additional outputs in the \verb|run_01_0i/Output_Indirect| folder, described in Sec.~\ref{sec:app3}. 
To understand better the naming of all columns we provide below two examples of sequential grid scans for the `fast' mode and for the `precise' mode respectively.

\paragraph {\it `Fast' Method}
We ask for a very simple sampling over a couple of dark matter mass MXd values:
\begin{verbatim}
   import model DMsimp_s_spin0_MD
   define darkmatter ~xd
   generate relic_density
   add direct_detection
   add indirect_detection 
   output test_y0y0_scalar
   launch test_y0y0_scalar
   indirect=flux_source
   set sigmav_method inclusive
   set indirect_flux_source PPPC4DMID
   set MXd scan:[100,200]
   set MY0 100
   set WY0 AUTO
\end{verbatim}
The output file is called \verb|scan_run_01.txt| and looks like:
{\small
\begin{verbatim}
# [01] : run
# [02] : mass#52   ---> dark matter mass (the scan reports the ID of the parameter card)
# [03] : Omegah^2      
# [04] : x_f
# [05] : sigmav(xf)
# [06] : xsi
# [07] : sigmaN_SI_p  --->  elastic SI cross-section dark matter - proton
# [08] : lim_sigmaN_SI_p  -->  90% CL upper limit for the SI cross-section dark matter - proton (XENON1T)
# [09] : sigmaN_SI_n  ---> elastic SI cross-section dark matter - neutron
# [10] : lim_sigmaN_SI_n ---> 90% CL upper limit for the SI cross-section dark matter - neutron (XENON1T)
# [11] : sigmaN_SD_p  ---> elastic SD cross-section dark matter - proton
# [12] : lim_sigmaN_SD_p  --->  90% CL upper limit for the SD cross-section dark matter - proton (PICO60) 
# [13] : sigmaN_SD_n  ---> elastic SD cross-section dark matter - neutron
# [14] : lim_sigmaN_SD_n  ---> 90% CL upper limit for the SD cross-section dark matter - neutron (LUX) 
# [15] : Nevents
# [16] : smearing
# [17] : xxdxxdb_ccx   ---> sigmav for annihilation into c cbar
# [18] : lim_xxdxxdb_ccx ---> 95% CL upper limit for sigmav from Fermi-LAT dSphs for final state c cbar
# [19] : xxdxxdb_ddx  ---> sigmav for annihilation into d dbar
# [20] : lim_xxdxxdb_ddx ---> 95% CL upper limit for sigmav from Fermi-LAT dSphs for final state d dbar
# [21] : xxdxxdb_uux  ---> sigmav for annihilation into u ubar
# [22] : lim_xxdxxdb_uux ---> 95% CL upper limit for sigmav from Fermi-LAT dSphs for final state u ubar
# [23] : xxdxxdb_ssx  ---> sigmav for annihilation into s sbar
# [24] : lim_xxdxxdb_ssx ---> 95% CL upper limit for sigmav from Fermi-LAT dSphs for final state s sbar
# [25] : xxdxxdb_ttx  ---> sigmav for annihilation into t tbar
# [26] : lim_xxdxxdb_ttx ---> 95% CL upper limit for sigmav from Fermi-LAT dSphs for final state t tbar
# [27] : xxdxxdb_bbx  ---> sigmav for annihilation into b bbar
# [28] : lim_xxdxxdb_bbx ---> 95% CL upper limit for sigmav from Fermi-LAT dSphs for final state b bbar
# [29] : xxdxxdb_y0y0  ---> sigmav for annihilation into  y0 y0
# [30] : lim_xxdxxdb_y0y0 ---> 95% CL upper limit for sigmav from Fermi-LAT dSphs for final state y0 y0 
# [31] : tot_Xsec --> total sigmav
# [32] : tot_SM_xsec ---> total sigmav for annihilation into all SM final states
# [33] : Fermi_sigmav ---> 95% CL Fermi-LAT  upper limit  for annihilation into SM final states
# [34] : pvalue_th --> pvalue for the `thermal' scenario
# [35] : like_th --> likelihood for the `thermal' scenario
# [36] : pvalue_nonth --> pvalue for the `all DM' scenario
# [37] : like_nonth --> likelihood for the `all DM'scenario
# [38] : flux_gammas --> total flux at earth for gamma rays [particle/(cm^2 sr)]
# [39] : flux_neutrinos_e --> total flux at earth for electron neutrinos [particle/(cm^2 sr)]
# [40] : flux_neutrinos_mu --> total flux at earth for muon neutrinos [particle/(cm^2 sr)] 
# [41] : flux_neutrinos_tau --> total flux at earth for tau neutrinos [particle/(cm^2 sr)]

run_01_01     1.00e+02     6.65e-03     2.40e+01     4.31e-08     5.55e-02     3.24e-15
1.37e-46      3.28e-15     1.37e-46     5.16e-35     5.43e-41     5.14e-35     2.77e-41
0.00e+00      0.00e+00     2.63e-41     2.31e-26     4.14e-46     2.39e-26     1.06e-46
2.39e-26      1.66e-43     2.39e-26     0.00e+00     1.00e-15     3.59e-40     2.59e-26        
5.93e-41     -1.00e+00     4.45e-40     3.85e-40     5.15e-26     0.00e+00    -1.76e+01       
1.65e-07     -1.76e+01     1.56e-25     4.96e-26     5.40e-26     4.86e-26
                
run_01_02     2.00e+02     1.93e-02     2.40e+01     1.17e-08     1.61e-01     3.27e-15        
2.60e-46      3.31e-15     2.60e-46     0.00e+00     9.13e-41     0.00e+00     5.24e-41        
0.00e+00      0.00e+00     4.21e-42     4.54e-26     6.63e-47     4.67e-26     1.70e-47        
4.67e-26      2.66e-44     4.67e-26     1.03e-38     6.08e-26     5.76e-41     5.00e-26        
1.63e-37     -1.00e+00     1.73e-37     1.03e-38     1.22e-25     6.73e-08    -1.76e+01       
5.34e-07     -1.76e+01     1.74e-23     5.26e-24     5.73e-24     5.16e-24 
\end{verbatim}
}
\normalsize
We notice that column \#[30] should be $-1$ since this channel is not available from \verb|ExpConstraints| class. Column \#[33] reports the limit calculated with the Fermi likelihood using the gamma-ray combined spectrum, obtained by summing the gamma-ray spectrum for each individual SM channel, scaled by the proper branching fraction. This limit must be compared with the total SM cross-section value in column \#[32]. Columns \#[34-37] reports the $p$-value and likelihood for the tested points, in the thermal and `all DM' scenario respectively.

\paragraph {\it `Precise' Method}
Similarly to the example above, we now run in the `precise' mode:
\small
\begin{verbatim}
   import model DMsimp_s_spin0_MD
   define darkmatter ~xd
   generate relic_density
   add direct_detection
   add indirect_detection 
   output test_y0y0_scalar
   launch test_y0y0_scalar
   indirect=flux_source
   set sigmav_method reshuffling     # (default MadDM option)
   set indirect_flux_source pythia8   # (default MadDM option)
   set nevents 50000
   set MXd scan:[100,200]
   set MY0 100
   set WY0 AUTO
\end{verbatim}
\normalsize
The output file is called \verb|scan_run_01.txt| and looks like: 
\small
{\
\begin{verbatim}
# [01] : run
# [02] : mass#52
# [03] : Omegah^2
# [04] : x_f
# [05] : sigmav(xf)
# [06] : xsi
# [07] : sigmaN_SI_p
# [08] : lim_sigmaN_SI_p
# [09] : sigmaN_SI_n
# [10] : lim_sigmaN_SI_n
# [11] : sigmaN_SD_p
# [12] : lim_sigmaN_SD_p
# [13] : sigmaN_SD_n
# [14] : lim_sigmaN_SD_n
# [15] : Nevents
# [16] : smearing
# [17] : xxdxxdb_ccx
# [18] : lim_xxdxxdb_ccx
# [19] : xxdxxdb_ddx
# [20] : lim_xxdxxdb_ddx
# [21] : xxdxxdb_uux
# [22] : lim_xxdxxdb_uux
# [23] : xxdxxdb_bbx
# [24] : lim_xxdxxdb_bbx
# [25] : xxdxxdb_xxcxxcb
# [26] : lim_xxdxxdb_xxcxxcb
# [27] : xxdxxdb_ssx
# [28] : lim_xxdxxdb_ssx
# [29] : xxdxxdb_ttx
# [30] : lim_xxdxxdb_ttx
# [31] : xxdxxdb_xxrxxr
# [32] : lim_xxdxxdb_xxrxxr
# [33] : xxdxxdb_y0y0
# [34] : lim_xxdxxdb_y0y0
# [35] : tot_Xsec
# [36] : tot_SM_xsec
# [37] : Fermi_sigmav
# [38] : pvalue_th
# [39] : like_th
# [40] : pvalue_nonth
# [41] : like_nonth
# [42] : flux_gammas
# [43] : flux_neutrinos_e
# [44] : flux_neutrinos_mu
# [45] : flux_neutrinos_tau



run_01_01     1.00e+02      6.65e-03      2.40e+01      4.31e-08      5.55e-02      3.24e-15      
1.37e-46      3.28e-15      1.37e-46      5.16e-35      5.43e-41      5.14e-35      2.77e-41      	
0.00e+00      0.00e+00      4.02e-37      2.31e-26      6.34e-42      2.39e-26	     1.62e-42      
2.39e-26      5.49e-36      2.59e-26      0.00e+00      -1.00e+00     2.55e-39      2.39e-26      	
0.00e+00      1.00e-15      0.00e+00	     -1.00e+00     1.45e-40      -1.00e+00     5.90e-36      
-3.46e-05     5.05e-26      1.09e-06      -1.76e+01     1.97e-05      -1.76e+01     2.06e-21      
6.60e-22      7.18e-22      6.46e-22	

run_01_02     2.00e+02      1.93e-02      2.40e+01      1.17e-08      1.61e-01      3.27e-15      
2.60e-46      3.31e-15      2.60e-46      0.00e+00      9.13e-41      0.00e+00	     5.24e-41      
0.00e+00      0.00e+00      7.89e-38      4.54e-26	     1.24e-42      4.67e-26      3.18e-43      
4.67e-26      1.08e-36      5.00e-26      0.00e+00      -1.00e+00     4.99e-40      4.67e-26	      
1.91e-34      6.08e-26      0.00e+00      -1.00e+00	    3.05e-33      -1.00e+00     3.24e-33      
-3.46e-05     4.76e-26      8.02e-05      -1.76e+01     4.98e-04      -1.76e+01     1.22e-17      
1.59e-19      1.73e-19      1.55e-19
\end{verbatim}
}
\normalsize

\subsection{\PyMN sampling of the model parameter space}\label{sec:appmn}

If \PyMN and \MN have been successfully installed by the user\footnote{For details on how to install both \MN and \PyMN we refer the reader to the appropriate websites~\cite{mn,pymn} and references therein.}, which has also added it to the common libraries, \maddm automatically detects \PyMN, without any needed changes within the \maddm setup. In order to run \PyMN, the user needs to turn on the \verb|run_multinest| flag (upon executing the launch command) typing 4 (the default value is OFF):
\begin{verbatim}
| 4. Run Multinest scan | nestscan = ON | OFF | 
\end{verbatim}
and configure the \MN card (by typing 7). For instance the user interface looks like:

\iffalse
\begin{verbatim}
The following switches determine which programs are run:
/============== Description ==========|==== values =======|==== other options =======\
| 1. Compute the Relic Density        |      relic = ON   |   OFF                    |
| 2. Compute direct(ional) detection  |    direct = OFF   |   Please install module  |
| 3. Compute indirect detection/flux  |   indirect = OFF  |   Please install module  |
| 4. Run Multinest scan               |   nestscan = ON   |   OFF                    |
\====================================================================================/
 You can also edit the various input card:
 * Enter the name/number to open the editor
 * Enter a path to a file to replace the card
 * Enter set NAME value to change any parameter to the requested value
 /====================================================\ 
 |  5. Edit the model parameters    [param]           |  
 |  6. Edit the MadDM options       [maddm]           |
 |  7. Edit the Multinest options  [multinest]        |
 =====================================================/
\end{verbatim}
\fi

 \begin{figure}[h]
\centering
\includegraphics[width=1\columnwidth,trim=0mm 0mm 0mm 0mm, clip]{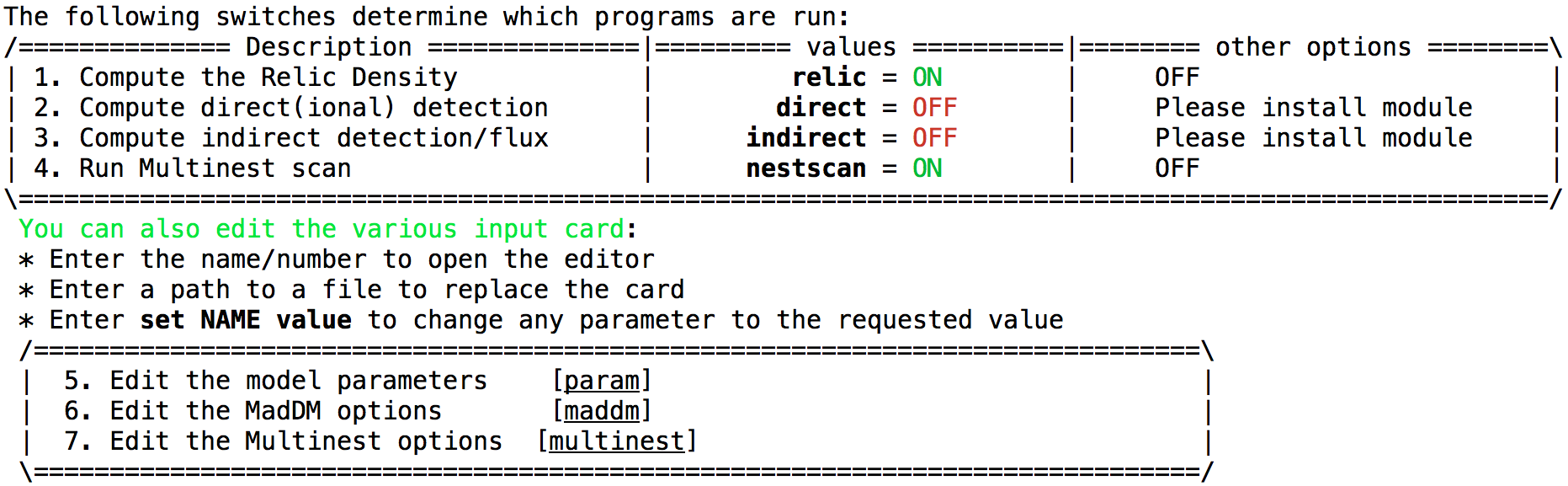}
\end{figure}

The \PyMN card contains the following options:
\begin{itemize}
\item \verb|prefix|: A string which is used to prefix all of the \PyMN output files. It is set to \verb|mnest_| by default.

\item \verb|prior|: Choice of prior distribution functions. By default uniform and log-uniform distributions over the range of parameters defined by the \verb|scan_parameter| command are available, meaning that the parameter space can be sampled according to a prior uniform on the linear scale, or uniform in the log scale. For parameter scans which span many orders of magnitude in dark matter parameters, we recommend using the log-uniform prior. Users can also implement their own choices inside the \verb|myloglikelihood| function of the \MN class.

\item \verb|loglikelihood|: This item sets the likelihood functions used in the \PyMN run for each observable separately (specifically, the code uses the logarithm of the likelihood function). The two implemented options are \verb|gaussian| and \verb|half_gaussian|. Users can add their own in the myloglikelihood function of the \PyMN class. In every case, the code will assume the experimental limits or observations provided in the \verb|ExpConstraints| class. For any observable $\mathcal{O}$ with a measured value  $\hat{\mathcal{O}}$, the two default loglikelihood options are defined as:
\begin{itemize}
\item  \verb|gaussian|:    \,\,\,\,\,\,\,   ${\rm log} (\mathcal{L})  = -\frac{1}{2} ( \mathcal{O} - \hat{\mathcal{O}})^2 / \sigma_\mathcal{O}^2$
\item  \verb|halfgaussian|:  \,\,\,   ${\rm log} (\mathcal{L}) =  \left\{
\begin{array}{cl}
    -\frac{1}{2} \left[ {\rm log}_{10} \left(\frac{\mathcal{O}}{\hat{\mathcal{O}}}\right)/   \sigma^\prime_\mathcal{O} \right]^2 ,  &\,\,\,\,\, \mathcal{O} > {\rm max}( \mathcal{O})_{\rm exp}  \\
           1 , &\,\,\,\,\, \mathcal{O} \leq {\rm max}( \mathcal{O})_{\rm exp} 
\end{array} 
\right. $
\end{itemize}
In case of the Gaussian likelihood the width $\sigma_\mathcal{O}$ is interpreted as the uncertainty in the observation, while in the case of the half gaussian, it is taken as the tolerance of the exclusion bound. In case of \verb|half_gaussian|, the user should specify the log of the gaussian width $\sigma'_\mathcal{O}$ as a parameter.

For instance, the dark matter relic density has been measured with high precision by the Planck satellite, hence for that observable our implementation of \PyMN takes a gaussian likelihood with parameters defined in Eq.~\ref{eq:planck}. On the other hand if the user is interested in dark matter candidates which can also be under abundant the choice of a half-gaussian likelihood function is more appropriate. For exclusion bounds, the default is the half-gaussian likelihood, except for gamma-ray constraints from dSphs. In that case the full Fermi-LAT likelihood function is used.

\item \verb|livepts|: Number of live points to be used by the \PyMN algorithm to estimate the likelihood map.

\item \verb|sampling_efficiency|: This is the efficiency of the sampling algorithm. The default is set to 0.3, which should ensure a good accuracy for parameter inference.

\item \verb|scan_parameter|: sets a parameter to be sampled over within a boundary defined by minimal and maximal value. For instance \verb|scan_parameter m_\chi 10 1000| will set up the parameter $m_\chi$ to be sampled between the boundary of (10-1000) GeV.

\item \verb|output_variables|: sets up observables (relic density, direct detection cross section etc.) which should be printed out by \PyMN. If left blank, the code will print out all of the available observables. 
\end{itemize}
For more details on a given setting the user can also look at the \verb|README| file within the \MN directory.

The output of \PyMN chain will be written in the \verb|multinest_chains| directory of the project folder. The data stored in these files allow the user to run both a Bayesian and a frequentist statistical analysis for model parameter inference.\footnote{If the likelihood function is very flat it might be necessary to set a very low efficiency and tolerance in the \PyMN settings to keep the algorithm running in order to sample efficiently all the parameter space~\cite{Cerdeno:2012ix,Arina:2013jya} for having both the posterior distribution function and for the profile likelihood function.} For instance \PyMN provides already some Python module for easy plotting of the \MN marginalised posterior distribution functions for the parameters under scrutiny. \MN creates several output
files; below we describe the more relevant for the user (\verb|<prefix>| mentioned below is defined in the \MN card, as described above):
\begin{itemize}
\item \verb|<prefix>.txt|: this is chain file to be used for parameter inference as it provides all the information to get the posterior marginalised distributions functions or the profile likelihood functions ( first column: sample probability, second column: $-2 \times$ log-likelihood, other columns: parameter values + additional observable saved);
\item \verb|<prefix>_info.log|: info file containing informations about the \PyMN run, including the column labels for the \verb|<prefix>.txt| file;
\item \verb|<prefix>post_equal_weights.dat|: it contains the equally weighted posterior samples. Columns have parameter values + additional observables values followed by the log-likelihood value. This file can not be used for statistical purposes however is well suited for mapping the allowed model parameter space in 2D or 3D plots.
\item \verb|<prefix>resume.dat|: \PyMN is able to do checkpointing. If this file is created, the program will resume from the last run. Delete this file is you do not want to resume from the past run.
\item \verb|<prefix>physlive.dat|: it contains the current set of live points. When \PyMN is in the burning phase, namely generating the live points, it doesn't write any output file but solely this monitoring file. The user can check here the progresses on the burning phase. When this phase is over the file is not updated anymore and the other output files appear. This file contains nPar+2 columns. The first nPar columns are the ndim parameter values along with the (nPar-ndim) additional observables or parameters that are being passed by the likelihood routine for MultiNest to save along with the additional ndim observables. The nPar+1 column is the log-likelihood value, while the last column is the node number.
\end{itemize}
To have additional information on the monitoring and output files the user is referred to the \verb|README| file within the \MN directory.

%\section*{References}
\bibliographystyle{elsarticle-num.bst} 
\bibliography{biblio}

\end{document}